\documentclass[11pt]{article}
\pdfoutput=1

\usepackage{amssymb,amsmath
}
\usepackage{cite}
\usepackage{graphicx}
\usepackage{slashed}
\usepackage{multirow}
\usepackage{float}
\usepackage{textcomp}
\usepackage{fancyhdr}
\usepackage{tikz}
\usepackage{pstricks}
\usepackage{color}
\definecolor{rosso}{cmyk}{0,1,1,0.4}
\definecolor{rossos}{cmyk}{0,1,1,0.55}
\definecolor{rossoc}{cmyk}{0,1,1,0.2}
\definecolor{blu}{cmyk}{1,1,0,0.3}
\definecolor{blus}{cmyk}{1,1,0,0.6}
\definecolor{bluc}{cmyk}{1,1,0,0.1}
\definecolor{verde}{cmyk}{0.92,0,0.59,0.25}
\definecolor{verdec}{cmyk}{0.92,0,0.59,0.15}
\definecolor{verdes}{cmyk}{0.92,0,0.59,0.4}

\usepackage{hyperref}%
\hypersetup{colorlinks,bookmarksopen,bookmarksnumbered,
linkcolor=blus,pdfstartview=FitH,urlcolor=rossos,citecolor=verde}

\restylefloat{table}

\newcommand{\be}{\begin{equation}}
\newcommand{\ee}{\end{equation}}
\newcommand{\bi}{\begin{itemize}}
\newcommand{\ei}{\end{itemize}}
\newcommand{\bea}{\begin{eqnarray}}
\newcommand{\eea}{\end{eqnarray}}
\newcommand{\bc}{\begin{center}}
\newcommand{\ec}{\end{center}}

\newcommand{\TeV}{\,\mathrm{TeV}}
\newcommand{\GeV}{\,\mathrm{GeV}}

\def\Tr{{\rm Tr\,}}

 \def\La{\Lambda}

\def\({\left(}
\def\){\right)}
\def\be{\begin{equation}}
\def\ee{\end{equation}}
\def\bes{\begin{subequations}}
\def\ees{\end{subequations}}
\def\bea{\begin{eqnarray}}
\def\eea{\end{eqnarray}}
\def\bry{\begin{array}}
\def\ery{\end{array}}
\def\bit{\begin{itemize}}
\def\eit{\end{itemize}}
\def\ben{\begin{enumerate}}
\def\een{\end{enumerate}}
\def\l{\label}

\def\t{\widetilde}

\def\pit{\widetilde{\pi}}

\def\ttil{\widetilde{t}}

\def\dst{\displaystyle}
\def\ovl{\overline}
\def\La{\mathcal{L}}
\def\f{\frac}

\def\demub{\partial_{\mu}}

\def\demua{\partial^{\mu}}

\def\dsl{\slashed{\partial}}

\def\Hd{H^{\dag}}
\def\mst{m_{*}}
\def\ovl{\overline}
\def\<{\langle}
\def\>{\rangle}

\def\O{{\mathcal{O}}}

\def\lsim{\mathrel{\rlap{\lower3pt\hbox{\hskip0pt$\sim$}}
   \raise1pt\hbox{$<$}}}         
\def\gsim{\mathrel{\rlap{\lower4pt\hbox{\hskip1pt$\sim$}}
   \raise1pt\hbox{$>$}}}         

\newcommand{\luv}{\Lambda_{\scriptscriptstyle{UV}}}

\newcommand{\lnp}{\Lambda_{\scriptscriptstyle{NP}}}
\newcommand{\gsm}{g_{\scriptscriptstyle{SM}}}
\newcommand{\tsm}{\widetilde{\rm SM}}
\newcommand{\tH}{\widetilde H}

\makeatletter
\@addtoreset{equation}{section}
\makeatother

\title{
\vspace{-1.5cm}
\vspace{1cm}
\vspace{0.0 cm}
{\huge
{{\bf{Precision Tests and Fine Tuning\\ in Twin Higgs Models}}}
\\ \vspace*{5pt} 
}}
\vspace{1cm}
\author{{\large
\text{Roberto Contino$^{1}$\,}\footnote{roberto.contino@sns.it}
\text{\,,\,\,Davide Greco$^{2}$\,}\footnote{davide.greco@epfl.ch}
\text{\,,\,\,Rakhi Mahbubani$^{3}$\,}\footnote{rakhi.mahbubani@cern.ch}}\vspace{2mm} ,\\ 
{\large
\text{Riccardo Rattazzi$^{2}$\,}\footnote{riccardo.rattazzi@epfl.ch}
\text{ and Riccardo Torre$^{2}$\,}\footnote{riccardo.torre@epfl.ch}} \vspace{4mm}\\ 
{\small\emph{$^{1}$ Scuola Normale Superiore and INFN, Pisa, Italy}}\\ \vspace{-10pt} 
{\small\emph{$^{2}$ Theoretical Particle Physics Laboratory, Institute of Physics, EPFL, Lausanne, Switzerland}}\vspace{4mm}\\ 
{\small\emph{$^{3}$ Theoretical Physics Department, CERN, Geneva, Switzerland}}
}
\date{}

\begin{document}
\begin{titlepage}
\vspace*{-2cm}
\begin{flushright}
CERN-TH-2017-026
\end{flushright}
\begin{center}
\vspace*{15mm}

{\LARGE \bf
Precision Tests and Fine Tuning\\[0.25cm] in Twin Higgs Models} \\
\vspace{1.4cm}

\renewcommand{\thefootnote}{\fnsymbol{footnote}}
{\large
\text{Roberto Contino$^{1}$\,}\footnote{roberto.contino@sns.it}
\text{\,,\,\,Davide Greco$^{2}$\,}\footnote{davide.greco@epfl.ch}
\text{\,,\,\,Rakhi Mahbubani$^{3}$\,}\footnote{rakhi.mahbubani@cern.ch}}\vspace{2mm} ,\\ 
{\large
\text{Riccardo Rattazzi$^{2}$\,}\footnote{riccardo.rattazzi@epfl.ch}
\text{ and Riccardo Torre$^{2}$\,}\footnote{riccardo.torre@epfl.ch}} \vspace{4mm}\\ 
{\small\emph{$^{1}$ Scuola Normale Superiore and INFN, Pisa, Italy}}\\ 
{\small\emph{$^{2}$ Theoretical Particle Physics Laboratory, Institute of Physics, EPFL, Lausanne, Switzerland}}\\
{\small\emph{$^{3}$ Theoretical Physics Department, CERN, Geneva, Switzerland}}

\end{center}

\vspace*{6mm}
\begin{abstract}\noindent\normalsize
We analyze the parametric structure of Twin Higgs (TH) theories and assess the gain in fine tuning which they enable
compared to extensions of the Standard Model with colored top partners.
Estimates show that, at least in the simplest realizations of the TH idea, the separation between the mass of new colored particles and the electroweak scale
is controlled by the coupling strength of the underlying UV theory, and that a parametric gain is achieved only for strongly-coupled dynamics.
Motivated by this consideration we focus on one of these simple realizations, namely composite TH theories, and study 
how  well such constructions can reproduce electroweak precision data.
The most important effect of the Twin states is found to be the infrared contribution to the Higgs quartic coupling, while direct corrections
to electroweak observables are sub-leading and negligible.
We perform a careful fit to the electroweak data including the leading-logarithmic corrections to the Higgs quartic up to three loops.
Our analysis shows that agreement with electroweak precision tests can
be achieved with only a moderate amount of tuning, in the range 5-10\%, in theories where colored states 
have mass of order 3-5 TeV and are thus out of reach of the LHC. 
For these levels of tuning, larger masses are excluded by a perturbativity bound, which makes these theories possibly discoverable, hence falsifiable, 
at a future 100~TeV collider.
\end{abstract}
\end{titlepage}

\newpage
\tableofcontents
\newpage

\section{Introduction}
The principle of naturalness offers arguably the main motivation for exploring physics at around the weak scale. According to naturalness, the plausibility of specific
parameter choices in quantum field theory must be assessed using
symmetries and
selection rules.
When viewing the
Standard Model (SM) as an effective field theory valid below a
physical cut-off scale and considering only the known interactions of
the Higgs boson, we expect the following corrections to its
mass\footnote{We take $m_h^2=2m_H^2=\lambda_hv^2/2$ with $\langle
  H\rangle=v/\sqrt{2}=174$ GeV, which corresponds to a potential  \begin{equation*} V=-m_H^2|H|^2+\frac{\lambda_h}{4} |H|^4\, .\label{convention}\end{equation*}}
\be
\delta m_h^2 =\frac{3 y_t^2}{4\pi^2}\Lambda_t^2 -\,\frac{9 g_2^2}{32 \pi^2}\,\Lambda_{g_2}^2\,-\,\frac{3 g_1^2}{32 \pi^2}\,\Lambda_{{g_1}}^2\,-\frac{3 \lambda_h}{8\pi^2}\Lambda_h^2+\,\dots,
\label{quadraticdiv}
\ee
where each $\Lambda$ represents the physical
cut-off scale in a different sector of the theory.  The above equation is simply dictated by symmetry:
dilatations (dimensional analysis) determine the scale dependence and
the broken shift symmetry of the Higgs field sets the coupling dependence. Unsurprisingly,
these contributions arise in any explicit UV completion of
the SM, although in some cases there may be other larger ones.
According to Eq.~\eqref{quadraticdiv},
any  given (large) value of the scale of new physics can be
associated with a (small) number $\epsilon$, which characterizes the
accuracy at which the different contributions to the mass must cancel
among themselves,
in order to reproduce
the observed value $m_h\simeq 125 \GeV$. 
 As the largest loop factor is due to the top
 Yukawa coupling, according to Eq.~\eqref{quadraticdiv} the scale $\lnp$ where new states must first appear is related to $m_h^2$ and $\epsilon$ via
\be
\lnp^2\sim \frac{4\pi^2}{3 y_t^2} \times \frac{m_h^2}{\epsilon} \quad\Longrightarrow\quad \lnp \sim 0.45 \sqrt{\frac{1}{\epsilon}} \TeV\, .
\label{tuning1}
\ee
The dimensionless quantity $\epsilon$ measures how finely-tuned
$m_h^2$  is, given $\lnp$, and can therefore be regarded as a measure of the tuning. Notice that the contributions from $g_2^2$
and $\lambda_h$ in Eq.~\eqref{quadraticdiv} correspond to $\lnp =
1.1\TeV/\sqrt{\epsilon}$ and $\lnp =1.3\TeV/\sqrt{\epsilon}$,
respectively. Although not significantly different from the relation
in the top sector, these scales would still be large enough to push new states out of  direct reach of the LHC for $\epsilon \sim 0.1$.
 
Indeed, for a given $\epsilon$, Eq.~\eqref{tuning1} only provides an
upper bound for $\lnp$; in the more fundamental UV
theory there can in principle exist larger corrections to $m_h^2$ which are not captured by Eq.~\eqref{quadraticdiv}. In particular, in the Minimal Supersymmetric SM (MSSM) with high-scale
mediation of the soft terms, $\delta m_h^2$ in Eq.~\eqref{quadraticdiv} is
logarithmically enhanced by RG evolution above the weak scale.
 In that case, Eq.~\eqref{tuning1} is modified as follows:
 \be
\lnp^2\sim \frac{2\pi^2}{3 y_t^2} \times \frac{1}{\ln \luv/\lnp}\times \frac{m_h^2}{\epsilon}\,,
\label{tuning2}
\ee
where $\lnp$ corresponds to the overall mass of the stops and
 $\luv\gg \lnp$ is the scale of mediation of the soft terms. However,
 for generic Composite Higgs (CH) models, as well as for supersymmetric
 models with low-scale mediation, Eq.~\eqref{tuning1} provides a fair
 estimate of the relation between the scale of new physics and the
 amount of tuning, the Higgs mass being fully generated by quantum corrections at around the weak scale.  If the origin of $m_h$ is normally termed {\it
   soft} in the MSMM with large $\luv$ (Eq.~\eqref{tuning2}), it should then be termed {\it
   supersoft} in models respecting Eq.~\eqref{tuning1}.  As is well known, and shown by Eq.~\eqref{tuning2}, the natural expectation in the MSSM for 
$\luv \gsim 100 \TeV$ is $\lnp \sim
m_Z\sim m_h$. In view of this, soft scenarios were already somewhat
constrained by direct searches at LEP and Tevatron, whereas the
natural range of the scale of supersoft models is only now being probed at the LHC.

Eq.~\eqref{tuning1} sets an absolute upper bound on $\lnp$ for a given fine
tuning $\epsilon$, but does not give any information on its nature.
In particular it does not specify the quantum numbers of the new
states that enter the theory at or below this scale. Indeed, the most relevant states associated with the top
sector, the so-called top partners, are  bosonic in standard supersymmetric models and fermionic in CH models.  Nonetheless, one common feature of these standard
scenarios is that the top partners carry SM quantum numbers, color in
particular. They are thus copiously produced in hadronic collisions,
making the LHC a good probe of these scenarios. Yet there remains the
logical possibility that the states that are primarily responsible for
the origin of the Higgs mass at or below $\lnp$ are not charged under
the SM, and thus much harder to produce and detect at the LHC.  The Twin Higgs (TH) is
probably the most interesting of the (few) ideas that take this
approach \cite{Chacko:2005pe,Barbieri:2005ri,Chacko:2005vw,Chacko:2005un,Chang:2006ra,Batra:2008jy,Craig:2013aa,Craig:2014aea,Geller:2014kta,Burdman:2014zta,Craig:2014roa,Craig:2015pha,Barbieri:2015aa,Low:2015aa,Curtin:2015bka,Csaki:2015gfd,Craig:2016kue,Harnik:2016koz,Barbieri:2016zxn,Chacko:2016hvu,Craig:2016lyx,Katz:2016wtw}.  This is primarily because the TH mechanism
can, at least in principle, be implemented in a SM extension valid up to ultra-high
scales.  The structure of TH models is such that the states at the threshold $\lnp$ in Eq.~\eqref{tuning1} carry quantum numbers under the gauge group of a copy, a twin, of the SM, but are neutral under the SM gauge group.
These twin states, of which the twin tops are particularly relevant,
are thus poorly produced at the LHC. The theory must also contain
states with SM quantum numbers, but their mass $m_*$ is boosted with
respect to $\lnp$ roughly by a factor $g_*/\gsm$, where $g_*$
describes the coupling strength of the new dynamics, while $\gsm$
represents a generic SM coupling.  As discussed in the next section,
depending on the structure of the model, $\gsm$ can be either the top
Yukawa or the square root of the Higgs quartic. As a result, given the tuning $\epsilon$, the squared mass of the new colored and charged states is roughly given by
\be
m_*^2\sim \frac{4\pi^2}{3 y_t^2} \times \frac{m_h^2}{\epsilon} \times \left (\frac {g_*}{\gsm}\right )^2\, .
\label{tuning3}
\ee
For $g_*> \gsm$, we could define these model as effectively {\it
  hypersoft}, in that, for fixed fine tuning, the gap between the SM-charged states and the weak scale
is even larger than that in supersoft models. In practice the above equation implies that, for strong $g_*$, the new states are out of reach of the LHC even for mild tuning (see Sec.~\ref{sec2.1} for a more precise statement). Eq.~\eqref{tuning3} synthesizes the potential relevance of the TH mechanism, and 
makes it clear that the new dynamics must be rather strong for the
mechanism to work. Given the hierarchy problem, it then seems almost
inevitable to make the TH a Composite TH (although it could also be a Supersymmetric Composite TH). Realizations of the TH mechanism within the paradigm of CH models with fermion partial compositeness \cite{1991NuPhB.365..259K} have already been proposed, both in the holographic and effective theory set-ups \cite{Batra:2008jy,Geller:2014kta,Barbieri:2015aa,Low:2015aa}. 

It is important to recognize that the factor that boosts the mass of the  states with SM gauge quantum numbers in Eq.~\eqref{tuning3} is the coupling $g_*$
itself. Because of this, strong-dynamics effects in the Higgs sector,
which are described in the low-energy theory by non-renormalizable
operators with coefficients proportional to powers of $g_*/m_*$, do
not ``decouple" when these states are made heavier, at fixed fine tuning $\epsilon$.  In the standard parametrics of the CH, $m_*/g_*$ is of the order of  
$f$, the decay constant of the $\sigma$-model within which the Higgs
doublet emerges as a pseudo Nambu-Goldstone Boson (pNGB). Then $\xi\equiv
  v^2/f^2$, as well as being a measure of the fine tuning through $\epsilon=2\xi$, also
measures the relative deviation of the Higgs couplings from the SM
ones, in the TH like in any CH model.\footnote{The factor of two difference between the fine tuning $\epsilon$ and $\xi$ is due to the $Z_{2}$ symmetry of the Higgs potential in the TH models, as shown in Section \ref{sec2.1} \cite{Craig:2013aa}.} Recent Higgs coupling
measurements roughly constrain $\xi\lesssim
10-20\%$ \cite{Khachatryan:2016vau}, and a sensitivity of order $5\%$ is expected in the
high-luminosity phase of the LHC \cite{Thamm:2015zwa}. However Higgs loop effects in
precision $Z$-pole observables measured at LEP already limit $\xi
\lsim 5\%$ \cite{Ciuchini:2013vb,deBlas:2016nqo}. Having to live with this few-percent tuning would somewhat
undermine the motivation for the clever TH construction. In ordinary
CH models this strong constraint on $\xi$ can in
principle be relaxed thanks to compensating corrections to the
$\widehat{T}$ parameter coming from the top partners. In the most natural
models, these are proportional to $y_t^4 v^2/m_*^2$ and thus, unlike the
Higgs-sector contribution, decouple when $m_*$ is increased. This
makes it hard to realize such a compensatory effect in the most
distinctive  range of parameters for TH models, where
$m_*\sim 5-10\TeV$.   Alternatively one could consider including custodial-breaking
couplings larger than $y_t$ in the top-partner sector. Unfortunately
these give rise to equally-enhanced contributions to the Higgs
potential, which would in turn require further ad-hoc cancellations.  

As already observed in the literature \cite{Low:2015aa,Barbieri:2015aa}
another important aspect of TH models is that
calculable IR-dominated contributions to the Higgs quartic coupling
almost saturate its observed value. Though a
welcome property in principle, this sets even stronger constraints on additional UV
contributions, such as those induced by extra sources of custodial
breaking.  In this paper we study the correlation between
these effects, in order to better assess the relevance of the TH construction as a valid alternative to more standard ideas about EW-scale physics.
Several such studies already exist for standard composite
Higgs scenarios \cite{Anastasiou:2009rv,Grojean:2013qca,Ghosh:2015wiz}. In extending these to the TH we shall encounter an
additional obstacle to gaining full benefit from the TH boost in Eq.~\eqref{tuning3}:
the model structure requires rather ``big" multiplets, implying a
large number of degrees of freedom. This results in an upper
bound for the coupling that is parametrically smaller than $4\pi$ by naive dimensional analysis (NDA); hence the boost factor is similarly depressed. We shall discuss in detail how serious and unavoidable a limitation this is.

The paper is organized as follows: in section \ref{sec2} we discuss the general
structure and parametrics of TH models, followed by section \ref{sec3} where we discuss the more specific class
of composite TH models we focus on for the purpose of our study. In
sections \ref{sec:effpot} and \ref{sec:EWPT} we present our computations of the basic physical
quantities: the Higgs potential and precision electroweak parameters
($\widehat S,\, \widehat T,\, \delta g_{Lb}$).  Section \ref{sec:results} is
devoted to a discussion of the resulting constraints on the model and an appraisal of the whole TH scenario. Our conclusions are presented in section \ref{sec7}.

\section{The Twin Higgs scenario}\label{sec2}
\subsection{Structure and Parametrics}\label{sec2.1}

In this section we outline the essential aspects of the TH mechanism. Up to details and variants which are not crucial for the present discussion,
the TH scenario involves an exact duplicate, $\widetilde{\rm SM}$, of the
SM fields and interactions, underpinned by a $Z_2$ symmetry. In practice
this $Z_2$ must be explicitly broken in order to obtain a realistic
phenomenology, and perhaps more importantly, a realistic
cosmology \cite{Barbieri:2016zxn,Chacko:2016hvu,Craig:2016lyx}. However the sources of $Z_2$ breaking can have a structure and size that makes them irrelevant in the discussion of naturalness in electroweak symmetry breaking, which is the main goal of this section.

Our basic assumption is that the SM and its Twin emerge from a more fundamental $Z_2$-symmetric theory at the scale $m_*$, at which new states with SM quantum numbers, color in particular, first appear.
In order to get a feel for the mechanism and its parametrics, it is
sufficient to focus on the most general potential for two Higgs
doublets $H$ and $\tH$, invariant under the gauge group $G_{SM}\times
\widetilde G_{SM}$, with $G_{SM}=SU(3)_c \times SU(2)_{L}\times U(1)_{Y}$, as well as a $Z_2$:
\be
V(H, \widetilde H)\,=\, -m_{\cal H}^2(|H|^2+|\widetilde
H|^2)\,+\,\frac{\lambda_{\cal H}}{4}(|H|^2+|\widetilde H|^2)^2+\,\frac{\hat \lambda_h}{8}(|H|^4+|\widetilde H|^4)\,.
\label{quartic}
\ee
Strictly speaking, the above potential does not have minima with
realistic ``tunable" $\langle H\rangle$.  This goal can be achieved
by the simple addition of a naturally small $Z_2$-breaking mass term
which, while changing the vacuum expectation value, does not affect the estimates of fine tuning, and hence will be
neglected for the purposes of this discussion.  Like for 
the SM Higgs, the most general potential is accidentally invariant
under a custodial $SO(4)\times{\widetilde{SO}(4)}$. Notice however
that in the limit $\hat \lambda_h\to 0$, the additional $Z_2$ enhances
the custodial symmetry to $SO(8)$,  where ${\cal H}\equiv H\oplus
\tH\equiv {\bf 8}$.  In this exact limit, if $\tH$ acquired an
expectation value $\langle \tH\rangle \equiv f/\sqrt{2}$, all 4 components of
the ordinary Higgs $H$ would remain exactly massless NGBs. Of
course the SM and $\tsm$ gauge and Yukawa couplings, along with $\hat
\lambda_h$, explicitly break the $SO(8)$ symmetry that protects the Higgs. Consider
however the scenario where these other couplings, which are known to
be weak, can be treated as small $SO(8)$-breaking perturbations of a
stronger $SO(8)$-preserving underlying dynamics, of which the
quartic coupling $\lambda_{\cal H}$ is a manifestation. In this situation we
can re-assess the relation between the SM Higgs mass, the amount of
tuning and the scale $m_*$ where new states charged under the SM are
first encountered, treating $\hat \lambda_h$ as a small perturbation
of $\lambda_{\cal H}$. At zeroth order, i.e.~neglecting $\hat \lambda_h$, we
can expand around the vacuum $\langle \tH\rangle^2=2m_{\cal
  H}^2/\lambda_{\cal H}\equiv f^2/2$, $\langle H\rangle =0$. The spectrum
consists of a heavy scalar $\sigma$, with mass $m_\sigma=\sqrt{2}m_{\cal H}=\sqrt{\lambda_{\cal H}}f/\sqrt{2}$,
corresponding to the radial mode, 3 NGBs eaten by the Twin gauge bosons, which get masses $\sim g f/2$ and the massless~$H$. When
turning on $\hat \lambda_h$, $SO(8)$ is broken explicitly and $H$
acquires a potential. At leading order in a $\hat
\lambda_h/\lambda_{\cal H}$
expansion the result is simply given by substituting $|\tilde
H|^2=f^2/2-|H|^2$ in
Eq.~\eqref{quartic}.\footnote{Notice that the effective Higgs quartic receives approximately equal contributions from $|H|^4$ and $|\widetilde H|^4$. This is a well-known and interesting property of the TH, see for instance ref.~\cite{Craig:2013aa}.} The quartic coupling
and the correction to the squared mass are then given by
\be 
\lambda_h\simeq \hat \lambda_h \qquad\qquad \delta
m_H^2\sim \hat \lambda_h f^2/8 \simeq ( \lambda_h/2 \lambda_{\cal H})m_{\cal H}^2\, .
\label{m_H}
\ee
As mentioned above, we assume that $m_H^2$ also receives an independent
contribution from a  $Z_2$-breaking mass term, which can be ignored in the estimates of tuning.
Note that in terms of the physical masses of the Higgs, $m_h$,
and of its heavy Twin, $m_\sigma$, we have precisely the same numerical relation
$\delta m_h^2=( \lambda_h/ 2\lambda_{\cal H})m_{\sigma}^2$. The amount of tuning $\epsilon$, defined as $m_h^2/\delta m_h^2$, is given by $\epsilon = 2\xi=
2v^2/f^2$.

Our estimate of $\delta m_H^2$ in Eq.~\eqref{m_H}
  is based on a simplifying approximation where the $SO(8)$-breaking
  quartic is taken $Z_2$-symmetric. In general we could allow
  different couplings $\hat \lambda_h$ and $\hat\lambda_{\tilde h}$
  for $|H|^4$ and $|\tilde H|^4$ respectively, constrained by the requirement $\hat \lambda_h+\hat \lambda_{\tilde h}\simeq  2\lambda_h$. As the estimate of $\delta m_H^2$ in Eq.~\eqref{m_H} is determined by the $|\tilde H|^4$ term, it is clear that a reduction of $\hat \lambda_{\tilde h}$, with $\lambda_h$ fixed, would improve the tuning, as emphasized in Ref.~\cite{Katz:2016wtw}. As discussed in Section \ref{sec:effpot}, however, a significant fraction of the contribution to
 $\hat \lambda_h$ and $\hat \lambda_{\tilde h}$ is coming from  RG
 evolution due to the top and Twin top. According to our analysis, $\hat \lambda_h/\hat \lambda_{\tilde h}$ varies between $1.5$ in the simplest models to $3$ in models
 where $\hat \lambda_{\tilde h}$ is purely IR-dominated as in in
 Ref.~\cite{Katz:2016wtw}. Though interesting, this gain does not
 change our parametric estimates.

The ratio $\lambda_h/\lambda_{\cal H}$ is the crucial parameter in the game. Indeed it is through Eq.~\eqref{m_H} that $m_H$ is sensitive to quantum corrections to the Lagrangian mass parameter $m_{\cal H}$, or, equivalently, that the physical Higgs mass $m_{h}$ is sensitive to the physical mass of the radial mode $m_{\sigma}$. In particular, what matters is the correlation of $m_{\sigma}$ with, and its  sensitivity to, $m_*$, where new states with SM quantum numbers appear. One can
 think of three basic scenarios for that relation, which we now illustrate, ordering them by increasing level of model-building ingenuity. Beyond these scenarios there is the option of tadpole-dominated electroweak symmetry breaking, which we shall briefly discuss at the end.

\subsubsection{Sub-Hypersoft Scenario}\label{sec:sub-hypersoft}
\noindent{The simplest} option is given by models with $m_{\sigma}\sim
m_*$. Supersymmetric TH models with medium- to high-scale soft-term
mediation belong to this class \cite{Craig:2013aa}, with $m_*$ representing the soft mass of the squarks. Like in the
MSSM, $m_{\cal H}$, and therefore $m_{\sigma}$, is generated via RG evolution: two decades of
running are sufficient to obtain $m_{\sigma} \sim m_{*}$. Another
example is composite TH models \cite{Low:2015aa,Barbieri:2015aa}. In their  simplest incarnation they are characterized by one overall mass scale $m_*$ and coupling $g_*$
\cite{Giudice:1024017}, so that by construction one has $m_{\sigma}
\sim m_{*}$ and $\lambda_{\cal H}\sim g_*^2$. As discussed  below
Eq.~\eqref{m_H}, in both these scenarios one then expects
 $\delta m_h^2 \sim (\lambda_h/ 2\lambda_{\cal H} )m_*^2$. It is interesting to compare this result to the leading top-sector contribution in Eq.~\eqref{quadraticdiv}. For that purpose it is worth noticing that, as discussed in Section \ref{sec:effpot}, in TH models the RG-induced contribution to the Higgs quartic coupling
$\Delta \lambda_h |_{\scriptscriptstyle {RG}}\sim (3y _t^4/\pi^2) \ln
m_*/m_t $ (more than) saturates its experimental value $\lambda_h\sim 0.5$
for $m_*\sim 3-10 \TeV$.~\footnote{For this naive estimate we have taken the Twin-top contribution equal to the top one, so that the result is just twice the SM one. For a more precise statement see Section \ref{sec:effpot}.}
We  can thus  write
\be
\delta m_h^2 \sim (\lambda_h/ 2\lambda_{\cal H} )m_*^2\sim \frac{3 y_t^4}{2\pi^2}\frac{1}{\lambda_{\cal H}} \ln (m_*/m_t)\, m_*^2\equiv \frac{3 y_t^2}{2\pi^2}\times \frac{y_t^2}{g_*^2}\times  \ln (m_*/m_t)\times  m_*^2
\label{eq:mh-subhyper}
\ee
which should be compared to the first term on the right-hand side of Eq.~\eqref{quadraticdiv}. Accounting for the possibility of tuning we then have
\be
m_*\,\,\sim\,\, 0.45 \times\frac{g_*}{\sqrt{2}y_t}\times\sqrt {\frac{1}{\ln (m_*/m_t)}}\times\sqrt{\frac{1}{\epsilon}}\,\,\TeV\,.
\label{boost1}
\ee
Compared to Eq.~\eqref{tuning1}, the mass of colored states is on one hand parametrically boosted by the ratio $g_*/(\sqrt{2}y_t)$, and  on the other it is mildly decreased by the logarithm. The motivation for, and gain in, the ongoing work on the simplest realization of the TH idea pivot upon the above $g_*/y_t$. The basic question is how high $g_*$ can be pushed without leading to a breakdown of the effective description.  One goal of this paper is to investigate to what extent
one can realistically gain from this parameter in more explicit CH realizations. Applying naive dimensional analysis (NDA) one would be tempted to say that $g_*$ as big as $\sim 4\pi$ makes sense, in which  case  $m_*\sim 10\TeV$ would only cost  a mild $\epsilon \sim 0.1$ tuning. However  such an estimate seems  quantitatively too naive. For instance,  by focusing on the simple toy model whose potential is given by Eq.~\eqref{quartic}, we can associate the upper bound on $\lambda_{\cal H}\equiv g_*^2$, to the point where perturbation theory breaks down. One possible way to proceed is to consider 
 the one loop beta function
\be
\mu \frac{d\lambda_{\cal H}}{d\mu}= \frac{N+8}{32\pi^2} \lambda_{\cal H}^{2}\,,
\label{eq:betafun}
\ee
and to estimate the maximum value of the coupling $\lambda_{\cal H}$ as that for which $\Delta\lambda_{\cal H}/\lambda_{\cal H}\sim O(1)$ through one e-folding of RG evolution. 
We find 
\be
\lambda_{\cal H}=\frac{2\, m_{\sigma}^{2}}{f^{2}}\lsim {32\pi^2 \over N+8}\quad\Longrightarrow \quad \frac{m_{\sigma}}{f}\lsim\pi\,, \text{ for }N=8\,\,,
\label{boundsigma}
\ee
which also gives $g_*\sim \sqrt{\lambda_{\cal H}}\lsim \sqrt{2}\pi$, corresponding to a significantly smaller maximal gain in Eq.~\eqref{boost1} with respect to the NDA estimate. In Section \ref{sec:perturbativity} we shall perform alternative estimates in more specific CH constructions, obtaining similar results. 

It is perhaps too narrow-minded to stick rigidly to such estimates to 
determine the boost that $g_*/(\sqrt{2}y_t)$ can give to $m_*$. Although it is parametrically true that the stronger the coupling $g_*$, the heavier the colored partners can be at fixed tuning, the above debate over factors of a few make it difficult to be more specific in our estimates. In any case the gain permitted by Eq.~\eqref{boost1} is probably less than one might naively have hoped, making it fair to question the motivation for the TH, at least in its ``sub-hypersoft'' realization. With this reservation in mind, we continue our exploration of the TH in the belief that the connection between naturalness and 
LHC signatures is so crucial that it must be analyzed in all its possible guises.

Concerning in particular composite TH scenarios one last important model building issue concerns the origin of the Higgs quartic $\lambda_h$.  In generic CH
 it is known that the contribution to $\lambda_h$ that arises at $O(y_t^2)$   is too large when $g_*$ is strong. Given that the TH mechanism demands $g_*$ as strong as possible
then composite TH models must ensure that the leading $O(y_t^2)$ contribution is absent so that $\lambda_h$ arises at $O(y_t^4)$. As discussed in Ref. \cite{Barbieri:2015aa}, this property is not guaranteed
but it can be easily ensured provided the couplings that give rise to $y_t$ via partial compositeness respect specific selection rules.

\subsubsection{Hypersoft scenario}
The second option corresponds to the  structurally robust situation where $m_{\sigma}^2$ is one loop factor smaller than $m_*^2$.
This is for instance  achieved if  ${\cal H}$ is a PNG-boson octet multiplet associated to the spontaneous breaking $SO(9)\to SO(8)$ in a model with fundamental scale $m_*$. Another option would be to have a supersymmetric model where supersymmetric masses of order $m_*$ are mediated to the stops at the very scale $m_*$ at which $\cal H$ is massless. Of course in both cases a precise computation of $m_{\sigma}^2$ would require the full theory. However a parametrically correct estimate can be given by considering the quadratically divergent 1-loop corrections in the low energy theory, in the same spirit of Eq.~\eqref{quadraticdiv}. As $y_t$ and $\lambda_{\cal H}$ are expected to be the dominant couplings the analogue of Eqs.~\eqref{quadraticdiv} and \eqref{m_H} imply
\be
\delta m_h^2 \sim \frac{\lambda_h}{2\lambda_{\cal H}} \left( \frac{3y_t^2}{4\pi^2}+\frac{5\lambda_{\cal H}}{16\pi^2}\right ) m_*^2=\left (\frac{y_t^2}{\lambda_{\cal H}}+\frac{5}{12}\right )\frac{3\lambda_h}{8\pi^2} m_*^2\, .
\label{fraternal}
\ee
Very roughly, for $\lambda_{\cal H}\gsim y_t^2$, top effects become sub-dominant and the natural value for    $m_h$ becomes
 controlled by $\lambda_h$, like the term induced by the Higgs quartic in  Eq.~\eqref{quadraticdiv}. In the absence of tuning this roughly corresponds to the technicolor limit $m_* \sim 4\pi v$, while allowing for fine tuning we have
\be
m_*\sim 1.4 \times \sqrt {\frac{1}{\epsilon}} \,\TeV\, .
\label{1.4tev}
\ee
It should be said that in this scenario there is no extra boost of $m_*$ at fixed tuning by taking $\lambda_{\cal H}> y_t^2$. Indeed the choice $\lambda_{\cal H}\sim y_t^2$ is preferable as concerns electroweak precision tests (EWPT). It is well known that RG evolution in the effective theory below $m_{\sigma}$ gives rise to corrections to the $\widehat S$ and $\widehat T$ parameters as discussed in Section \ref{sec:EWPT} \cite{Barbieri:2007bh}. In view of the relation $\epsilon = 2v^2/f^2$ this gives a direct connection between
fine-tuning electroweak precision data, and the mass of the Twin Higgs $m_{\sigma}$. At fixed $v^2/f^2$, EWPT then favor the smallest possible $m_{\sigma} = \sqrt \lambda_{\cal H} f/\sqrt{2}$, that is the smallest $\lambda_{\cal H}\sim y_t^2$. The most plausible spectrum in this class of models is roughly the following: the Twin scalar $\sigma$ and the Twin tops appear around the same scale $\sim y_t f /\sqrt{2}$, below the colored partners who live at $m_*$. The presence of the somewhat light scalar $\sigma$ is one of interesting features of this class of models.

\subsubsection{Super-Hypersoft Scenario}
This option is a clever variant of the previous one, where below the scale $m_*$ approximate supersymmetry survives in the Higgs sector
in such a way that the leading contribution to $\delta m_{\cal H}^2$  proportional to $\lambda_{\cal H}$ is purely due to the top sector \cite{Craig:2013aa}. In that way Eq.~\eqref{fraternal} reduces to
\be
\delta m_h^2 \sim \frac{\lambda_h}{2\lambda_{\cal H}} \left( \frac{3y_t^2}{4\pi^2}\right ) m_*^2=\frac{y_t^2}{\lambda_{\cal H}}\times \frac{3\lambda_h}{8\pi^2} m_*^2\, .
\label{super}
\ee
so that by choosing $g_*> y_t$ one can push the scale $m_*$ further up with fixed fine tuning $\epsilon$
\be
m_*\sim 1.4 \times \frac{g_*}{\sqrt{2}y_t}\times \sqrt {\frac{1}{\epsilon}} \,\TeV\, .
\label{highest}
\ee
In principle even under the conservative assumption that $g_*\sim \sqrt{2}\pi$ is the maximal allowed value, this scenario seemingly allows $m_*\sim 14\TeV$ with a mild $\epsilon\sim 0.1$ tuning.

It should be said that in order to realize this scenario one would need to complete $\cal H$ into a pair of chiral superfield octets ${\cal H}_u$ and $
{\cal H}_d$, along the lines of Ref.~\cite{Craig:2013aa}, as well as add a singlet superfield $S$ in order to generate the Higgs quartic via the superpotential trilinear $g_* S {\cal H}_u {\cal H}_d$.
Obviously this is a very far-fetched scenario combining all possible ideas to explain the smallness of the weak scale: supersymmetry, compositeness and the Twin Higgs mechanism.

\subsubsection{Alternative vacuum dynamics: tadpole induced EWSB}
In all the scenarios discussed so far the tuning of the Higgs vacuum expectation value (\textsc{vev}) and
that of the Higgs mass coincided: $\epsilon $, which controls the
tuning of $m_h^2$ according to Eqs.~\eqref{boost1}, \eqref{1.4tev} and
\eqref{highest}, is equal to $2v^2/f^2$, which measures the tuning of
the \textsc{vev}. This was because the only tuning in the Higgs potential was
associated with the small quadratic term, while the quartic was
assumed to be of the right size without the need for further
cancellations (see e.g. the discussion in
Ref.\cite{DeSimone:2012ul}). Experimentally however, one can
distinguish between the need for tuning that originates from
measurements of Higgs and electroweak observables, which are
controlled by $v^2/f^2$, and that coming from direct searches for top
partners. Currently, with bounds on colored top partners at just
around 1 TeV \cite{ATLAS:2016qlg,CMS:2016dmr}, but with Higgs couplings already bounded to lie
within $10-20\%$ of their SM value \cite{Khachatryan:2016vau}, the only reason for tuning in all
TH scenarios is to achieve a small $v^2/f^2$. It is then fair
to consider options that reduce or eliminate only the tuning of
$v^2/f^2$. As argued in Ref.~\cite{Harnik:2016koz}, this can be
achieved by modifying the $H$ scalar vacuum dynamics, and having its
\textsc{vev} induced instead by a tadpole mixing with an additional
electroweak-breaking technicolor (TC) sector
\cite{Kagan,Galloway:2013dma,Chang:2014ida}. In order to preserve the
$Z_2$ symmetry one adds two Twin TC sectors, both characterized by a mass scale
$m_{TC}$ and a decay constant $f_{TC}\sim  m_{TC}/4\pi$ (i.e.~it is parametrically convenient to assume $g_{TC}\sim 4\pi$). Below the TC scale the dynamics in the visible and Twin sectors is complemented by Goldstone triplets $\pi_a$ and $\tilde \pi_a$ which can be embedded into doublet fields according to
\be
\Sigma =f_{TC} e^{i\pi_a \sigma_a}\left(\begin{array}{c}
0 \\1 \end{array}\right)\,,\qquad\qquad
\tilde \Sigma =f_{TC} e^{i\tilde \pi_a \sigma_a}\left(\begin{array}{c}
0 \\1
\end{array}\right)\,,
\label{png}
\ee
and are assumed to mix with $H$ and $\tilde H$ via the effective potential terms
\be
V_{\text{tadpole}} = M^2 (H^\dagger \Sigma + \tilde H^\dagger \tilde \Sigma)+{\rm {h.c.}}\, .
\ee
Assuming $m_{TC}\ll m_{\cal H}$ the $\tH$ vacuum dynamics is not significantly  modified, but, for  $m_{TC}> m_h$,  $V_{\text{tadpole}}$ acts like a rigid
tadpole term for $H$. The expectation value $\langle H \rangle $ is thus determined by balancing such a tadpole against the gauge-invariant  $|H|^2$
mass term; the latter will then roughly coincide with $m_h^2$.  In
order for this to work, by Eq.~\eqref{m_H} the $SO(8)$-breaking
quartic $\hat \lambda_h$ should be negative, resulting in $v\sim
(M^2/m_h^2) f_{TC}$. It is easy to convince oneself that the corrections to Higgs couplings are $O(f_{TC}^2/v^2)$: present bounds can then be satisfied for $f_{TC}\sim v/\sqrt{10}\simeq 80$ GeV. In turn, the value of $v/f$ is controlled by $f$ and can thus be naturally small.
The TC scale is roughly $m_{TC}\sim 4\pi f_{TC}\sim 600-800$ GeV,  while the non-eaten pNGB $\pi$ in Eq.~\eqref{png}
have  a mass $m_\pi^2\sim M^2 v/f_{TC}\sim m_h^2(v/f_{TC})^2\sim 400$
GeV. The latter value, although rather low, is probably large enough to satisfy constraints from direct searches. In
our opinion, what may be  more problematic are EWPT, in view of the effects from the TC sector, which shares some of the vices of ordinary TC.
The IR contributions to $\widehat S$ and $\widehat T$, associated with the splitting  $m_{\pi_a}< m_{TC}$, are here smaller
than the analogues of ordinary technicolor (there associated with the splitting $m_W\ll m_{TC}$). However the UV
 contribution to $\widehat{S}$ is parametrically the same as in ordinary TC, in particular it is enhanced at large $N_{TC}$.
 Even at $N_{TC}=2$, staying within the
allowed $(\widehat{S},\widehat{T})$ ellipse still requires a correlated
contribution from $\Delta \widehat{T}$, which in principle should also be counted as  tuning. 
In spite of this, models with tadpole-induced EWSB represent a clever
variant where, technically, the dynamics of EWSB does not currently appear tuned. A thorough analysis of the constraints is certainly warranted.

\section{The Composite Twin Higgs}\label{sec3}

In this section and in the remainder of the paper, we will focus on
the CH realization of the TH, which belongs to the
sub-hypersoft class of models.  
In this simple and well-motivated context we shall discuss EWPT, fine
tuning and structural consistency of the model.

Our basic structural assumption is that at a generic UV scale $\Lambda_{UV} \gg m_*$, our theory can be decomposed into two sectors: a strongly-interacting Composite Sector and a weakly-interacting Elementary Sector. The Composite Sector is assumed to be endowed with the global symmetry $G=SO(8) \times U(1)_X \times Z_2$ and to be approximately scale- (conformal) invariant down to the  scale $m_*$, at which it develops a mass gap. 
We assume the overall interaction strength at the resonance mass scale
$m_*$ to be roughly described by one parameter $g_*$ \cite{Giudice:1024017}.
The large separation of mass scales $\Lambda_{UV}\gg m_*$ is assumed to arise naturally, in that the occurrence of the mass gap $m_*$ is controlled by 
either a marginally-relevant deformation, or by a relevant deformation
whose smallness is controlled by some global symmetry.
 At the scale $m_*$, $SO(8) \times U(1)_X \times Z_2$ is spontaneously broken to the
 subgroup $H=SO(7) \times U(1)_X$, giving rise to seven NGBs in the $\bf{7}$ of $SO(7)$ with decay constant $f\sim m_*/g_*$. 
The subgroup $U(1)_X$ does not participate to the spontaneous breaking, but its presence is needed to reproduce the hypercharges of the SM fermions, similarly to CH models.
  The Elementary Sector consists in turn of two separate
  weakly-interacting sectors: one containing the visible SM fermions
  and gauge bosons, corresponding to the SM gauge group $G_{SM} = SU(3)_{c}\times SU(2)_{L}\times U(1)_{Y}$; the other containing the Twin SM with the same fermion
  content and a $\widetilde{SM}$ gauge group $\widetilde G_{SM} = \widetilde{SU}(3)_{c}\times
  \widetilde{SU}(2)_{L}$. The external $Z_2$ symmetry, or Twin parity,
  interchanges these two copies.  For simplicity, and following~\cite{Barbieri:2015aa}, we choose not to
  introduce a mirror hypercharge field.  This is our only source of
  explicit Twin-parity breaking, and affects neither our discussion of
  fine tuning, nor that of precision electroweak measurements.

The Elementary and Composite sectors are coupled according to the paradigm of partial compositeness \cite{1991NuPhB.365..259K}. The elementary EW gauge bosons couple to the strong dynamics as a result of the weak gauging of the $SU(2)_{L} \times U(1)_Y \times \widetilde{SU}(2)_{L}$ subgroup of the global $SO(8)\times U(1)_X$. 
A linear mixing with the global conserved currents is thus induced:
\begin{equation}\label{MixLagrV}
\mathcal{L}^V_{\text{mix}}\supset g_2\, W_\mu^\alpha J^\mu_\alpha + g_1 \, B_\mu J_B^\mu + \widetilde{g}_2\, \widetilde{W}_\mu ^\alpha \widetilde{J}^\mu_\alpha,
\end{equation}
where $g_{1,2}$ and $\widetilde{g}_2$ denote the SM and Twin weak gauge couplings, $J_B^\mu \equiv J_{3R}^\mu + J_X^\mu$ and $J^\mu$, $\tilde J^\mu$ and
$J_X^\mu$ are the currents associated respectively to the $SU(2)_L$, $\widetilde{SU}(2)_L$ and $U(1)_X$ generators. The elementary fermions mix analogously with various
operators transforming as linear representations of $SO(8)$ that are generated in the far UV by the strongly-interacting dynamics.
The mixing Lagrangian takes the schematic form:
\begin{equation}\label{MixLagrF}
\mathcal{L}^F_{\text{mix}} \supset \bar{q}_L^\alpha \Delta_{\alpha A}\mathcal{O}_R^A+ \bar{t}_R\Theta_{ A}\mathcal{O}_L^A + \bar{\widetilde{q}}_L^\alpha \widetilde{\Delta}_{\alpha A}\mathcal{\widetilde{O}}_R^A+  \bar{t}_R\widetilde{\Theta}_{ A}\mathcal{\widetilde{O}}_L^A +\text{h.c.},
\end{equation}
where, following e.g.~Ref.~\cite{Mrazek:2011iu},  we introduced spurions $\Delta_{\alpha A}$, $\widetilde{\Delta}_{\alpha A}$, $\Theta _A$ and $\widetilde{\Theta}_A$ in order to 
uplift the elementary fields to linear representations of $SO(8)$, and match
the quantum numbers of the composite operators. The left-handed
mixings $\Delta_{\alpha A}$, $\widetilde{\Delta}_{\alpha A}$
necessarily break $SO(8)$ since $q_L$ only partially fills a multiplet
of $SO(8)$. The right-handed mixings, instead, may or may not break
$SO(8)$. The breaking of $SO(8)$ gives rise to a potential for the
NGBs at one loop and the physical Higgs is turned into a
pNGB. We conclude by noticing that $g_{1,2}$ and $\tilde g_2$ correspond to
quasi-marginal couplings which start off weak in the UV,  and remain
weak down to $m_*$. The fermion mixings could be either relevant or
marginal, and it is possible that some may correspond to interactions that grow as strong as $g_*$ at the IR scale $m_*$~\cite{Contino:2004cj}. 
In particular, as is well known, there is some advantage as regards tuning in considering the right mixings $\Theta _A$ and $\widetilde{\Theta}_A$ to be strong. 
In that case one may even imagine the IR scale
to be precisely generated by the corresponding deformation of the fixed point. While this latter option may be interesting from a top-down perspective, it would play no appreciable role in our low-energy phenomenological discussion.

\subsection{A simplified model }\label{sec:EffLag}

In order to proceed we now consider a specific realization of the
composite TH and  introduce a concrete simplified effective
Lagrangian description of its dynamics. Our model captures the most
important features of this class of theories, like the pNGB nature of
the Higgs field, and provides at the same time a simple framework for
the interactions between the elementary fields and the composite
states, vectors and fermions. We make use of this effective model
as an example of a specific scenario in which we can compute EW
observables, and study the feasibility of the TH idea as a new paradigm for physics at the EW scale. 

We write down an effective Lagrangian for the Composite TH model using the Callan-Coleman-Wess-Zumino (CCWZ) construction \cite{Coleman:1969p1798, Callan:1969p1799}, and generalizing the simpler case of a two-site model developed in Ref.~\cite{Barbieri:2015aa}. According to the CCWZ technique, a Lagrangian invariant under the global $SO(8)$ group can be written following the rules of a local $SO(7)$ symmetry. The basic building blocks are the Goldstone matrix $\Sigma({\bf{\Pi}})$, which encodes the seven NGBs, ${\bf{\Pi}}$, present in the theory, and the operators $d_\mu({\bf{\Pi}})$ and  $E_\mu({\bf{\Pi}})$ resulting from the Maurer-Cartan form constructed with the Goldstone matrix. An external $U(1)_X$ group is also added to the global invariance in order to reproduce the correct fermion hypercharges \cite{Barbieri:2015aa}. The CCWZ approach is reviewed and applied to the $SO(8)/SO(7)$ coset in Appendix \ref{AppCCWZ}.

Before proceeding, we would like to recall the simplified model philosophy
 of Ref.~\cite{Contino:2011np}, which we essentially employ. In a generic composite theory, the mass scale $m_*$ would control both the cut-off of the low energy $\sigma$-model and the mass of the resonances. In that case no effective Lagrangian method is expected to be applicable to describe the resonances. So, in order to produce a manageable effective Lagrangian we thus consider a Lagrangian for resonances that can, at least in principle, be made lighter that $m_*$. One more structured way to proceed could be to consider a deconstructed extra-dimension where the mass of the lightest resonances, corresponding to the inverse compactification length, is parametrically separated from the 5D cut-off, interpreted as $m_*$.
Here we do not go that far and simply consider a set of resonances that happen to be a bit lighter than $m_*$. We do so to
give a structural dignity to our effective Lagrangian, though at the end, for our numerical analysis, we just take the resonances a factor of 2 below $m_*$. We believe that is a fair procedure given our purpose of estimating the parametric consistency of the general TH scenario.

We start our analysis of the effective Lagrangian with the bosonic sector. Together with the elementary SM gauge bosons, the $W$'s and $B$, we introduce the Twin partners $\widetilde{W}$ to gauge the $\widetilde{SU}(2)_L$ group. As representative of the composite dynamics, we restrict our interest to the heavy spin-1 resonances transforming under the adjoint of $SO(7)$ and to a vector singlet. We therefore introduce a set of vectors $\rho_\mu ^a$ which form a ${\bf{21}}$ of $SO(7)$ and the gauge vector associated with the external $U(1)_X$, which we call $\rho^X_\mu$. The Lagrangian for the bosonic sector can be written as
\be\label{LBosonic}
\mathcal{L}_{\text{bosonic}} = \mathcal{L}_{\pi}+\mathcal{L}^V_{\text{comp}} + \mathcal{L}^V_{\text{elem}} + \mathcal{L}^V_{\text{mix}}\,.
\ee
The first term describes the elementary gauge bosons masses and the NGBs dynamics and is given by
\be\label{Lsigmamodel}
\mathcal{L}_{\pi}= {f^2\over 4} \text{Tr} \left[d_\mu d^\mu \right]\,.
\ee
The second term, $\mathcal{L}_{\text{comp}}^V$, is a purely composite term, generated at the scale $m_*$ after confinement; it reduces to the kinetic terms for the $\rho$ vectors, namely:
\begin{equation}\label{LCompV}
\mathcal{L}_{\text{comp}}^V = -{1\over 4 g_\rho^2} \rho_{\mu \nu}^a \rho^{\mu \nu a} -{1\over 4 g_{\rho^X}^2} \rho_{\mu \nu}^X \rho^{X\mu \nu}\, ,
\end{equation} 
where $\rho_{\mu\nu}^{a}=\partial_{\mu}\rho_{\nu}^{a}-\partial_{\nu}\rho_{\mu}^{a}-f_{abc}\rho_{\mu}^{b}\rho_{\nu}^{c}$, $\rho_{\mu\nu}^{X}=\partial_{\mu}\rho_{\nu}^{X}-\partial_{\nu}\rho_{\mu}^{X}$ and $g_\rho$ and $g_{\rho^X}$ are the coupling strengths for the composite spin-1 bosons.
The third term in Eq.~\eqref{LBosonic}, $\mathcal{L}_{\text{elem}}^V$, is a purely elementary interaction, produced at the scale $\Lambda_{UV}$ where the elementary fields are formally introduced. 
Also this Lagrangian can contain only the kinetic terms for the elementary fields:
\begin{equation}\label{LElemV}
\mathcal{L}_{\text{elem}}^V= - {1 \over 4 g_1^2 }B_{\mu \nu}B^{\mu \nu}  -{1 \over 4 g_2^2} W_{\mu \nu }^a W ^{a \mu \nu}-{1 \over 4 \widetilde{g}_2^2} \widetilde{W}_{\mu \nu }^a \widetilde{W} ^{a \mu \nu}\, ,
\end{equation} 
where $g_1$, $g_2$ and $\widetilde{g}_2$ denote the weak gauge couplings.
The last term in the Lagrangian \eqref{LBosonic}, $\mathcal{L}_{\text{mix}}^V$, is a
mixing term between the elementary and composite sectors originating from partial compositeness. We have:\footnote{\label{footnotegrho}Notice that in the Lagrangian \eqref{LMixV}, the parameters $f$, $M_{\rho}$, $M_{\rho^{X}}$, $g_{\rho}$ and $g_{\rho^{X}}$ are all independent. It is common to define the parameters $a_{\rho}=M_{\rho}/(g_{\rho}f)$ and $a_{\rho^{X}}=M_{\rho^{X}}/(g_{\rho^{X}}f)$, which are expected to be $O(1)$. In our analysis we set $a_{\rho}=1/\sqrt{2}$ corresponding to the two-site model value (see the last paragraph of this section) and $a_{\rho^{X}}=1$. }
\begin{equation}\label{LMixV}
\mathcal{L}_{\text{mix}}^V = {M_{\rho} ^2 \over 2 g_\rho^2}\left (\text{Tr} \!\left[ \rho_\mu^a T^{\bf{21}}_a - E_\mu \right] \right ) ^2
+ {M_{\rho^{X}}^2 \over 2 g_{\rho^X}^2 }(\rho_\mu^X - B_\mu)^2\, ,
\end{equation}
where $T^{\bf{21}}_a$ are the $SO(8)$ generators in the adjoint of $SO(7)$ (see appendix~\ref{AppCCWZ}).

We now introduce the Lagrangian for the fermionic sector. 
This depends on the choice of quantum numbers for the composite operators in 
Eq.~\eqref{MixLagrF}. The minimal option is to choose $\mathcal{O}_R$
and $\mathcal{\widetilde{O}}_R$ to be in the fundamental representation of $SO(8)$, whereas the 
operators $\mathcal{O}_L$ and $\mathcal{\widetilde{O}}_L$ are singlets of the global group. Therefore, the elementary SM doublet and its Twin must be embedded into 
fundamental representations of $SO(8)$, whereas the $t_R$ and the $\widetilde{t}_R$ are complete singlets under the global $SO(8)$ invariance. This choice is particularly useful to generalize our discussion to the case of a fully-composite right-handed top. From the low-energy perspective, the linear mixing between composite operators and elementary fields translates into a linear coupling between the latter and a layer of fermionic resonances excited from the vacuum by the operators in the fundamental and singlet representations of the global group. Decomposing the $\bf{8}$ of $SO(8)$ as ${\bf{8}} = {\bf{7}}+{\bf{1}}$ under $SO(7)$, we introduce a set of fermionic resonances filling a complete fundamental representation of $SO(7)$ and another set consisting of just one singlet.\footnote{Notice that in general we should introduce two different singlets in our Lagrangian. One corresponds to a full $SO(8)$ singlet, while the other is the $SO(7)$ singlet appearing in the decomposition $\mathbf{8}=\mathbf{7}+\mathbf{1}$ of the fundamental of $SO(8)$ under the $SO(7)$ subgroup. We will further simplify our study identifying the two singlets with just one composite particle.} We denote with $\Psi_{\bf{7}}$ the fermionic resonances in the septuplet and with $\Psi_{\bf{1}}$ the singlet, both charged under $SU(3)_{c}$. Together with them, we must introduce  analogous composite states charged under $\widetilde{SU}(3)_{c}$; we use the corresponding notation  $\widetilde{\Psi}_{\bf{7}}$ and $\widetilde{\Psi}_{\bf{1}}$. We refer to Ref.~\cite{Barbieri:2015aa} for the complete expression of ${\Psi_{\bf{7}}}$ and $\widetilde{\Psi}_{\bf{7}}$ in terms of the constituent fermions. 

The fermionic effective Lagrangian is split into three parts, which have the same meaning as the analogous distinctions we made for the bosonic sector of the theory:
\be
\mathcal{L}_{\text{fermionic}} = \mathcal{L}^F_{\text{comp}} + \mathcal{L}^F_{\text{elem}} + \mathcal{L}^F_{\text{mix}}\,.
\ee
The fully composite term is given by:
\begin{equation}\label{LCompF}
\begin{split}
\mathcal{L}_{\text{comp}}^F = & \, \overline{\Psi}_{\bf{7}} (i \slashed D_{\bf{7}} - M_{\Psi})\Psi_{\bf{7}}+\overline{\Psi}_{\bf{1}}( i \slashed D_{\bf{1}} -M_S)\Psi_{\bf{1}}+ \overline{\widetilde{\Psi}}_{\bf{7}} (i \slashed \nabla - \widetilde{M}_{\Psi})\widetilde{\Psi}_{\bf{7}}+\overline{\widetilde{\Psi}}_{\bf{1}}( i \slashed \partial -\widetilde{M}_S)\widetilde{\Psi}_{\bf{1}} \\[0.1cm]
&  \displaystyle + \left(i c_L  \overline{\Psi}_{{\bf 7} L}^i\, 
\slashed{d}_i \Psi_{{\bf 1} L} +i c_R \overline{\Psi}_{{\bf 7} R}^i\, \slashed{d}_i \Psi_{{\bf 1} R} +i \widetilde{c}_L \overline{\widetilde{\Psi}}_{{\bf 7} L}^i\, \slashed{d}_i \widetilde{\Psi}_{{\bf 1} L} +i \widetilde{c}_R \overline{\widetilde{\Psi}}_{{\bf 7} R}^i\, \slashed{d}_i \widetilde{\Psi}_{{\bf 1} R}+ \text{h.c.} \right)\, , 
\end{split}
\end{equation}
where $D_{{\bf 7}\mu} = \nabla_\mu + i X B_\mu$, $D_{{\bf 1}\mu} = \partial_\mu + i X B_\mu$, and  $\nabla_\mu = \partial_\mu + i E_\mu$.
We have introduced two sets of $O(1)$ coefficients, $c_L$ and $c_R$ and their Twins, for the interactions mediated by the $d_\mu$ operator. 
Considering the elementary part of the Lagrangian, it comprises just the kinetic terms for the doublets and right-handed tops:
\begin{equation}
\mathcal{L}_{\text{elem}}^F = \overline{q}_L i \slashed D q_L + \overline{t}_R i \slashed D t_R +  \overline{\widetilde{q}}_L i \slashed D \widetilde{q}_L  +  \overline{\widetilde{t}}_R i \slashed \partial \widetilde{t}_R.
\end{equation}
The final term in our classification is the elementary/composite mixing that we write again following the prescription of partial compositeness. 
With our choice of quantum numbers for the composite operators, the spurions in Eq.~\eqref{MixLagrF} can be matched to dimensionless couplings
according to
\begin{equation}\label{SpurionSM}
\Delta_{\alpha A} = 
\begin{pmatrix}
 0 & 0 & i y_L & -y_L & 0 \times 4\\
 i y_L & y_L & 0 & 0 & 0 \times 4 \\
\end{pmatrix}, \qquad \Theta_A = y_R\,,
\end{equation}
and
\begin{equation}\label{SpurionTwin}
\widetilde{\Delta}_{\alpha A} =
\begin{pmatrix}
 0 \times 4& 0 & 0 & i \widetilde{y}_L & -\widetilde{y}_L \\
0 \times 4 & i \widetilde{y}_L & \widetilde{y}_L & 0 & 0 \\
\end{pmatrix}, \qquad \widetilde{\Theta}_A = \widetilde{y}_R,
\end{equation}
where we have introduced the elementary/composite mixing parameters $y_L$, $y_R$ and their Twin counterparts. These dimensionless $y$'s control the strength of the interaction between 
the elementary and composite resonance fields, according to the Lagrangian:
\begin{equation}\label{LMixF}
\begin{split}
\mathcal{L}_{\text{mix}}^F = & \, f \left( \bar{q}_L^\alpha \Delta_{\alpha A} \Sigma_{A i} \Psi_{\bf{7}}^i+ \bar{q}_L^\alpha \Delta_{\alpha A} \Sigma_{A 8} \Psi_{\bf{1}} 
 + y_R \bar{t}_R \Psi_{\bf{1}}  + \text{h.c.} \right) \\[0.1cm]
& + f \left( \bar{\widetilde{q}}_L^\alpha \widetilde{\Delta}_{\alpha A} \Sigma_{A i} \widetilde{\Psi}_{\bf{7}}^i
 + \bar{\widetilde{q}}_L^\alpha \widetilde{\Delta}_{\alpha A} \Sigma_{A 8} \widetilde{\Psi}_{\bf{1}} 
 + \widetilde{y}_R \bar{\widetilde{t}}_R \widetilde{\Psi}_{\bf{1}} + \text{h.c.} \right)\, .
\end{split}
\end{equation}
 Depending on the UV boundary condition and the relevance or
 marginality of the operators appearing in Eq.~\eqref{MixLagrF}, the
 $y$'s can vary from weak to $O(g_*)$.  Correspondingly the light fermions vary from being completely elementary (for $y$ weak) to
 effectively fully composite (for $y\sim g_*$).
 For reasons that will become clear, given $y_{t}\sim y_{L}y_{R}/g_{*}$, it is convenient to take $y_L\simeq \widetilde{y}_L\sim y_t$, i.e.~weak left mixing, and $y_R\simeq \widetilde{y}_R\sim g_*$. For such strong right-handed mixing the right-handed tops can be practically considered part of the strong sector.

The last term that we need to introduce in the effective Lagrangian describes the interactions between the vector and fermion resonances and originates completely in the Composite Sector. We have:
\begin{equation}\label{LCompVF}
\begin{split}
\mathcal{L}_{\text{comp}}^{VF} = & \displaystyle\sum_{i= L,R} \Big[ 
      \alpha_i \, \overline{\Psi}_{{\bf{7}}i} (\slashed \rho- \slashed E) \Psi_{{\bf{7}}i}
   + \alpha_{{\bf{7}}i}\, \overline{\Psi}_{{\bf{7}}i} (\slashed \rho^X - \slashed B )\Psi_{{\bf{7}}i} 
   + \alpha_{{\bf{1}}i} \, \overline{\Psi}_{{\bf{1}}i} (\slashed \rho^X - \slashed B ) \Psi_{{\bf{1}}i} \\
 &\hspace{1.05cm} +\widetilde{\alpha}_i \, \overline{\widetilde{\Psi}}_{{\bf{7}}i} (\slashed \rho- \slashed E) \widetilde{\Psi}_{{\bf{7}}i}    
   + \widetilde{\alpha}_{{\bf{7}}i} \, \overline{\widetilde{\Psi}}_{{\bf{7}}i} (\slashed \rho^X - \slashed B )\widetilde{\Psi}_{{\bf{7}}i} 
   +\widetilde{\alpha}_{{\bf{1}}i} \, \overline{\widetilde{\Psi}}_{{\bf{1}}i} (\slashed \rho^X - \slashed B ) \widetilde{\Psi}_{{\bf{1}}i} \Big] ,
\end{split}
\end{equation} 
where all the coefficients $\alpha_i$ appearing in the Lagrangian are $O(1)$ parameters. 

We conclude the discussion of our effective Lagrangian by clarifying its two-site model limit \cite{Panico:1359049} (see also Ref.~\cite{Contino:2015mha}). 
This is obtained by combining the singlet and the septuplet into a
complete representation of $SO(8)$, so that the model enjoys an
enhanced $SO(8)_L\times SO(8)_R$ global symmetry. This is achieved by
setting $c_L = c_R = \widetilde{c}_L = \widetilde{c}_R = 0$ and all
the $\alpha_i$ equal to 1. Moreover, we have to impose $M_{\rho} =g_\rho
f/\sqrt{2}$, so that the heavy vector resonances can be reinterpreted as
gauge fields of $SO(7)$. As shown in Ref.~\cite{Panico:1359049}, with this
choice of the free parameters the Higgs potential becomes calculable
up to only a logarithmic divergence, that one can regulate by imposing
just one renormalization condition. In the subsequent sections, we
will extensively analyze the EW precision constraints in the general
case, as well as in the two-site limit.

\subsection{Perturbativity of the simplified model}
\label{sec:perturbativity}

In Section \ref{sec:sub-hypersoft} it was noted that a TH construction typically involves large multiplicities of states and, as a consequence, the dynamics 
responsible for its UV completion cannot be maximally strongly coupled.  This in turn limits the improvement in fine tuning that can be achieved
compared to standard scenarios of EWSB.  In our naive estimates of Eqs.~\eqref{eq:mh-subhyper}, \eqref{fraternal} and \eqref{super} the interaction strength of the UV theory was 
controlled  by the $\sigma$-model quartic coupling $\lambda_{\cal H}$ or, equivalently, by $m_{\sigma}/f$. By considering the $\lambda_{\cal H}$ one-loop \mbox{$\beta$-function}  (Eq.~\eqref{eq:betafun}) we estimated the maximal value of $\lambda_{\cal H}$ as the one  corresponding to an $O(1)$ relative change through one e-folding of RG evolution.  
For an $SO(8)/SO(7)$ $\sigma$-model  this led to $\sqrt{\lambda_{\cal H}} \lesssim \sqrt{2}\pi$, or, equivalently, $m_{\sigma}/f\lesssim \pi$.

Alternatively, the limit set by perturbativity on the UV interaction
strength may also be estimated in the effective theory described by the
non-linear $\sigma$-model by determining the energy scale at which
tree-level scattering amplitudes become non-perturbative.  For
concreteness, we considered the following two types of scattering processes: $\pi\pi \to \pi\pi$
and $\pi\pi \to \bar\psi \psi$, where $\pi$ are the NGBs and $\psi = \{ \Psi_7, \tilde\Psi_7\}$ denotes a composite fermion transforming in the fundamental of $SO(7)$.
Other processes can (and should) be considered, with the actual bound being given by the strongest of the constraints obtained in this way.

Requiring that the process $\pi\pi \to \pi\pi$ stay perturbative up the cutoff scale $m_*$ gives the bound
\begin{equation}
\label{FirstLimit}
\frac{M_{\rho}}{f} \sim \frac{M_\Psi}{f} \lesssim \frac{m_*}{f} < \frac{4\pi}{\sqrt{N-2}} \simeq 5.1\, ,
\end{equation}
where the second inequality is valid in a generic $SO(N)/SO(N-1)$ non-linear $\sigma$-model, and we have set $N=8$ in the last step.
More details on how this result was obtained can be found in appendix~\ref{Apppertbound}.
Eq.~\eqref{FirstLimit} in fact corresponds to a limit on the
interaction strength of the UV theory, given that the couplings among 
fermion and vector resonances are of order $M_\Psi/f$ and $M_{\rho}/f$, respectively.
Perturbativity of the scattering amplitude for $\pi\pi \to \bar\psi\psi$ instead gives (see appendix~\ref{Apppertbound} for details)
\begin{equation}
\label{SecondLimit}
\frac{M_{\rho}}{f} \sim \frac{M_\Psi}{f} \lesssim \frac{m_*}{f} < {\sqrt{12\sqrt{2}}\pi\over \sqrt{N_f}}\simeq {4\pi \over \sqrt{N_f}}\, ,
\end{equation}
where $N_f$ is the  multiplicity of composite fermions (including the number of colors and families). 
Our simplified model with one family of composite fermions has $N_f = 6$, which gives a limit similar
to Eq.~\eqref{FirstLimit}: $M_\Psi/f \lesssim 5.3$. A model with three families of composite quarks and leptons has instead $N_f = 24$, from which follows 
the stronger bound $M_\Psi/f \lesssim 2.6$.

As a third alternative, one could analyze when 1-loop corrections to a given observable
become of the same order as its tree-level value. We applied this
criterion to our simplified model by considering the $\hat S$
parameter, the new physics contribution to which includes a tree-level correction from heavy vectors given by
Eq.~\eqref{eq:dSrho}, and a one-loop correction due to heavy fermions,
which can be found in Appendix~\ref{AppFormulae}. By requiring that the one-loop term be smaller than the tree-level correction, we obtain a bound on the strong coupling constant $g_\rho$.
As an illustration, we consider the two-site model limit $c_L=c_R=0$ and $M_{\rho}=g_{\rho}f/\sqrt{2}$ and keep the dominant UV contribution to $\hat{S}$ in Eq.~\eqref{DSPsi} which is logarithmically 
sensitive to the cut-off. By setting $m_* =2 M_\Psi$, we find:
\begin{equation}
\label{eq:limitgrho}
{\Delta \hat{S}_\text{1-loop} \over \Delta \hat{S}_\text{tree}} < 1 \implies {M_{\rho} \over f}  = \frac{g_\rho}{\sqrt{2}} < {\pi \over \sqrt{2\log 2}} \simeq 2.7\,.
\end{equation}

The perturbative limits obtained from Eqs.~\eqref{FirstLimit},~\eqref{SecondLimit} and \eqref{eq:limitgrho} are comparable to that on $\lambda_{\cal H}$ derived in 
Sec.~\ref{sec:sub-hypersoft}.
As already discussed there, one could take any of these results as
indicative of the maximal interaction strength in the underlying UV dynamics, though none of them 
should be considered as a sharp exclusion condition. In our analysis of EW observables 
we will make use of Eq.~\eqref{FirstLimit} with $N=8$ and of Eq.~\eqref{SecondLimit} with $N_f = 24$ to highlight the regions of parameter space where our perturbative 
calculation is less reliable.  We use both limits as a measure of the intrinsic uncertainty which is inevitably associated
with this type of estimation.

\section{Higgs Effective Potential}
\label{sec:effpot}

As anticipated in the general discussion of Section \ref{sec2.1}, a potential for the Higgs boson is generated at the
scale $m_*$ by loops of heavy states  through the $SO(8)$-breaking couplings of the elementary fields to the strong sector.
Once written in terms of the Higgs boson $h$ (where $H^\dagger H = f^2 \sin^2(h/f)/2$, $\tilde H^\dagger \tilde H = f^2 \cos^2(h/f)/2$), 
at 1-loop this UV threshold contribution has the form~\cite{Barbieri:2015aa}:
\begin{equation}
\label{eq:potentialUV}
\frac{V(m_*)}{f^4} = \frac{3}{32\pi^2} \left[ \frac{1}{16} g_1^2 g_\rho^2 L_1 + (y_L^2 - \tilde y_L^2) g_\Psi^2 L_2 \right] \sin^2\frac{h}{f} 
+ \frac{3 y_L^4}{64\pi^2} F_1 \left( \sin^4\frac{h}{f} + \cos^4\frac{h}{f} \right)\, ,
\end{equation}
where $g_\Psi \equiv M_\Psi/f$, $L_1$, $L_2$, $F_1$ are $O(1)$
dimensionless functions of the masses and couplings of the theory and
the explicit expression of the function $F_1$ is reported in
Eq.~\eqref{FFunctions} of Appendix \ref{AppBetaFunc}.
The first term in the above equation originates from $Z_2$-breaking
effects.\footnote{Sub-leading $Z_2$-breaking terms have been neglected
  for simplicity. The complete expressions are given in Ref.~\cite{Barbieri:2015aa}.}
The second term, generated by loops of fermions, is $Z_2$ symmetric and explicitly violates the $SO(8)$ invariance; it thus corresponds
to the (UV part of the) last term of Eq.~\eqref{quartic}.  Upon electroweak symmetry breaking, Eq.~\eqref{eq:potentialUV} contributes to the physical Higgs 
mass an amount equal to
\begin{equation}
\label{eq:deltamHUV}
\delta m_h^2|_{UV} = \frac{3 y_L^4}{4\pi^2} F_1 f^2 \xi (1-\xi)\, ,
\end{equation}
where $\xi$ controls the degree of vacuum misalignment: 
\begin{equation}
\xi \equiv \sin^2\frac{\langle h \rangle}{f}=\frac{v^{2}}{f^{2}} .
\end{equation}

Below the scale $m_*$ an important contribution to the potential arises from loops of light states, in particular from the
top quark and from its Twin. The bulk of this IR contribution is captured by  the RG evolution of the Higgs potential from the scale $m_*$ down to the electroweak scale. As noted in previous studies (see e.g. Ref.~\cite{Barbieri:2015aa}), for sufficiently
large $m_*$ this IR effect dominates over the UV threshold correction
and can reproduce the experimental Higgs mass almost entirely.
An analogous IR correction to the Higgs quartic arises in SUSY theories
with large stop masses, from loops 
of top quarks. The distinctive feature  of any TH
scenario, including our model, is  the additional Twin top contribution.

The Higgs effective action, including the leading $O(\xi)$ corrections associated with operators of dimension 6, was computed at 1-loop in Ref.~\cite{Barbieri:2015aa}; the resulting IR contribution to $m_h^2$  was found to be
\begin{equation}
\label{eq:deltamHLL}
\delta m_h^2|_{\text{IR}}^\text{1-loop} = \frac{3 y_t^4}{8\pi^2} f^2 \xi (1-\xi) \left( \log\frac{m_*^2}{m_t^2} + \log\frac{m_*^2}{\widetilde{m}_{ t}^2} \right)\, ,
\end{equation}
where $y_t$ denotes the top Yukawa coupling. The two single-log terms in parentheses correspond to the IR contributions to the effective Higgs quartic $\lambda_h$
from the top quark and Twin top respectively.
Leading-logarithmic corrections of the form $(\alpha \log)^n$, arising at higher loops have however an important numerical impact.\footnote{Here $\alpha=g_{\text{SM}}^{2}/4\pi$, with $g_{\text{SM}}$ being any large SM coupling, i.e.~$g_{S}$ and $y_{t}$.}
For example, $(\alpha \log)^2$ corrections 
generated by 2-loop diagrams (mostly due to the running of the top and
Twin top Yukawa couplings, that are induced by respectively QCD and Twin QCD)
 are expected to give a $\sim 30\%$ reduction 
in the Higgs mass for $m_*\simeq 5\TeV$, 

We have computed the IR contribution to the Higgs mass in a combined expansion in $\xi$ and $(\alpha \log)$. We have included the LO electroweak contribution, all terms up to NLO in $\alpha_{t}$ and $\alpha_{S}$, and some contributions, expected to be leading for not too large $m_{*}$, at NNLO in $\alpha_{S}$. 
We report the details in Appendix~\ref{AppBetaFunc}.

\begin{figure}[t!]
\begin{center}
\includegraphics[width=0.6\textwidth]{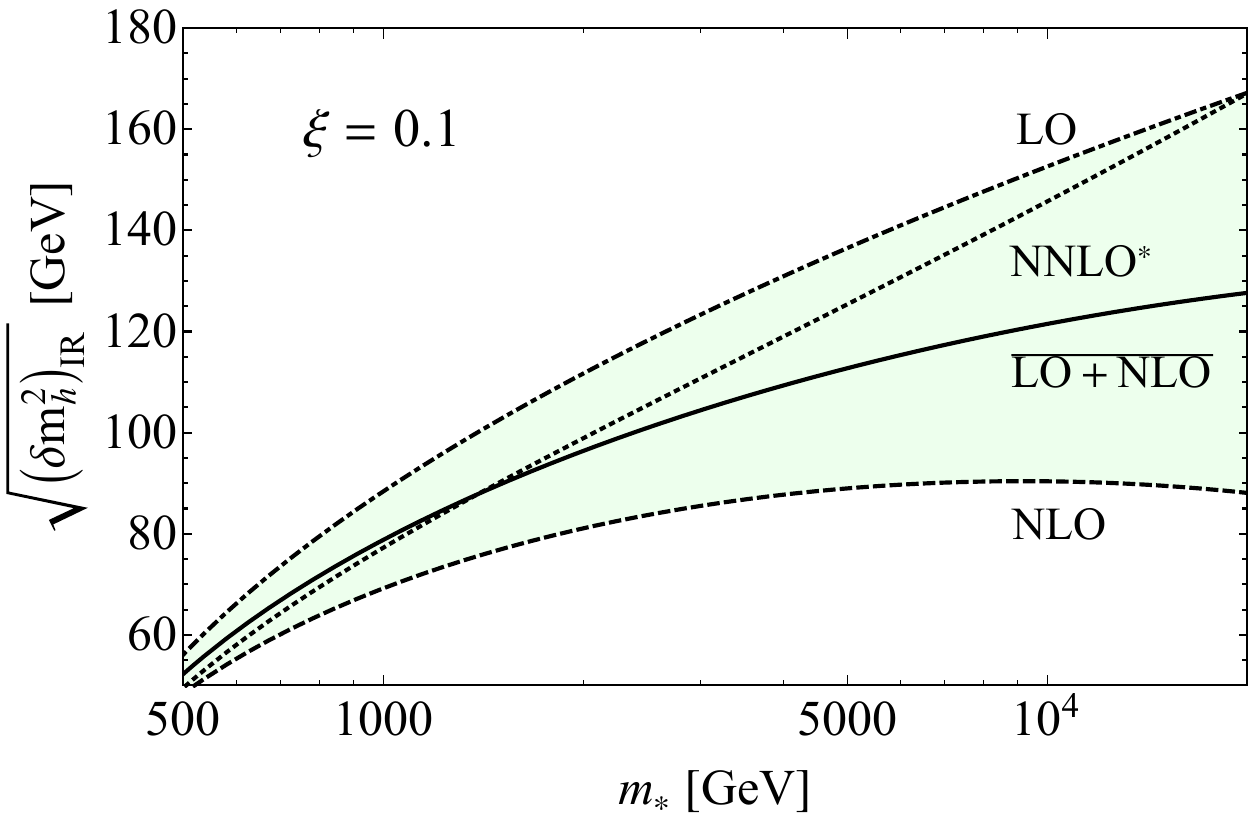}
\end{center}
\caption{{\it IR contribution to the Higgs mass as a function of the
    scale $m_*$ for $\xi=0.1$. The dot-dashed (upper) and dashed (lower) curves denote
    the LO and NLO result in a combined perturbative expansion in $(\alpha
\log)$ and $\xi$. The dotted curve contains the pure QCD part of
the NNLO correction (see Appendix~\ref{AppBetaFunc} for details). The shaded region represents the uncertainty in our estimate. The central value of this band, given by the solid black line and computed as the average of the LO and NLO results, is the value that we use as our numerical estimate throughout the paper.
}}
\label{fig:deltamHIR}
\end{figure}

The value of $(\delta m_h^2|_{IR})^{1/2}$ is shown in Figure
\ref{fig:deltamHIR} as a function of $m_*$ for $\xi=0.1$. The upper
and lower curves represent respectively the LO and NLO calculation in
our combined $(\xi,\alpha\log)$ expansion. Numerically, the naive
expectation is confirmed, as the NLO correction decreases $(\delta
m_h^2|_{IR})^{1/2}$ by $\sim 35\%$ for $m_{*}=5$ TeV. The dotted
curve, indicated as NNLO$^{*}$, includes NNLO contributions of order
$\alpha_{t}\xi^{2}\log$, $\alpha_{t}\alpha_{S}\xi\log^{2}$, and
$\alpha_{t}\alpha_{S}^{2}\log^{3}$ (see Appendix
\ref{AppBetaFunc}). Additional contributions of order
$\alpha_{t}^{2}\xi\log^{2}$, $\alpha_{t}^{2}\alpha_{S}\log^3$ and
$\alpha_{t}^{3}\log^3$ are not included. An attempt to include these
contributions has been made in Ref.~\cite{Greco:2016zaz}. However, the
calculation presented there misses some additional contributions from
the Twin GB and does not represent the full NNLO calculation. The
picture that we get from our NNLO$^{*}$ calculation and the incomplete
result of Ref.~\cite{Greco:2016zaz}, is that the NNLO$^{*}$ gives an
overestimate of the IR contribution to the Higgs mass as the
aforementioned neglected effects are expected to give a reduction.
Given the lack of a complete NNLO calculation, we estimate our
uncertainty on the IR Higgs mass as the green shaded region lying
between the LO and NLO results in Figure \ref{fig:deltamHIR}. In order
to choose, within this uncertainty, a value that is as close as
possible to the full NNLO result, in the rest of the paper we take as
input value for the IR correction to the Higgs mass the average of the LO and NLO results, corresponding to the solid black line in the Figure.
%

The plot of Figure \ref{fig:deltamHIR} illustrates one of the
characteristic features of TH models: the IR contribution to the Higgs
mass largely accounts for its experimental value and is completely
predicted by the theory in terms of the low-energy particle content
(SM plus Twin states). In particular, considering the aforementioned
input value and choosing as a benchmark values
$m_{*}=5~\text{TeV}$ and $\xi=0.1$, we find that the contributions of
the SM and Twin light degrees of freedom account for around 50\% and 40\% of the Higgs mass squared, respectively, so that almost the entire experimental value of the Higgs mass can be due only to the IR degrees of freedom. Threshold effects arising at the UV matching scale, on the other hand, are model dependent but give a sub-leading correction.
An accurate prediction of the Higgs mass and an assessment of the
plausibility of the model thus requires a precise determination of its
IR contribution. Indeed the difference between the IR contribution of
Figure \ref{fig:deltamHIR} and the measured value $m_h=125$ GeV must
be accounted for by the UV threshold contribution in
Eq.~\eqref{eq:deltamHUV}; for our previous benchmark choice of $m_*$,
about 12\% of the Higgs mass should be generated by the UV physics. This translates into a generic constraint on the size of $y_L$, a parameter upon which electroweak precision observables (EWPO) crucially depend, thus creating a  non-trivial correlation between the Higgs mass, EWPO and naturalness.

\section{Electroweak Precision Observables}
\label{sec:EWPT}

In this section we compute the contribution of the new states described by our simplified model
to the EWPO. Although it neglects the effects of the heavier resonances,
our calculation is expected to give a fair assessment of the size of the corrections due to the full strong dynamics,
and in particular to reproduce the correlations among different observables.

It is well known that, under the assumption of quark and lepton universality, short-distance corrections to the electroweak observables
due to heavy new physics can be expressed in terms of four parameters, $\widehat S$, $\widehat T$, $W$, $Y$,  defined in Ref.~\cite{Barbieri:2004p1607} (see also Ref.~\cite{Grinstein:1991cd} for an equivalent analysis)
as a generalization of the parametrization introduced by Peskin and Takeuchi in Refs.~\cite{Peskin:1990zt,Peskin:1992p1662}.
Two additional parameters, $\delta g_{Lb}$ and $\delta g_{Rb}$, can be
added to account for the modified couplings of the $Z$ boson to left-
and right-handed bottom quarks respectively.\footnote{We define $\delta g_{Lb}$ and $\delta g_{Rb}$ in terms of the following effective Lagrangian in the
unitary gauge:
\begin{equation}
{\cal L}_{eff} \supset \frac{g_2}{2c_W} Z_\mu \, \bar b \gamma^\mu \!\left[ (g_{Lb}^{SM} + \delta g_{Lb}) (1-\gamma_5) +
(g_{Rb}^{SM} + \delta g_{Rb}) (1+\gamma_5)\right] \! b + \dots
\end{equation}
where the dots stand for higher-derivative terms and $g_{Lb}^{SM} = -1/2 + s_W^2/3$, $g_{Rb}^{SM} = s_W^2/3$.
}
A naive estimate shows that in CH theories, including our TH model, $W$ and $Y$ are sub-dominant in an expansion 
in the weak couplings~\cite{Giudice:1024017} and can thus be neglected.
The small coupling of the right-handed bottom quark to the strong dynamics makes also $\delta g_{Rb}$ small and negligible in our model.
We  thus focus on $\widehat S$, $\widehat T$ and $\delta g_{Lb}$, and compute them
by including effects from the exchange of vector and fermion resonances, and from Higgs compositeness.

We work at the 1-loop level and at leading order in the electroweak couplings 
and perform an expansion in inverse powers of the new physics scale.
In this limit, the Twin states do not affect the EWPO as a consequence of their being neutral under the SM gauge group. 
Deviations from the SM predictions arise only from heavy states with SM quantum numbers and are
parametrically the same as in ordinary CH models with singlet $t_R$.
This can be easily shown by means of naive dimensional analysis and symmetries as follows. 
Twin tops interact with the SM fields only through higher-dimensional operators. The operators relevant for the EWPO are those
involving either a SM current or a derivative of the hypercharge field strength:
\begin{equation} \label{eq:Twinoperators}
O_{B\tilde t} = \frac{g'}{m_{W}^{2}} \partial^\mu B_{\mu\nu}  \, \bar {\tilde t} \gamma^\nu \tilde t \, , \qquad 
O_{q\tilde t} = \frac{1}{v^{2}}\bar q_L \gamma_\mu q_L\, \bar {\tilde t} \gamma^\mu \tilde t  \, , \qquad 
O_{H\tilde t} = \frac{i}{v^{2}}H^\dagger \overleftrightarrow {D_\mu} H \, \bar {\tilde t} \gamma^\mu \tilde t \, ,
\end{equation}
where  $\tilde t$ indicates either a right- or left-handed Twin top.\footnote{Notice that $O_{H\tilde t}$ can be 
rewritten in terms of the other two operators by using the equations of motion, but it is still useful to consider it in our discussion.}
The first two operators of Eq.~\eqref{eq:Twinoperators} are generated at the scale $m_*$ by the tree-level exchange of the $\rho_X$. Their coefficients (in a basis with canonical kinetic terms) are
respectively of order $(m_W^2/m_*^2)(\tilde y/g_*)^2$ and $(y_L^2 v^2/m_*^2) (\tilde y/g_*)^2$, where $\tilde y$ equals either $\tilde y_L$ or $\tilde y_R$ depending 
on the chirality of 
$\tilde t$. The third operator breaks custodial isospin and the only way it can be generated is via the exchange of weakly coupled elementary fields at loop level. 
Given that the contribution to EWPO is further suppressed by $\tilde t$ loops, the third operator can affect EWPO only at, at least, two loops and is thus clearly negligible.
By closing the $\tilde t$ loops the first two operators can give rise to effects that are schematically of the form $BB$, $B\bar qq$ or $(\bar q q)^2$. 
The formally quadratically divergent piece of the loop integral renormalizes the corresponding dimension-6 operators. For instance the second structure gives 
\begin{equation}
C \frac{g'}{16\pi^2}\frac{y_L^2}{m_*^2}\left (\frac{\tilde y}{g_*}\right )^4 \,\partial_\nu B^{\mu\nu} \bar q_L \gamma_\mu q_L
\end{equation}
with $C$ an $O(1)$ coefficient which depends on the details of the physics at the scale $m_*$. Using the equations of motion for $B$, 
the above operator gives rise to a correction to the $Zb\bar b$ vertex of relative size
\begin{equation}
\frac{\delta g_{Lb}}{g_{Lb}}\sim \frac{g'^2}{16\pi^2}\frac{y_L^2 v^2}{m_*^2} \left (\frac{\tilde y}{g_*}\right )^4
\end{equation}
which, even assuming $\tilde y \sim g_*$, is $O(g'/y_t)^2$ suppressed with respect to the leading visible sector effect we discuss below.
Aside the quadratically divergent piece there is also  a logarithmic divergent piece whose
overall coefficient is calculable. The result is further suppressed with respect to the above contribution by a factor $(m_{\tilde t}^2/m_*^2)\ln (m_{\tilde t}^2/m_*^2)$. 

An additional contribution could in principle come from loops of the extra
three ``Twin" NGBs contained in the coset $SO(8)/SO(7)$. 
Simple inspection however shows that there is no corresponding 1-loop diagram contributing to the EWPO.
In the end we conclude that the effect of Twin loops is negligible.

Since the effects from the Twin sector can be neglected, the corrections to $\widehat S$, $\widehat T$ and $\delta g_{Lb}$ are parametrically the same
as in ordinary CH models. We now give a concise review of the contributions to each of these quantities, distinguishing between 
the threshold correction generated at the scale~$m_*$ and the contribution arising from the RG evolution down to the electroweak scale.
For recent analyses of the EWPO in the context of $SO(5)/SO(4)$ CH models see for example Refs.~\cite{Grojean:2013qca,Contino:2015mha,Ghosh:2015wiz}.

\subsection{$\widehat S$ parameter}
\label{sec:Sparameter}

The leading contribution to the $\widehat S$ parameter arises at  tree level from the exchange of spin-1 resonances.
Since only the $(3,1)$ and $(1,3)$ components of the spin-1 multiplet contribute, its expression is the same as
in $SO(5)/SO(4)$ composite-Higgs theories:\footnote{We neglect for simplicity a contribution from the operator $E_{\mu\nu} \rho^{\mu\nu}$, which also arises at tree level.
See for example the discussion in Refs.~\cite{Contino:2015mha,Contino:2015gdp}.}
\begin{equation}
\label{eq:dSrho1}
\Delta \widehat S_{\rho} = \frac{g_2^2}{2 g_\rho^2} \xi \, .
\end{equation}
In our numerical analysis presented in Section \ref{sec:results} we use the two-site model relation $M_{\rho}=g_{\rho}f/\sqrt{2}$ to rewrite
\begin{equation}
\label{eq:dSrho}
\Delta \widehat S_{\rho} = \frac{m_{W}^2}{M_{\rho}^{2}} \, .
\end{equation}
The 1-loop contribution from loops of spin-1 and fermion resonances is sub-dominant (by a factor $g_*^2/16\pi^2$) 
and will be neglected for simplicity in the following. Nevertheless, we explicitly computed the fermionic contribution (see Appendix \ref{AppFormulae})  to monitor
the validity of the perturbative expansion and estimate the limit of strong coupling in our model (a discussion on this aspect was given in Section \ref{sec:perturbativity}). 
An additional threshold correction to $\widehat S$, naively of the same order as Eq.~\eqref{eq:dSrho},
arises from the exchange of cutoff modes at $m_*$.
As already anticipated, we neglect this correction in the following. In this respect
our calculation is subject to an $O(1)$ uncertainty and should rather be considered as an estimate, possibly more refined than a naive one, which takes 
the correlations among different observables into account.

Besides the UV threshold effects described above, $\widehat S$ gets an IR contribution from RG evolution down to the electroweak scale.
The leading effect of this type comes from the compositeness of the Higgs boson, and is the same as in $SO(5)/SO(4)$ CH models~\cite{Barbieri:2007bh}:
\begin{equation} \label{eq:dShiggs}
\Delta \widehat S_{h} = \frac{g_2^2}{192\pi^2} \xi \log\frac{m_*^2}{m_h^2}\, .
\end{equation}
In the effective theory below $m_*$ this corresponds to the evolution of the
\mbox{dimension-6} operators 
\be
O_W = \frac{ig}{2m_W^2}  \, H^\dagger \sigma^i \overleftrightarrow {D^\mu} H \, D^\nu  W_{\mu \nu}^i\,,\qquad O_B = \frac{ig'}{2m_W^2}\, H^\dagger  \overleftrightarrow {D^\mu} H \, \partial^\nu  B_{\mu \nu}
\ee
induced by a 1-loop insertion of
\be
O_H = \frac{1}{2v^2}\, \partial^\mu(H^\dagger H) \partial_\mu ( H^\dagger H)\,.
\ee
Denoting with $\bar c_i$ the coefficients of the effective operators and working at leading order in the SM couplings, the RG evolution can be expressed as
\begin{equation}
\label{eq:RGci}
\bar c_i(\mu) = \left( \delta_{ij} +  \gamma_{ij}\, \log\frac{\mu}{M} \right) \bar c_j(M) \, ,
\end{equation}
where $\gamma_{ij}$ is the anomalous dimension matrix (computed at leading order in the SM couplings).
The $\widehat S$ parameter gets a correction $\Delta \widehat S = (\bar c_W(m_Z) + \bar c_B(m_Z)) \xi$,
and one has $\gamma_{W, H} + \gamma_{B, H} = -g_2^2/(96\pi^2)$.
An additional contribution to the running arises from insertions of the current-current operators
\begin{equation} \label{eq:OHpsi}
O_{Hq} =  \frac{i}{v^2} \bar q_L \gamma^\mu q_L \,  H^\dagger \overleftrightarrow{D_\mu} H\, , \quad\ \
O_{Hq}' =  \frac{i}{v^2} \bar q_L \gamma^\mu \sigma^i q_L\,  H^\dagger\sigma^i \overleftrightarrow{D_\mu} H \, , \quad\ \
O_{Ht} =   \frac{i}{v^2}  \bar t_R \gamma^\mu t_R\,  H^\dagger \overleftrightarrow{D_\mu} H
\end{equation}
in a loop of top quarks.
This is however suppressed by a factor $y_L^2/g_*^2$ compared to Eq.~\eqref{eq:dShiggs} and will be neglected.
The suppression arises because the current-current operators are generated at the matching scale with coefficients 
proportional to~$y_L^2$.

The total correction to the $\widehat S$ parameter in our model is $\Delta \widehat S = \Delta \widehat S_\rho + \Delta \widehat S_{h}$, with the two contributions given by Eqs.~\eqref{eq:dSrho} and \eqref{eq:dShiggs}.

\subsection{$\widehat T$ parameter}

Tree-level contributions to the $\widehat T$ parameter are forbidden in our model by the $SO(3)$ custodial symmetry preserved 
by the strong dynamics,  and can only arise via loops involving the elementary states.
 A non-vanishing effect arises at the 1-loop level corresponding to a violation of custodial isospin by two units.
The leading contribution comes from loops of  fermions and is proportional to $y_L^4$, given that 
the spurionic transformation rule of  $y_L$ is that of a doublet,  while $y_R$ is a singlet. We find:
\begin{equation} \label{eq:dTpsi}
\Delta \widehat T_\Psi = a_{UV} N_c \frac{y_L^2}{16\pi^2} \frac{y_L^2v^2}{M_\Psi^2} 
                                + a_{IR} N_c \frac{y_t^2}{16\pi^2} \frac{y_L^2v^2}{M_\Psi^2} \log\frac{M_1^2}{m_t^2} \, ,
\end{equation}
where $a_{UV,IR}$ are $O(1)$ coefficients whose values are reported in Appendix \ref{AppFormulae} and we have defined $M_1 \equiv \sqrt{M_S^2 + y_R^2 f^2}$.
The result is finite and  does not depend on the cutoff scale $m_*$. 
The first term corresponds to the UV threshold correction generated at the scale $\mu = M_1 \sim M_\Psi$.
The second term instead encodes the IR running from the threshold scale down to low energy, due to loops of top quarks.
In the effective theory below $M_1$ it corresponds to the RG evolution of the dimension-6 operator 
\be
O_T = \frac{1}{2v^2} (H^\dagger {\overleftrightarrow { D^\mu}} H)^2
\ee due to insertions of the current-current operators of Eq.~\eqref{eq:OHpsi}.
In particular, $\Delta\widehat T = \widehat c_T(m_Z) \xi$ and one has $\gamma_{T,Ht} = - \gamma_{T, Hq} = 3y_t^2/4\pi^2$, $\gamma_{T,Hq'} = 0$.
Notice that the size of the second contribution with respect to the first is $O[(y_t/y_L)^2  \log(M_1^2/m_t^2)]$:
for $y_t\sim y_L$, that is for fully composite $t_R$, the IR dominated contribution is formally logarithmically enhanced and dominant.

Further contributions to $\widehat T$ come from loops of spin-1 resonances, the exchange of cutoff modes and Higgs compositeness.
The latter is due to the modified couplings of the composite Higgs to vector bosons and reads~\cite{Barbieri:2007bh}:
\begin{equation}\label{eq:dTH}
\Delta \widehat T_{h} = -\frac{3g_1^2}{64\pi^2} \xi \log\frac{m_*^2}{m_h^2}\, .
\end{equation}
In the effective theory it corresponds to the running of $O_T$ due to the insertion of the operator $O_H$ in a loop with hypercharge. The contribution is
of the form of Eq.~\eqref{eq:RGci} with $\gamma_{T,H} = 3 g_1^2/32\pi^2$.
The exchange of spin-1 resonances gives a UV threshold correction which is also proportional to $g_1^2$ (as a spurion, the hypercharge coupling transforms as an 
isospin triplet), but without any log enhancement. It is thus subleading compared to Eq.~(\ref{eq:dTH}) and we will neglect it for simplicity
(see Ref.~\cite{Contino:2015mha} for the corresponding computation in the context of $SO(5)/SO(4)$ models).
Finally, we also omit the effect of the cutoff modes because it is incalculable, 
although naively this is of the same order as the contribution from states included in our simplified model. Our result is thus subject to an $O(1)$ uncertainty.

The total contribution to the $\widehat T$ parameter in our model is therefore $\Delta \widehat T = \Delta \widehat T_{h} + \Delta \widehat T_\Psi$\, with the two contributions given by Eqs.~\eqref{eq:dTpsi} and \eqref{eq:dTH}.

\subsection{$\delta g_{Lb}$}\label{sec:deltagLb}

In the limit of vanishing transferred momentum, tree-level corrections to $\delta g_{Lb}$ are forbidden by 
the $P_{LR}$ parity of the strong dynamics that exchanges $SU(2)_L$ with $SU(2)_R$ in the visible $SO(4)$ and 
$\widetilde{SU}(2)_L$ with $\widetilde{SU}(2)_R$ in the Twin $\widetilde{SO}(4)$ (see Appendix \ref{AppCCWZ} for details). This is a simple extension of the $P_{LR}$ symmetry of CH models which protects the $Zb\bar b$ coupling from large corrections~\cite{Agashe:2006at}.
In our case $P_{LR}$ is an element of the unbroken $SO(7)$ and keeps the vacuum unchanged. It is thus an exact invariance of the strong dynamics,
differently from  $SO(5)/SO(4)$ models where it is accidental at $O(p^2)$.
The gauge couplings $g_{1,2}$ and $y_L$ explicitly break it, while $y_R$ preserves it.
At finite external momentum $\delta g_{Lb}$ gets a non-vanishing tree-level contribution:
\begin{equation}\label{DeltaTree}
(\delta g_{Lb})_\text{tree} = \frac{ f^2 \xi}{8 M_\rho^2} \left[ g_1^2 (\alpha_L + \alpha_{{\bf{7}}L} ) - g_2^2 \alpha_L \right] \frac{y_L^2 f^2}{M_\Psi^2 + y_L^2 f^2}\, .
\end{equation}
In the effective theory below $M_1$, this correction arises from the dimension-6 operators 
\be
O_{Bq}=\frac{g'}{m_{W}^{2}}\partial^\mu B_{\mu\nu} \bar q_L \gamma^\nu q_L\,,\qquad O_{Wq}=\frac{g}{m_{W}^{2}} D^\mu W^a_{\mu\nu} \bar q_L \gamma^\nu \sigma^a q_L,.
\ee
It is of order $(y_L^2/g_*^2) (g^2/g_*^2) \xi$, hence a factor $g^2/g_*^2$ smaller than the naive expectation in absence of the $P_{LR}$ protection.

At the 1-loop level, corrections to $\delta g_{Lb}$ arise from the virtual exchange of heavy fermion and vector states. 
The leading effect comes at $O(y_L^4)$ from loops of heavy fermions 
(the corresponding diagrams are those of figs.~\ref{fig:DeltaGLoop} and~\ref{fig:DeltaGDivergent}) and reads
\begin{equation} \label{eq:dgLb1loop}
(\delta g_{Lb})_\Psi = \frac{y_L^2}{16\pi^2} N_c \frac{y_L^2v^2}{M_\Psi^2} \left( b_{UV} + c_{UV} \log \frac{m_*^2}{M_\Psi^2} \right)
                                + b_{IR} \frac{y_t^2}{16\pi^2} N_c \frac{y_L^2v^2}{M_\Psi^2} \log\frac{M_1^2}{m_t^2} \, .
\end{equation}
The expressions of the $O(1)$ coefficients $b_{UV,IR}$ and $c_{UV}$ are reported in Appendix \ref{AppFormulae}. 
The first term is logarithmically divergent and encodes the UV threshold correction at the matching scale. The divergence comes, in particular, from diagrams where
the fermion loop is connected to the $b$-quark current through the exchange of a spin-1 resonance~\cite{Grojean:2013qca}. 
A simple operator analysis shows that the threshold contribution from the vector resonances 
in the adjoint of $SO(7)$ identically vanishes in our model  (see appendix~\ref{AppVectorDeltag} for details). 
An additional UV threshold contribution to $\delta g_{Lb}$ arises from diagrams where the spin-1 resonances
circulate in the loop. For simplicity we will not include such effect in our analysis (see Ref.~\cite{Ghosh:2015wiz} for the corresponding computation 
in the context of $SO(5)/SO(4)$ models).
It is however easy to show that  there is no possible diagram with $\rho_X$ circulating in the loop as a consequence of its quantum numbers, while 
the corresponding contribution from vector  resonances in the adjoint of $SO(7)$ is non-vanishing in this case.

The second term in Eq.~\eqref{eq:dgLb1loop} accounts for the IR running down to the electroweak scale. In the effective theory below $M_1$ one has
$\delta g_{Lb} = -(\bar c_{Hq}(m_Z) + \bar c'_{Hq}(m_Z))/2$, hence the IR correction arises from the evolution of the operators $O_{Hq}$ and $O_{Hq}'$
due to loops of top quarks. In this case the operators that contribute to the running via their 1-loop insertion are those of Eq.~\eqref{eq:OHpsi} 
as well as the following  four-quark operators~\cite{Elias-Miro:2013mua}:
\begin{equation}
O_{LR} = \left(\bar q_L \gamma^\mu q_L\right) \left(\bar t_R \gamma_\mu t_R\right) , \quad
O_{LL} = \left(\bar q_L \gamma^\mu q_L\right)  \left(\bar q_L \gamma_\mu q_L\right), \quad
O_{LL}' = \left(\bar q_L \sigma^a \gamma^\mu q_L\right)  \left(\bar q_L \sigma^a \gamma_\mu q_L\right)\, .
\end{equation}
In fact, the operators contributing at  $O(y_L^2 y_t^2)$ to Eq.~\eqref{eq:RGci} are only those generated at $O(y_L^2)$ at the matching scale; these are
$O_{Ht}$,  the linear combination $O_{Hq} - O_{Hq}'$ (even under~$P_{LR}$), and $O_{LR}$ (generated via the exchange of $\rho_X$).\footnote{The operators $O_{LL}$
and $O_{LL}'$ are generated at $O(y_L^4)$ by the tree-level exchange of both $\rho_X$ and $\rho$.}
Notice finally that the relative size of the IR and UV contributions to $\delta g_{Lb}$ is $O[(y_t/y_L)^2  \log(M_1^2/m_t^2)]$ precisely like in the case of $\Delta \widehat T_\Psi$.

It is interesting that in our model the fermionic corrections  to $\delta g_{Lb}$ and $\widehat T$ are parametrically of the same order 
and their signs tend to be correlated. It is for example well known that a heavy fermion with the quantum numbers of $t_R$ gives a positive 
correction to both quantities~\cite{Rattazzi:1989qm,Carena:2007ua,Gillioz:2008hs,Anastasiou:2009rv}. We have verified that this is also the case in our model for 
$M_S \ll M_\Psi \sim M_\rho$ (light singlet).\footnote{In this limit one has $\Delta \widehat T_\Psi \simeq 3 (\delta g_{Lb})_\Psi$.}
Conversely, a light septuplet ($M_\Psi \ll M_S \sim M_\rho$) gives a negative contribution to both $\delta g_{Lb}$ and $\widehat T$.\footnote{The existence of 
a similar sign correlation in the limit of a light $(2,2)$ has been pointed out in the context of $SO(5)/SO(4)$ CH models, 
see Ref.~\cite{Grojean:2013qca}.}
Although in general the expressions for $\Delta \widehat T_\Psi$ and $(\delta g_{Lb})_\Psi$ are uncorrelated, their signs tend to be  the same whenever the 
contribution from $\rho_X$ to Eq.~\eqref{eq:dgLb1loop} is subleading. The sign correlation can instead be broken if $\rho_X$ contributes significantly to 
$\delta g_{Lb}$ (in particular,  $(\delta g_{Lb})_\Psi$ can be negative for $\alpha_{iL} = - \alpha_{iR}$).
The importance of the above considerations lies in the fact that EW precision data prefer a positive $\widehat T$ and a negative $\delta g_{Lb}$. Situations when both
quantities have the same sign are thus experimentally disfavored.

Considering that no additional correction to $\delta g_{Lb}$ arises from Higgs compositeness, and that we neglect as before the incalculable effect due to cutoff states, 
the total contribution in our model is $\delta g_{Lb} = (\delta g_{Lb})_\text{tree} + (\delta g_{Lb})_\Psi$, with the two contributions given by eqs.~\eqref{DeltaTree} and \eqref{eq:dgLb1loop}.

\section{Results and Discussion}\label{sec:results}

We are now ready to translate the prediction for the Higgs mass and the EWPO into bounds on the parameter space of our simplified model and for the composite TH in general. 
We are interested in quantifying the degree of fine tuning that our construction suffers when requiring the mass scale of the heavy fermions to lie above 
the ultimate experimental reach of the LHC. As discussed in Section 2, this scale receives the largest boost from the TH mechanism (without a corresponding increase in the fine tuning of the Higgs mass) in the regime of parameters corresponding to a fully strongly-coupled theory, where no quantitatively precise EFT description is allowed. Our computations of physical quantities in this most relevant regime should then be interpreted as an {\it {educated Naive Dimensional Analysis} } (eNDA) estimate, where one hopes to capture the generic size of effects beyond the naivest $4\pi$ counting, and including factors of a few  related to multiplet size, to spin and to numerical accidents. In the limit where $M_\rho/f$ and $M_\Psi/f$ are significantly below their perturbative upper bounds our computations are well defined. eNDA then corresponds to assuming that the results do not change by more than $O(1)$ (i.e.~less than $O(5)$ to be more explicit) when extrapolating to a scenario where the resonance mass scale sits at strong coupling. In practice we shall consider the resonant masses up to their perturbativity bound and vary the  $\alpha_i$ and $c_i$ parameters within an $O(1)$ range.\footnote{Notice indeed that $(\alpha=1,c=0)$ and $(\alpha=0,c=1/\sqrt{2})$ correspond to specific limits at weak coupling, namely the 
two-site model and the linear sigma model respectively. This suggests that their natural range is $O(1)$.} In view of the generous parameter space that we shall explore our analysis should be viewed as conservative, in the sense that a realistic TH model will never do better.

Let us now describe the various pieces of our analysis. Consider first the Higgs potential, where the dependence on physics at the resonance mass scale is encapsulated in the function $F_1$ (Eq.~\eqref{eq:deltamHUV}) which controls the UV threshold correction to the Higgs quartic. It is calculable in our simplified  model and the result is $O(1)$
(its expression is reported in Eq.~(\ref{FFunctions})), but it can easily be made a bit smaller at the price of some mild tuning  by varying the  field content  or the representations of the heavy fermions. In order to account for these options and thus broaden the scope of our analysis we will treat $F_1$ as a free $O(1)$ parameter.
The value of $F_1$ has a direct impact on the size of the left-handed top mixing $y_L$, since $\delta m_h^2|_{UV} \sim y_L^4 F_1$, and hence controls the 
interplay between the Higgs potential and EWPO. Specifically, as we already stressed, a smaller $F_1$ implies a larger value of $y_L$, which in turn gives a larger $\Delta \widehat{T}\propto y_{L}^{4} v^{2}/M_{\Psi}^{2}$.  This could help  improve the compatibility with EWPT even for large $M_\Psi$, at the cost of a small additional tuning due to the need for a clever maneuver in the $\widehat S,\widehat T$ plane to get back into the ellipse, as well as the fact that  $F_1$ is generically expected to be $O(1)$.
In the following we will thus treat $F_1$ as an input parameter and use Eqs.~\eqref{eq:deltamHUV} and \eqref{MassZero} to fix $y_L$ and $y_{R}$ in terms of the Higgs and top quark experimental masses.
Our final results will be shown for two different choices of $F_1$, namely $F_1 =1$ and $F_1 = 0.3$, in order to illustrate how the bounds are affected by changing the
size of the UV threshold correction to the Higgs potential.

The EWPO and the Higgs mass computed in the previous sections depend on several parameters, in particular on the mass spectrum of resonances (see Appendix \ref{AppSpectrum}), 
the parameters $c_i$, $\alpha_i$ of Eqs.~\eqref{LCompF},~\eqref{LCompVF}, and the parameter $F_{1}$  discussed above. In order to focus on the situation where resonances can escape detection at the LHC,  we will assume that their masses are all comparable and that they lie at or just below
the cutoff scale. In order to simplify the numerical analysis we thus set $M_\Psi = M_S=\widetilde{M}_{\Psi}= \widetilde{M}_S=M_{\rho}=M_{\rho^{X}}=m_{*}/2$.
The factor of two difference between $M_{\Psi}$ and $m_{*}$ is chosen
to avoid setting all UV logarithms of the form $\log(m_{*}/M_{\Psi})$
to zero, while not making them artificially large. As a further simplification 
we set $c_{L}=c_{R}\equiv c$, $\alpha_{{\bf{7}}L}=\alpha_{{\bf{1}}L}$ and $\alpha_{{\bf{7}}R}=\alpha_{{\bf{1}}R}$. 
The parameter $\alpha_L$ appears only in the tree-level contribution to $\delta g_{Lb}$, see Eq.~\eqref{DeltaTree}, and we fix it equal to~1 for simplicity. Even though the above choices represent a significant reduction of the whole available parameter space, for the purpose of our analysis they represent a sufficiently rich set where EWPT can be successfully passed.

\begin{figure}[t!]
\begin{center}
\includegraphics[width=0.43\textwidth]{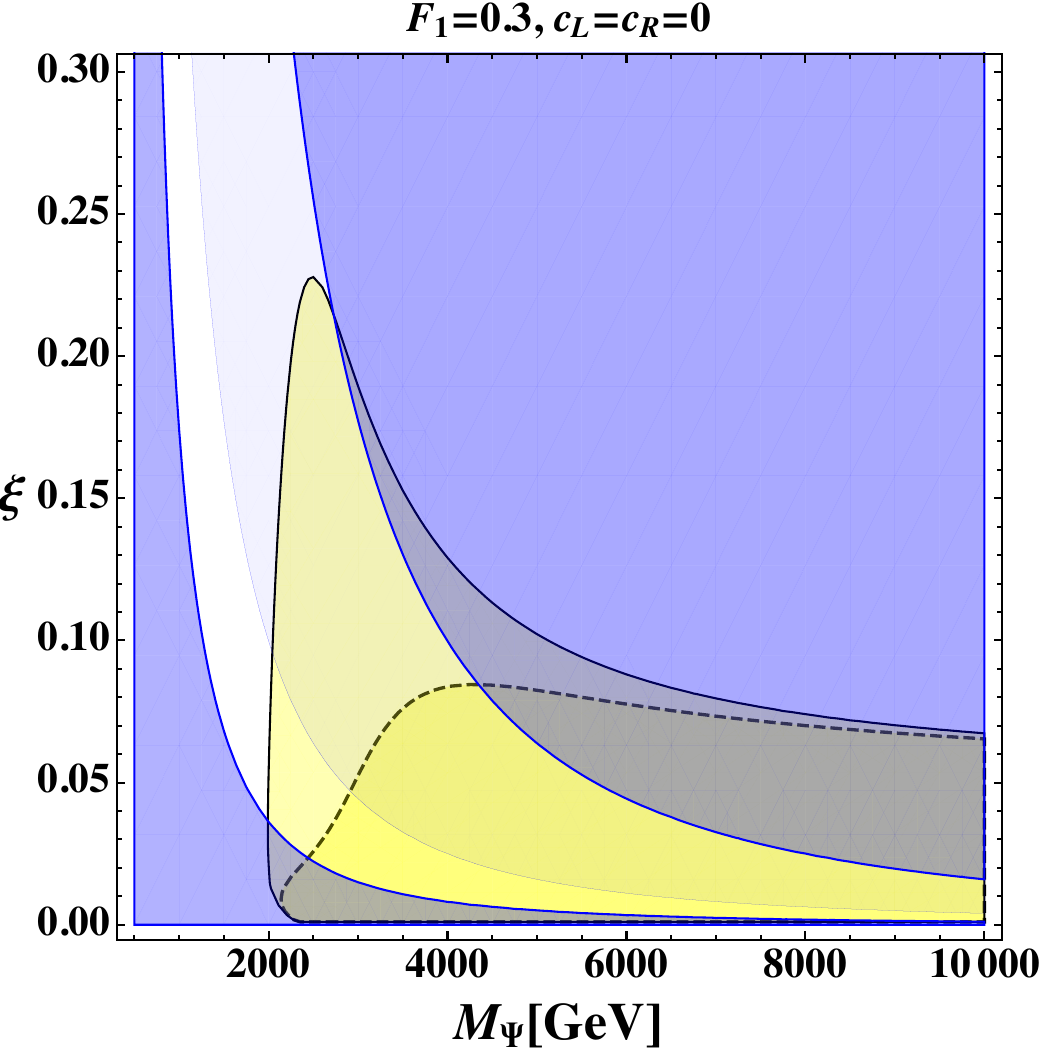}
\hspace{1cm}
\includegraphics[width=0.43\textwidth]{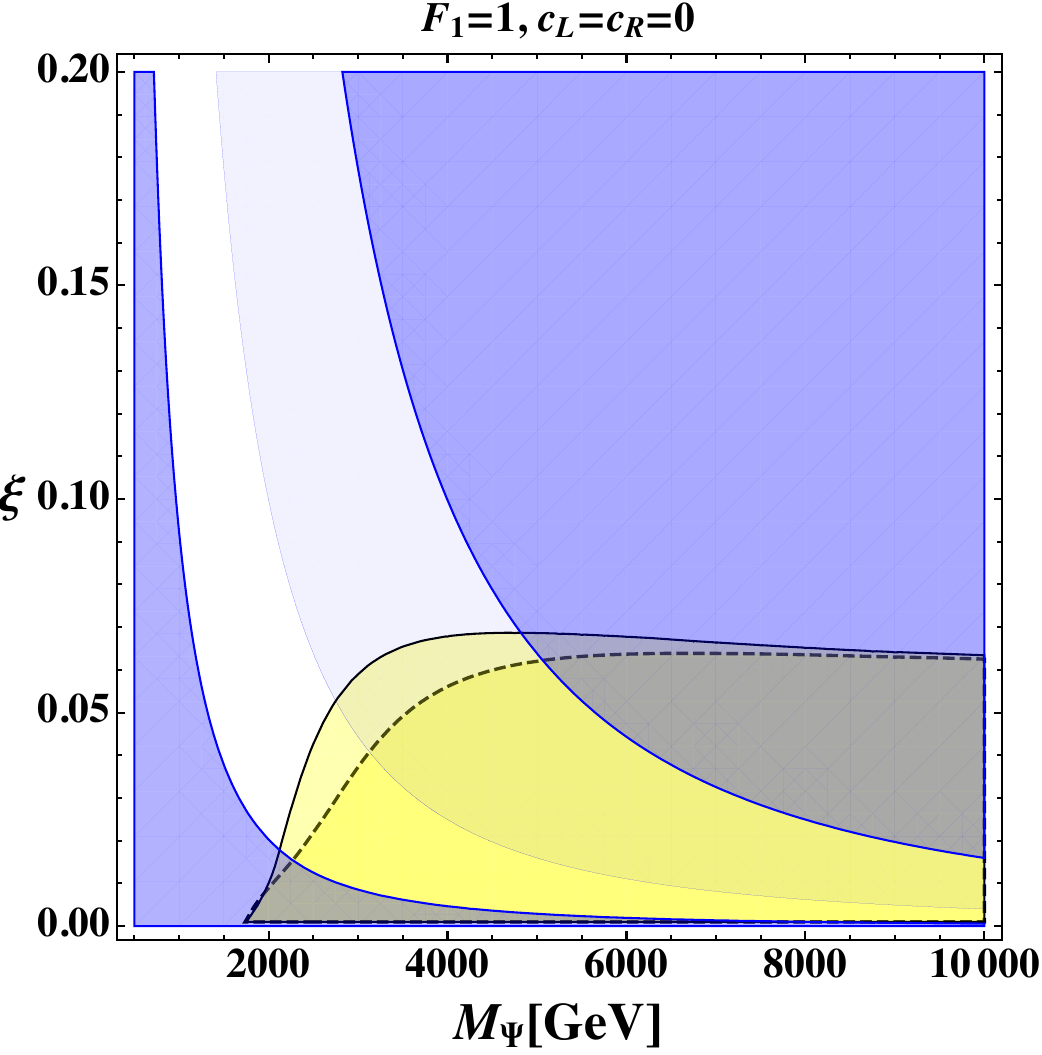}
\end{center}
\caption{\it Allowed regions in the $(M_\Psi, \xi)$ plane for $F_1=0.3$ (left panel) and $F_{1}=1$ (right panel). See the text for an explanation of the different regions 
and of the choice of parameters.}
\label{fig:mpsixi}
\end{figure}

Let us now discuss the numerical bounds on the parameter space of our simplified model. They have been obtained by fixing the top and Higgs masses to their 
experimental value and performing the numerical fit described in Appendix \ref{ewfit}.
As experimental inputs, we use the PDG values of the top quark pole mass $m_t = 173.21\pm 0.51\pm 0.71$
(see last paragraph of Appendix \ref{AppBetaFunc}), and of the Higgs mass, $m_h = 125.09 \pm 0.24\,$GeV~\cite{Olive:2016xmw}.
Figure \ref{fig:mpsixi} shows the results of the fit in the $(M_{\Psi},\xi)$ plane for $F_1=0.3$ (left panel) and $F_{1}=1$ (right panel). 
In both panels we have set $c=0$, which corresponds to the two-site model limit of our simplified Lagrangian.
%
%
The yellow regions correspond to the points 
that pass the $\chi^2$ test at $95\%$ confidence level (CL), see Appendix \ref{ewfit} for details. Solid black contours denote the regions for which 
$\alpha_{{\bf{1}}L}= - \alpha_{{\bf{1}}R}=1$, while dashed contours surround the regions obtained with $\alpha_{{\bf{1}}L}=\alpha_{{\bf{1}}R}=1$. The areas in blue are theoretically inaccessible.
The lower left region in dark blue, in particular, corresponds to $M_{\Psi}/f \equiv g_\Psi <y_L$.
The upper dark- and light-blue regions correspond instead to points violating the perturbative limits on $g_\Psi$ given by Eq.~\eqref{FirstLimit} with $N=8$,
and Eq.~\eqref{SecondLimit} with $N_f =24$, respectively (see Section \ref{sec:perturbativity} for a discussion).
The difference between these two regions can be taken as an indication of the uncertainty associated with the perturbative bound.\footnote{ Notice that because of our choice $m_*= 2 M_\Psi$, the scale $m_*$ lies a factor of 2 above the pertubative cut-off. This is compatible with the semiquantitative nature of our estimates. As we stated previously we insisted in keeping $m_*= 2 M_\Psi$ because it implies a more generic contribution to electroweak precision observables.}

In the left panel of Figure \ref{fig:mpsixi} the allowed (lighter yellow) region extends up to $\xi\simeq 0.2$ for masses $M_{\Psi}$ in the $2-3$ TeV range. Such large values of $\xi$ are possible in this case because the fermionic contribution to $\Delta \widehat T_\Psi$ turns out to be sufficiently large and positive to compensate for both the negative $\Delta \widehat T_h$ in Eq.~\eqref{eq:dTH} and the positive $\Delta \widehat S_{\rho}$ and $\Delta \widehat S_{h}$ 
in Eqs.~\eqref{eq:dSrho},~\eqref{eq:dShiggs}. 
For larger $M_\Psi$ the fermionic contribution $\Delta \widehat T_\Psi$ becomes too small and this compensation no longer occurs.
In this case, however, the strongest bound comes from the perturbativity limit (blue region), which makes points with large $M_\Psi$ at fixed $\xi$ theoretically inaccessible.
Notice that large values of $\xi$ become excluded if one considers the choice $\alpha_{{\bf{1}}L}=\alpha_{{\bf{1}}R}=1$ leading to the dashed contour.
The large difference between the solid and dashed curves (i.e.~lighter and darker yellow regions) depends on the sign correlation between $\Delta \widehat T_\Psi$ 
and $\delta g_{Lb}$.
In the case of the solid line, the signs are anti-correlated (e.g.~positive $\Delta \widehat T_\Psi$ and negative $\delta g_{Lb}$), allowing for the 
compensation effect by $\Delta \widehat T_\Psi$. In the case of the dashed line, instead, the signs of the two parameters are correlated (both positive), so that when 
$\Delta \widehat T_\Psi$ is large, $\delta g_{Lb}$ is also large and positive. This makes it more difficult to pass the $\chi^2$ test, since data prefer a negative $\delta g_{Lb}$. 

In the right panel of Figure \ref{fig:mpsixi}, obtained with $F_{1}=1$, the allowed yellow region shrinks because the larger value of $F_{1}$ implies a smaller $y_{L}$ hence
a smaller $\Delta \widehat T_\Psi$. In this case the $\chi^2$ test is passed only for $\xi< 0.06$, and the difference between the solid and dashed lines is small since the 
large and positive $\Delta\widehat{S}$ always dominates the fit. Masses $M_\Psi$ larger than $\sim 5\,$TeV are excluded by the perturbative bound, unless one considers
smaller values of $\xi$. 

\begin{figure}[t!]
\begin{center}
\includegraphics[width=0.43\textwidth]{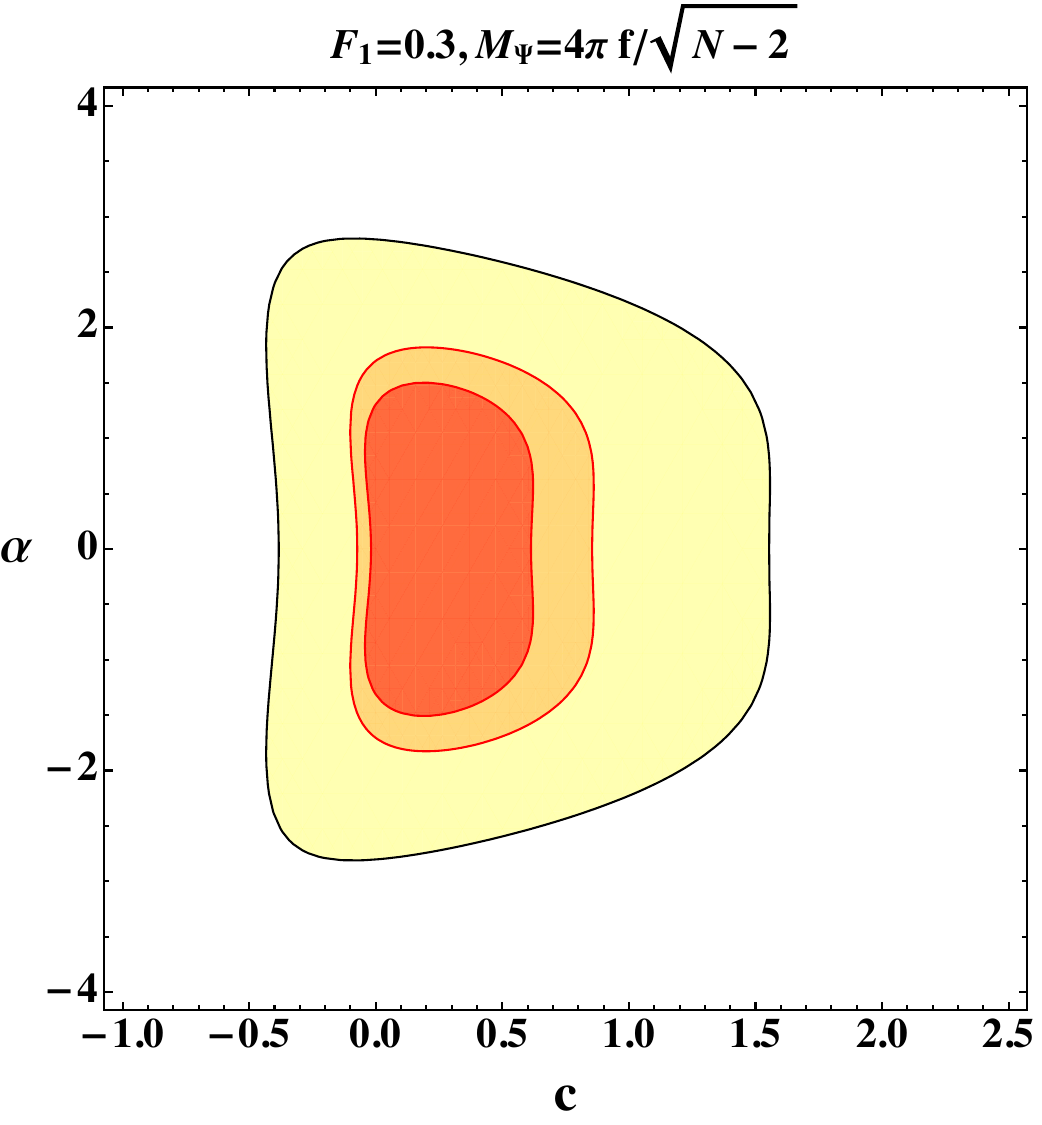}
\hspace{1cm}
\includegraphics[width=0.43\textwidth]{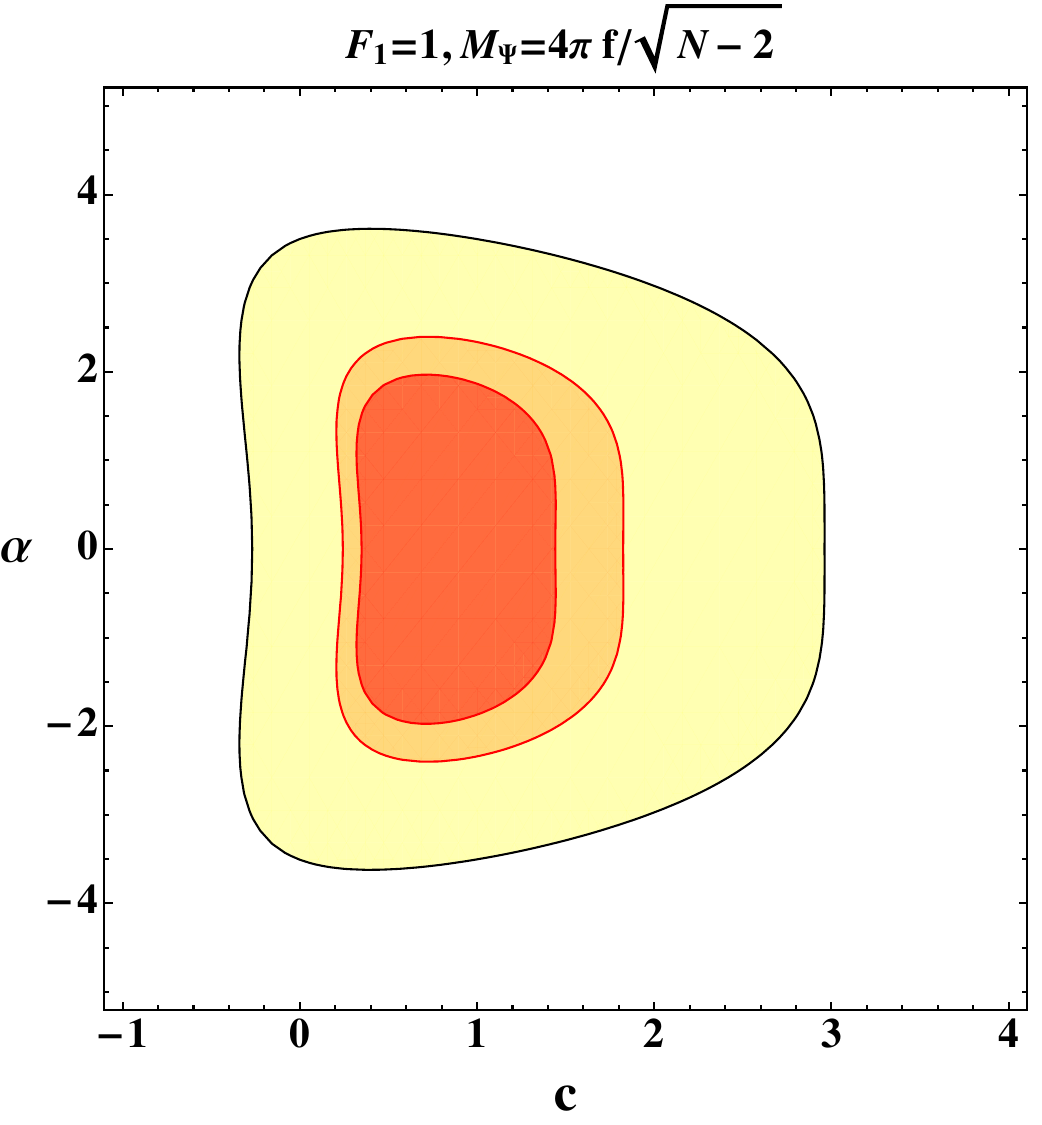}
\end{center}
\caption{\it Allowed regions in the $(c, \alpha)$ plane, with $c=c_{L}=c_{R}$, for $F_1=0.3$ (left panel) and $F_{1}=1$ (right panel). 
The yellow, orange and red regions correspond to $\xi = 0.05$, $0.1$ and $0.15$ respectively.
See the text for an explanation of the choice of the other parameters.}
\label{fig:calpha}
\end{figure}

The results of Figure \ref{fig:mpsixi} can change significantly if the parameter $c$ is allowed to be different from zero. In particular, as one can verify from our
formulae in Appendix \ref{AppFormulae}, positive values of $c$ increase $\Delta \widehat{T}$ and as a result the allowed regions in Figure \ref{fig:mpsixi} shift to the right
towards larger values of $M_\Psi$. In this case the perturbative bound excludes a large portion of the region passing the $\chi^2$ test.
The effect of varying $c$ is illustrated in Figure \ref{fig:calpha}, which shows the 95$\%$ CL allowed regions in the plane $(c,\alpha)$ for 
$F_{1}=0.3$ (left panel) and $F_{1}=1$ (right panel).
%
%
In both panels we have set $\alpha \equiv \alpha_{{\bf{1}}L}=-\alpha_{{\bf{1}}R}$ (ensuring positive $\Delta \widehat T_\Psi$ and negative $\delta g_{Lb}$).
The yellow, orange and red regions correspond, respectively, to $\xi=0.05$, $\xi=0.1$ and $\xi=0.15$, with the masses
of the resonances fixed at their perturbative upper bound $ M_\rho=M_\Psi =4\pi f/\sqrt{N-2}$ (which for $N=8$ gives $ \sim 6/ 4 /3.2 ~\text{TeV}$). 
Note that increasing $F_{1}$ (reducing $y_{L}$) shifts the allowed region towards positive values of~$c$, since as mentioned above, smaller $y_{L}$ requires a larger positive $c$ to get a large enough $\Delta \widehat T_\Psi$. Obviously, larger values of $\xi$ correspond to smaller allowed regions, as is clear from Figure \ref{fig:mpsixi}. Finally, notice that the vertically-symmetric structure of the allowed regions is due to the quadratic dependence
of $\delta g_{Lb}$ on $\alpha$.
From these plots, one can see that for resonances conceivably out of direct reach of the LHC and  for $\xi\sim 0.1$, corresponding to about $20\%$ tuning of the Higgs mass,  both $\alpha$ and $c$ are allowed to span a good fraction of their expected  $O(1)$ range. No dramatic extra tuning in these parameters seems therefore necessary to meet the constraints of EWPT. In particular, considering the plot for $F_1=1$ (right panel), one notices that the bulk of the allowed region is at positive $c$. For instance by choosing $c\sim 0.2-0.5$ the plot in the $(\xi, M_\Psi)$ plane for $F_1=1$ becomes quite similar to the one at the left of Figure 3 valid for $F_1=0.3$: there exists a ``peak" centered at $M_\Psi \sim 2-4$ TeV and extending up to $\xi\sim 0.2$. The specific choice $c=0$ is thus particularly restrictive for $F_1=1$ (right panel of Figure \ref{fig:mpsixi}), but this restriction is lifted for positive $c$. Overall we  conclude that for $\xi \sim 0.1$ and for resonances just beyond the LHC reach, the correct value of the Higgs quartic can be obtained and EWPT passed with only a mild additional tuning associated with a sign correlation $\alpha_{{\bf{1}}L}=-\alpha_{{\bf{1}}R}$, and a correlation between $c$ and $F_1$ (e.g. $c>0$ for $F_1=1$). These correlations allow the various contributions to $\widehat T$ and $\delta g_{Lb}$ to compensate for each other, achieving agreement with EWPT. If forced to quantify the tuning inherent in these effects, we could estimate it to be around $1/4=(1/2)\times (1/2)$, given about $1/2$ of the plausible choices for both $\alpha_i$ and $c_i$ are allowed.

\section{Summary}\label{sec7}
In this paper we tried to assess how plausible a scenario yielding no
new particles at the LHC can be obtained using the TH construction. We
distinguished three possible classes of models: the sub-hypersoft,
the hypersoft  and the super-hypersoft, with increasing degree of technical
complexity and decreasing (technical) fine tuning. We then focused on
the CH incarnation of the simplest option, the sub-hypersoft
scenario, where the
boost factor for the mass of colored partners (Eq.~\eqref{boost1}) at
fixed tuning is roughly given by  
\begin{equation}
\frac{g_*}{\sqrt 2 y_t}\times \frac{1}{\sqrt{\ln (m_*/m_t)}}\;.
\label{boostfinal}
\end{equation}
Here the gain derives entirely from the relative coupling strength $g_*/\sqrt 2 y_t$, making the marriage of Twinning and compositeness practically obligatory.  We attempted a more precise estimate of the upper limit on  $g_*/\sqrt 2 y_t$, as compared with previous studies (e.g.~Ref.~\cite{Barbieri:2015aa}). We found by independent but consistent estimates, that the bound ranges from 
$ \sim 3$ in a toy sigma model (Eq.~(\ref{boundsigma})) to $ \sim 5$ in a simplified CH model (Eq.~(\ref{FirstLimit})), with both limits somewhat below  the NDA estimate of $ 4\pi\sim 12$. Consequently for a mild tuning $\epsilon \sim 0.1$ the upper bound on the mass of the resonances with SM quantum numbers is closer to the $3-5$ TeV range than it is to $10$ TeV.
This gain, despite being less spectacular than naively expected, is
still sufficient to push these states out of direct reach of the LHC,
provided we resort to full strong coupling. In practice this implies no real computational advantage from considering holographic realizations of composite TH constructions: since the boost
factor is controlled by the KK coupling, the 5D description breaks down in precisely the most interesting regime, where the KK coupling is strong. In this situation computations based on an explicit 5D
construction, such as the ones studied in
Refs.~\cite{Geller:2014kta,Csaki:2015gfd} for instance, are no better than numerical estimates made in our simplified model.  Indeed we have checked that EWPT can be satisfied in a sizable portion of the
parameter space, given some interplay among the various
contributions. 
In particular the IR corrections to $\widehat T$ and
  $\widehat S$ are enhanced by $\ln (m_*/m_h)$, and for $\xi>0.1$ the
  compensating contribution to $\widehat T$, which  decreases like
  $1/m_*^2$, is necessary. Given that perturbativity limits $m_*$ to be
  below $5$ TeV for $\xi >0.1$ (see the upper blue exclusion region in
  Fig.~\ref{fig:mpsixi}) this compensation in EWPT can still take
  place at the price of a moderate extra tuning. For $\xi$ of order a
few percent on the other hand, EWPT would be passed without any
additional tuning, while the masses of SM-charged resonances would be
pushed up to the 10 TeV range, where nothing less than a 100 TeV
collider would be required to discover them, and that barely so \cite{Golling:2016gvc,Matsedonskyi:2015dns}. 

Although EWPT work similarly in the CH and composite TH frameworks,
the two are crucially different when it comes to contributions to the
Higgs quartic.  In the CH these are enhanced when $g_*$, i.e. $m_*/f$,  is strong and,
as discussed for instance in Ref.~\cite{DeSimone:2012ul}, in order to
avoid additional tuning of the Higgs quartic  $g_*$ cannot be too large. According to the study in Refs.~\cite{Matsedonskyi:2012ws,Marzocca:2012tt,Pomarol:2012vn,Pappadopulo:2013wt}
the corresponding upper bound on the mass of the colored top partners in CH reads roughly $m_*/f \lsim 1.5$; this should be compared to 
the upper bound $m_*/f \lsim 5$ from strong coupling we found in Eq.~(\ref{FirstLimit}). The Higgs quartic protection afforded by the TH  mechanism
allows us to take $m_*/f$ as large as possible, allowing  the colored partners to be heavier at fixed $f$, hence at fixed fine tuning $\xi$. In the end the gain is about a factor of $5/1.5\sim 3$, not impressive, but sufficient to place the colored partners outside of LHC reach for a mild tuning  $\xi \sim
0.1$.

Finally, we comment on the classes of models not covered in this
paper: the hypersoft and super-hypersoft scenarios. The latter
requires combining supersymmetry and compositeness with the TH
mechanism, which, while logically possible, does not correspond to any
existing construction.  Such a construction would need to be rather
ingenious, and we currently do not feel compelled to provide it, given
the already rather epicyclic nature of the TH scenario.  The simpler
hypersoft scenario, though also clever, can by contrast be
implemented in a straightforward manner, via e.g.~a tumbling $SO(9)\to
SO(8)\to SO(7)$ pattern of symmetry breaking. The advantage of this
approach is that it allows us to remain within the weakly-coupled
domain, due to the presence of a relatively light Twin Higgs scalar mode
$\sigma$, whose mass can be parametrically close to that of the Twin
tops, $\sim y_t f$ (around $1$ TeV for $\xi \sim 0.1$). As well as
giving rise to distinctive experimental signatures due to mixing with
the SM Higgs \cite{Buttazzo:2015bka}, the mass of the light $\sigma$
acts as a UV cut-off for the IR contributions to $\widehat S$ and
$\widehat T$ in Eqs.~\eqref{eq:dShiggs} and \eqref{eq:dTH})
\cite{Craig:2015pha}.  For sufficiently light $\sigma$ then, less or
no interplay between the various contributions is required in
order to pass EWPT.  Together with calculability, this property
may well single out the hypersoft scenario as the most plausible TH construction.

\label{5}

\section*{Acknowledgments}
We would like to thank
Andrey Katz, Alberto Mariotti, Kin Mimouni, Giuliano Panico, Diego Redigolo, 
Matteo Salvarezza and Andrea Wulzer
for useful discussions.
The Swiss National Science Foundation partially supported the work of D.G. and R.R. under contracts 200020-150060 and 200020-169696,
the work of R.C. under contract 200021-160190, and the work of  R.T. under the Sinergia network CRSII2-160814.
The work of R.C. was partly supported by the ERC Advanced Grant No. 267985 
\textit{Electroweak Symmetry Breaking, Flavour and Dark Matter: One
  Solution for Three Mysteries (DaMeSyFla)}. R.M. is supported by ERC
grant 614577 “HICCUP — High Impact Cross Section Calculations for Ultimate
Precision”.

\appendix


\section{$SO(8)$ generators and CCWZ variables}
\label{AppCCWZ}

In this appendix we define the generators of the $SO(8)$ algebra and describe the $SO(8)/SO(7)$ symmetry-breaking pattern, introducing the CCWZ variables for our model. 
We refer the reader to Refs.~\cite{Coleman:1969p1798} and~\cite{Callan:1969p1799} for a detailed analysis of this procedure and we closely follow 
the notation of Ref.~\cite{Barbieri:2015aa}.

We start by listing the twenty-eight generators of $SO(8)$ and decomposing them into irreducible representations of the unbroken subgroup $SO(7)$: 
$\bf{28} = \bf{7} \oplus \bf{21}$. They can be compactly written as:
\begin{equation}\label{Gener}
(T_{ij})_{kl} = {i \over \sqrt{2}} (\delta_{ik}\delta_{jl}-\delta_{il}\delta_{jk}),
\end{equation}
with $i,j,k,l = 1, \cdots, 8$. We choose to align the vacuum expectation value responsible for the spontaneous breaking of $SO(8)$ to $SO(7)$ along the 8-th
component: $\phi_0 = f (0,0,0,0,0,0,0,1)^t$. With this choice, the broken and unbroken generators, transforming, respectively, in the $\bf{7}$ and $\bf{21}$ of $SO(7)$, are:
\begin{equation}
(T^{\bf {7}}_\beta)_{\gamma\rho} =  {i \over \sqrt{2}} (\delta_{8\gamma}\delta_{\beta \rho}-\delta_{8\rho}\delta_{\beta \gamma})\, ,  \qquad\quad 
(T^{\bf {21}}_{\alpha \beta})_{\gamma \rho} =  {i \over \sqrt{2}} (\delta_{\alpha \gamma}\delta_{\beta \rho}-\delta_{\alpha \rho}\delta_{\beta \gamma})\, ,
\end{equation}
where  $\alpha, \beta= 1,\cdots , 7$ and $\gamma, \rho =1, \cdots , 8$.
It is useful to identify the subgroups $SO(4)\sim SU(2)_L \times SU(2)_R$ and $\widetilde{SO}(4)\sim \widetilde{SU}(2)_L \times \widetilde{SU}(2)_R$
of $SO(8)$; they are generated by:
\begin{equation}\label{GenGauged}
(T_{\bf L})^\alpha =  \begin{pmatrix}  t_{\bf L}^\alpha & 0 \\  0 & 0  \end{pmatrix},  \quad
(T_{\bf R})^\alpha =  \begin{pmatrix} t_{\bf R}^\alpha & 0 \\  0 & 0  \end{pmatrix}, \quad
(\widetilde{T}_{\bf L})^\alpha = \begin{pmatrix}  0 & 0 \\ 0 & t_{\bf L}^\alpha \end{pmatrix}, \quad
(\widetilde{T}_{\bf R})^\alpha =  \begin{pmatrix} 0 & 0 \\ 0 & t_{\bf R}^\alpha \end{pmatrix}\, ,
\end{equation}
where $t_{\bf L}^\alpha$ and $t_{\bf R}^\alpha$ are $4\times 4$ matrices defined as
\begin{equation}\label{GenGauged2}
(t_{\bf L, R}^\alpha)_{ij} = -{i \over 2}\left[{1 \over 2} \epsilon^{\alpha \beta \gamma} \left( \delta ^\beta_i \delta ^\gamma_j - \delta ^\beta_j \delta ^\gamma_i \right) \pm \left(\delta ^\alpha_i \delta^4_j - \delta^\alpha_j \delta^4_i  \right)\right]
\end{equation}
with $\alpha = 1, 2,3$ and $i,j = 1, \cdots ,4$.
The elementary SM and Twin vector bosons gauge, respectively, a subgroup $SU(2)_L\times U(1)_Y$ of $SO(4)$, with $U(1)_Y \equiv U(1)_{R3}$,
and the subgroup $\widetilde{SU}(2)_L$ inside $\widetilde{SO}(4)$. This choice corresponds to having zero vacuum misalignment at tree level.

The spontaneous breaking of $SO(8)$ to $SO(7)$ delivers seven NGBs,
that we collect in the vector ${\bf{\Pi} }=(\pi_2, \pi_1, -\pi_3, \pi_4, \pit_2,\pit_1,-\pit_3)^t$. 
The first four transform as a fundamental of $SO(4)$ and form the Higgs doublet; the remaining three are singlets
of $SO(4)$ and are thus neutral under the SM gauge group. 
All together they can be arranged in the Goldstone matrix
\begin{equation}\label{GoldMat}
\Sigma({\bf{\Pi}})=e^{i{\sqrt{2} \over f} {\bf{\Pi}} \cdot T^{\bf 7}}=
\begin{bmatrix}
\mathbb{I}_7- {{\bf{\Pi}} {\bf{\Pi}}^t \over {{\bf{\Pi}}^t \cdot {\bf{\Pi}}}} \left( 1 - \cos\left( {{ \sqrt{ {\bf{\Pi}}^t \cdot {\bf{\Pi}} } } \over f}\right) \right ) 
&   {{\bf{\Pi}} \over {  \sqrt{ {\bf{\Pi}}^t \cdot {\bf{\Pi}}} } } \sin\left( {{ \sqrt{ {\bf{\Pi}}^t \cdot {\bf{\Pi}}  } } \over f}\right)  \\[0.2cm]
 -{{\bf{\Pi}} \over { \sqrt{  {\bf{\Pi}}^t \cdot {\bf{\Pi}}}  }} \sin\left( {{ \sqrt{ {\bf{\Pi}}^t \cdot {\bf{\Pi}} } } \over f}\right)   
&  \cos\left( {{  \sqrt{ {\bf{\Pi}}^t \cdot {\bf{\Pi}} } } \over f}\right) 
\end{bmatrix}\, .
\end{equation}
The latter transforms non-linearly under the action of an $SO(8)$ group element $g$, according to the standard relation:
\begin{equation}\label{SigmaTrans}
\Sigma({\bf{\Pi}}) \rightarrow g \cdot \Sigma ({\bf{\Pi}}) \cdot h^\dagger({\bf{\Pi}},g),
\end{equation}
where $h({\bf{\Pi}}, g) \in SO(7)$ depends on $g$ and ${\bf \Pi}(x)$. We identify the Higgs boson with the NGB along the generator $T^{\bf{7}}_4$; 
in the unitary gauge, all the remaining NGBs are non-propagating fields and the ${\bf{\Pi}}$ vector becomes
\begin{equation}\label{PiUniGauge}
{\bf{\Pi}}|_{\begin{subarray}{l} \text{unitary} \\ \text{gauge} \end{subarray}} = (0, \cdots, \pi_4 = { \left\langle h \right\rangle + h(x)}, \cdots, 0)\, ,
\end{equation}
where $\xi = \sin^2(\langle h\rangle/f) = v^2/f^2$.
In this case the $\Sigma$ matrix simplifies to:
\begin{equation}\label{GoldMatUnGauge}
\Sigma({\bf{\Pi}})\big|_{\begin{subarray}{l} \text{unitary} \\ \text{gauge} \end{subarray}}=e^{i{\sqrt{2} \over f}\pi_4 T^{\bf 7}_4}=
\begin{bmatrix}
\mathbb{I}_3 & 0 & 0 & 0 \\
0 &  \cos \frac{\pi_4}{f}  & 0 & \sin \frac{\pi_4}{f}  \\
 0 & 0 & \mathbb{I}_3 &  0 \\
 0 & -\sin \frac{\pi_4}{f}  &  0 & \cos \frac{\pi_4}{f}  
\end{bmatrix}\, .
\end{equation}

Given the above symmetry breaking pattern, it is possible to define a LR parity,
\begin{equation}
P_{LR} = \text{diag} (-1,-1,-1,+1,-1,-1,-1,+1)\, ,
\end{equation}
which exchanges $SU(2)_L$ with $SU(2)_R$ inside $SO(4)$ and $\widetilde{SU}(2)_L$ with $\widetilde{SU}(2)_R$ inside $\widetilde{SO}(4)$.
The corresponding action on the fields is such that $\pi_4$
is even, while all the other NGBs are odd.
As already noticed in Sec.~\ref{sec:deltagLb}, $P_{LR}$ is  an element of both $SO(8)$ and  $SO(7)$, which means that it is an exact symmetry of the strong
dynamics and acts linearly on the physical spectrum of fields.

The CCWZ variables $d_\mu$ and $E_\mu$ are defined as usual through the Maurer-Cartan form,
\begin{equation}\label{MaurerCartan}
\Sigma^\dagger({\bf{\Pi}}) D_\mu \Sigma({\bf{\Pi}}) \equiv i d_\mu^i({\bf{\Pi}}) T^{\bf{7}}_i + i E_\mu^a({\bf{\Pi}}) T^{\bf{21}}_a,
\end{equation}
as the components along the broken and unbroken generators of $SO(8)$ respectively.
The derivative $D_\mu$ is covariant with respect to the external SM and Twin gauge fields:
\begin{equation}\label{CovDev}
D_\mu = \partial_\mu - i A_\mu ^A T^A , \qquad \text{with} \quad 
A_\mu ^A T^A = g_2 W_\mu ^\alpha (T_{\bf{L}})^\alpha +g_1  B_\mu (T_{\bf{R}})^3+\widetilde{g}_2 \widetilde{W}_\mu ^ \alpha (\widetilde{T}_{\bf{L}})^\alpha\, .
\end{equation} 
Under the action of a global element $g \in SO(8)$, $d_\mu$ and $E_\mu$ transform with the rules of a local $SO(7)$ transformation:
\begin{equation}\label{dETrans}
d_\mu \equiv d_\mu^i T^{\bf{7}}_i \rightarrow h({\bf{\Pi}},g)d_\mu h^\dagger({\bf{\Pi}},g) ,\qquad 
E_\mu \equiv E_\mu ^a T^{\bf{21}}_a \rightarrow  h({\bf{\Pi}},g) (E_\mu - i \partial_\mu ) h^\dagger({\bf{\Pi}},g)\, .
\end{equation}
It is straightforward to derive the explicit expressions of $d_\mu$ and $E_\mu$ from the Maurer-Cartan relation in Eq.~\ref{MaurerCartan}. They are however lengthy and not very illuminating, so we do not report them here. We can easily obtain the mass spectrum of the gauge sector of our theory from the NGB kinetic term; in the unitary gauge and after rotating to the mass eigenstate basis, we find:
\begin{equation}\label{GaugeMasses}
\mathcal{L}_{mass} = {f^2 \over 4} (d_\mu^i)^2 \supset {g_2^2 \over 4}  f^2 \xi ~W_\mu^+W^{\mu-}+{(g_1^2+g_2^2)\over 8}f^2 \xi ~ Z^\mu Z_\mu + {\widetilde{g}_2^2 \over 8}f^2 (1-\xi) ~\sum_{i=1}^3 (\widetilde{W}_\mu^i)^2.
\end{equation}

\section{Mass matrices and spectrum}\label{AppSpectrum}

In this appendix, we briefly discuss the mass matrices of the different charged sectors in the Composite TH model and the related particle spectrum. We refer to 
Ref.~\cite{Barbieri:2015aa} for the expressions of the $\Psi_{\bf{7}}$ and $\widetilde{\Psi}_{\bf{7}}$ multiplets in terms of their component heavy fermions.

We start by considering the fields that do not have the right quantum numbers to mix with the elementary SM and Twin quarks and whose mass is therefore independent 
of the mixing parameters $y_{L,R}$ and $\widetilde{y}_{L,R}$. These are the composite fermions $X_{5/3}$, $\widetilde{D}_1$ and $\widetilde{D}_{-1}$, with charges $5/3$, 1 and 
$-1$ respectively; their mass is exactly given by the Lagrangian
parameters $M_\Psi$ (for the first one), and $\widetilde{M}_\Psi$ (for
the last two).

The remaining sectors have charge $-1/3$, $0$ and $2/3$ and because of the elementary/composite mixing the associated mass matrices are in general non-diagonal and must be diagonalised by a proper field rotation. The simplest case is the $(-1/3)$-charged sector, containing the bottom quark and the heavy $B$ field; the mass matrix in the 
$\{ b, B\}$ basis is
\begin{equation}\label{MassOneThird}
M_{-1/3}= \begin{pmatrix} 0 & f {y_L} \\  0 & -{M_\Psi}  \end{pmatrix}\, .
\end{equation}
After rotation, we find a massless bottom quark (it has no mass since we are not including the $b_R$ in the model), and a massive $B$ particle with 
$m_B^2= {M_\Psi^2 + y_L^2 f^2}$.

As regards the sector of charge $2/3$, it contains seven different particles: the top quark, the top-like heavy states $T$ and $X_{2/3}$ and four composite fermions that 
do not participate in the SM weak interactions, $S^1_{2/3}, \cdots S^4_{2/3}$. In the $\{t, T, X_{2/3}, S^1_{2/3}, \cdots, S^4_{2/3}\}$ basis, the mass matrix is given by:
\begin{equation}\label{MassTwoThird}
M_{2/3}= \begin{pmatrix}
 0 & \frac{1}{2} f {y_L} \left(\sqrt{1-\xi }+1\right) & -\frac{1}{2} f {y_L} \left(\sqrt{1-\xi }-1\right) & 0 &  \displaystyle -\frac{f {y_L} \sqrt{\xi }}{\sqrt{2}} \\
 0 & -{M_\Psi } & 0 & 0 &  0 \\
 0 & 0 & -{M_\Psi } & 0 &  0 \\
 0 & 0 & 0 & -{M_\Psi }  \times \mathbb{I}_3&   0 \\
 f {y_R} & 0 & 0 & 0  & -{M_S} 
\end{pmatrix}\, .
\end{equation}
The  states $S^1_{2/3}, S^2_{2/3}, S^3_{2/3}$ completely decouple from the elementary sector and  do not mix with the top quark. Their mass is therefore exactly given 
by the Lagrangian parameter $M_\Psi$. The remaining $4\times 4$ matrix is in general too complicated to be analytically diagonalised, but one can easily find the spectrum 
in perturbation theory by expanding $M_{2/3}$ for $\xi \ll 1$, which is in general a phenomenologically viable limit. The leading order expression for the masses is then:
\begin{equation}\label{Mass23}
\begin{split}
m_t^2 & \simeq \displaystyle {f^4 \over 2}{y_L^2 y_R^2 \over {M_S^2+y_R^2 f^2}}{\xi}+ O(\xi^{2}), \\
m_{X_{2/3}} ^2 & = M_\Psi^2, \\
m_T^2 & \simeq\displaystyle M_\Psi^2 + y_L^2 f^2 \left(1-\frac{ \xi }{2}\right)+ O(\xi^2), \\
m_{S^4_{2/3}}^2 & \simeq \displaystyle { M_S^2+  y_R^2 f^2}+ \frac{  y_L^2 f^2 M_S^2 }{2 \left(M_S^2+ y_R^2 f^2\right)}\xi + O(\xi^2)\, . 
\end{split}
\end{equation}
The $X_{2/3}$ fermion can be also decoupled and its mass is exactly equal to $M_\Psi$. On the contrary, the other three particles mix with each other and their mass gets 
corrected after EWSB (as expected, the top mass is generated for non-zero values of $\xi$).

Finally, we analyze the neutral sector of our model. It comprises eight fields, the Twin top and bottom quarks, and six of the composite fermions contained in the 
$\widetilde{\Psi}_{\bf{7}}$ multiplet. Working in the field basis
$\{\widetilde{t}, \widetilde{b}, \widetilde{D}_0^1, \widetilde{D}_0^2, \widetilde{U}_0^1 , \cdots, \widetilde{U}_0^4 \}$, the mass matrix reads:
\begin{equation}\label{MassZero}
M_0 = \begin{pmatrix}
 0 & 0 & -\frac{1}{2} f \sqrt{\xi } \widetilde{y}_L & \frac{1}{2} f \sqrt{\xi } \widetilde{y}_L & 0 & 0 & -\frac{i f \widetilde{y}_L}{\sqrt{2}} & -\frac{f \sqrt{1-\xi } \widetilde{y}_L}{\sqrt{2}} \\
 0 & 0 & 0 & 0 & -\frac{i f \widetilde{y}_L}{\sqrt{2}} & \frac{f \widetilde{y}_L}{\sqrt{2}} & 0 & 0 \\
 0 & 0 & -\widetilde{M}_{\Psi } & 0 & 0 & 0 & 0 & 0 \\
 0 & 0 & 0 & -\widetilde{M}_{\Psi } & 0 & 0 & 0 & 0 \\
 0 & 0 & 0 & 0 & -\widetilde{M}_{\Psi } & 0 & 0 & 0 \\
 0 & 0 & 0 & 0 & 0 & -\widetilde{M}_{\Psi } & 0 & 0 \\
 0 & 0 & 0 & 0 & 0 & 0 & -\widetilde{M}_{\Psi } & 0 \\
 f \widetilde{y}_R & 0 & 0 & 0 & 0 & 0 & 0 & -\widetilde{M}_S 
\end{pmatrix}\, .
\end{equation}
After diagonalisation, we find one massless eigenvalue corresponding to the Twin bottom quark (it does not acquire mass since we are not introducing the $\widetilde{b}_R$ 
field). Four of the neutral heavy fermions completely decouple and acquire the following masses
\begin{equation}\label{Mass0Exact}
m_{\widetilde{U}^1_0}=m_{\widetilde{U}^3_0}=m_{\widetilde{D}^2_0}=\widetilde{M}_\Psi , \qquad\quad m_{\widetilde{U}^2_0}^2={\widetilde{M}_\Psi^2 + \widetilde{y}_L^2 f^2}.
\end{equation}
The elementary/composite mixing induces instead corrections to the masses of the remaining neutral particles; at leading order in $\xi$ we find:
\begin{equation}\label{Mass0}
\begin{split}
m_{\widetilde{t}}^2 & \simeq \displaystyle {f^4 \over {2} }{\widetilde{y}_L^2 \widetilde{y}_R^2 \over {\widetilde{M}_S^2+\widetilde{y}_R^2 f^2}}(1-\xi)+ O(\xi^2),\\
m_{\widetilde{D}^1_0}^2 & \simeq\displaystyle {\widetilde{M}_\Psi}^2+\frac{1}{2} {\widetilde{y}_L}^2f^2 \left(1+\xi \right) +O(\xi^2),\\
m_{\widetilde{U}^4_0}^2 & \simeq \displaystyle {\widetilde{M}_S}^2+ \widetilde{y}_R^2 f^2 
                                    +\frac{ \widetilde{y}_L^2 f^2 \widetilde{M}_S^2} {2 \left(\widetilde{M}_S^2+\widetilde{y}_R^2 f^2\right)} \left( 1- \xi \right)+O(\xi^2)\, .
\end{split}
\end{equation}

We conclude by noticing that the masses of the particles in different charged sectors are not unrelated to each other, but must be connected according to the action of the 
Twin Symmetry. In particular, it is obvious that the two singlets $S^4_{2/3}$ and $\widetilde{U}^4_0$ form an exact Twin pair, as it is the case for each SM quark and the 
corresponding Twin partner. The remaining pairs can be easily found from the spectrum and correspond to the implementation of the Twin Symmetry in the Composite Sector 
as defined in \cite{Barbieri:2015aa}.

\section{RG-improvement of the Higgs effective potential}
\label{AppBetaFunc}

In this appendix we describe the computation of the Higgs effective
potential and its RG-improvement. First of all, the UV threshold correction can be 
computed with a standard Coleman-Weinberg (CW) procedure, from which we can easily derive the function $F_1$ of Eq.~\eqref{eq:deltamHUV}. We find:
\begin{equation}\label{FFunctions}
\begin{split}
F_1 =\displaystyle 
-\frac{1}{4}\bigg[ & 1 + \frac{M_S^{4}}{\left(M_S^{2}+f^2 y_R^2\right){}^2} 
                             - \frac{M_S^{2}+M_{\Psi }^{2}-f^2 y_R^2}{M_S^{2}-M_{\Psi }^{2}+f^2 y_R^2}\,\log {m_* ^2 \over M_{\Psi}^2} \\[0.15cm]
& \displaystyle    +\frac{M_S^{2} \left(M_S^{2} \left(M_{\Psi }^{2}+f^2 y_R^2\right)+2 f^2 y_R^2 M_{\Psi }^{2}+M_S^{4}\right)}{\left(M_S^{2}+f^2 y_R^2\right){}^2 
                            \left(M_S^{2}-M_{\Psi }^{2}+f^2 y_R^2\right)} \,\log \frac{m_* ^2}{M_S^{2}+f^2 y_R^2} \bigg].
\end{split}
\end{equation}

The IR contribution to the Higgs mass can be organized using a joint expansion in $\xi$ and
$(\alpha_i \log)$, where 
$\alpha_i=g_i^2/(4\pi)$ for couplings $g_i$. Schematically:
\begin{equation}
\begin{split}
\label{potexpansion}
\delta m_h^{2}|_\text{IR} =
m_t^2\f{\alpha_{t}}{4\pi} t\bigg[& {\red a_{1}}+{\color{verdes}  b_{1}\xi+b_{2}\f{\alpha_{t}}{4\pi}t+b_{3}\f{\alpha_{S}}{4\pi}t}+ {\blue c_{1}\xi^{2}+ c_{2}\xi \f{\alpha_{S}}{4\pi}t+c_{3}\f{\alpha_{S}^{2}}{16\pi^{2}}t^2}\\
 & +c_{4}\xi \f{\alpha_{t}}{4\pi}t
 +c_{5}\f{\alpha_{t}^{2}}{16\pi^{2}}t^2+
 c_{6}\f{\alpha_{t}}{4\pi}\f{\alpha_{S}}{4\pi}t^2\bigg]
-{\red a_{2}  m_{W}^{2}\frac{\alpha_{\text{EW}}}{4\pi} t} +\text{Twin}\,,
\end{split}
\end{equation}
where $t=\log m_{*}^{2}/m_t^{2}$ and all the couplings are evaluated
at the scale $m_*$. Here $a_{i},b_{i},c_{i}$ are the $O(1)$ coefficients of
the LO, NLO and NNLO terms respectively.
The Twin contribution is obtained by substituting all $\alpha_i$,
$a_{i},b_{i},c_{i}$ with the corresponding tilded quantities and $t$
with $\tilde t = \log m_{*}^{2}/m_{\tilde t}^{2}$.
The calculation of the LO and NLO terms is straightforward and all
these contributions are included in our calculation. The NNLO terms
indicated in blue are also simple to evaluate, and are included in our
final result, indicated as NNLO$^{*}$. The calculation of the
remaining NNLO terms is more complicated, since it involves the
running of several higher dimensional operators involving both the SM
and Twin fields. An attempt to compute the full potential at NNLO has
been presented in Ref.~\cite{Greco:2016zaz}. However, while that
result includes the contribution of the SM Goldstone bosons,
additional contributions induced by the Twin Goldstones, which are
expected to arise at NNLO, are not included. Since the full
calculation at NNLO is beyond the scope of this paper, we limit
ourselves to only include the colored contributions in Eq.~\eqref{potexpansion}. 
Comparing with Ref.~\cite{Greco:2016zaz} suggests that our NNLO$^{*}$ result
gives an overestimate of the IR Higgs mass; that this is the case at
large $m_*$ is also evident from Figure \ref{fig:deltamHIR}.
For this reason we consider the NNLO$^{*}$ curve only as an indication
of the importance of the NNLO correction,
and use as final input for our numerical analysis the average of the LO and NLO results as described in Section \ref{sec:effpot}.

We present now a procedure for computing the aforementioned
contributions to the Higgs mass based on the approach of
Ref.~\cite{Greco:2016zaz}. 
The RG-improvement of the Higgs effective
potential can be obtained by solving the one-loop $\beta$-function of
the vacuum energy in the background of the Higgs field. The fermion
contribution to the vacuum energy is given by
\be\l{CWimproveda}
\f{d V_{f}(h_{c},t)^{\text{impr}}}{dt}=\f{N_{c}}{16\pi^{2}}\Big[M_{t}(h_{c},t)^{4}+M_{\ttil}(h_{c},t)^{4}\Big]\,,
\ee
whereas the leading gauge contribution is given by
\be
V_{g}(h_{c},t)=\f{3}{16\pi^2}\left[2M_W(h_c)^2 +
M_Z(h_c)^2 + 3M_{\t{W}}(h_c)^2 \right]t\,.
\ee

In order to compute the improved potential to NNLO$^*$
accuracy, we must input the gauge boson masses at tree-level and the renormalized top and Twin top masses
with leading-$\log$ accuracy in the corresponding Yukawa couplings and
with next-to-leading-log accuracy in the strong gauge couplings. After solving equation
\eqref{CWimproveda} for the effective potential, one must properly
account for the Higgs wavefunction renormalization before expanding
around the minimum of the potential in order to compute the physical
Higgs mass.  We fill in the salient details below.

On integrating out the heavy degrees of freedom, the Composite Twin
Higgs model at the scale $\mst $ can be represented by an an effective
Lagrangian that is invariant under a global $SU(3)_{c}\times
SU(2)_{L}\times U(1)_{Y}$ symmetry, identified with the gauge group of
the SM, as
well as an additional $\widetilde{SU(3)}_{c}\times
\widetilde{SU(2)}_{L+R}$.  $\widetilde{SU(3)_{c}}$ is the gauged Twin color
and $\widetilde{SU(2)}_{L+R}$ is a global Twin custodial group, broken
only by fermion interactions, 
under which the Twin gauge bosons ($\widetilde{W}$) and
Goldstone bosons (GBs) transform as triplets. We work in the gaugeless
limit $g=g'=\tilde{g}=0$, since we are only interesting in capturing the leading
contributions to
the Higgs potential that parametrically depend on $y_{t}$,
$\widetilde{y}_{t}$, $g_{S}$ and $\widetilde{g}_{S}$. 

The relevant light degrees of freedom are the SM fermion doublet,
$Q_{L}$, and singlet, $t_{R}$, and their Twin counterparts $\ttil_L$
and $\ttil_R$, the gluons and their Twins; the Higgs doublet (including
the massless GBs) and the Twin GBs.
For economy of notation we define a Twin Higgs doublet in analogy with
the SM one,
\be\l{Hdefs}
\dst \widetilde{H}
=\f{1}{\sqrt{2}}\(\bry{c}\sqrt{2}\pit^{+}\\ \widetilde{h}
+i\pit^{0}
\ery
\)\,,\qquad\textrm{with
   }\quad\widetilde{h}=f\sqrt{1-\f{2H^\dag H +\pit^{2}}{f^{2}}}\, ,
\ee
making explicit the dependence on the Twin Higgs which is integrated out at tree-level, thus realising the $SO(8)$ symmetry
nonlinearly.

Since the background field calculation at leading-log order requires corrections to the
fermion masses and Higgs wavefunction at only one-loop, we
can neglect in the effective Lagrangian at the scale $m_{*}$, any
operator that has vanishing coefficients at tree-level. We also omit all operators that cannot be predicted solely in
terms of the low-energy parameters of the theory ($y_{t}$ and
$\widetilde{y}_{t}$) and $f$, including products of Higgs and fermion
currents that arise from integrating out composite
fermions\footnote{In any case, these do not contribution to the
  effective potential at this order.}.
We make the following field redefinition:
\be\l{fredef}
\dst f\f{\sin\(\f{|\Pi|}{f}\)}{|\Pi|}\Pi^{i}\to \Pi^{i}\,
,\qquad\textrm{for}\quad|\Pi|=\sqrt{2\Hd H+\pit^2}
\ee
which allows the effective Lagrangian, containing a
pure scalar sector $\La_{S}$ and a fermionic sector $\La_{F}$, to be expressed in an especially
compact form.  The relevant
operators in each sector are given below:
\be\l{Lsigmabis}
\La_{S}= D_{\mu}\Hd D^{\mu} H+\f{1}{2}\demub \pit\demua\pit +\f{c_{H}\O_{H}+c_{H\pit}\O_{H\pit}+c_{\pit}\O_{\pit}}{2f^2} + 2 d_H\frac{\Hd H}{f^2}\O_H\,,
\ee
and
\begin{equation}
\l{LaF1bis}
\begin{split}
\La_{F}
=&\, i\ovl{b}_{L}\dsl b_{L}+i\ovl{b}_{R}\dsl b_{R}+i\ovl{t}_{L}\dsl t_{L}+i \ovl{t}_{R}\dsl t_{R} -y_{t}\left[\ovl{Q}_{L}H^{c} t_{R}+\text{h.c.}\right]\\
  &+i\ovl{\widetilde{b}}_{L}\dsl \widetilde{b}_{L}+i\ovl{\widetilde{b}}_{R}\dsl \widetilde{b}_{R}+i\ovl{\ttil }_{L}\dsl \ttil _{L}+i \ovl{\ttil }_{R}\dsl \ttil _{R}- \widetilde{y}_{t}\left[\ovl{\widetilde{Q}}_{L}\widetilde{H}^{c} \ttil _{R}+\text{h.c.}\right]\,.
\end{split}
\end{equation}
where
\be
\bry{l}
\dst \O_{H}=\demub(\Hd H)\demua (\Hd H)\,,\vspace{1mm}\\
\dst \O_{\pit}=\demub\pit^2\demua\pit^2\,,\vspace{1mm}\\
\dst \O_{H\pit}=\demub \(\Hd H\)\demua \pit^2\,.
\ery
\ee
We have defined a Twin fermion doublet
$\widetilde{Q}_{L}=(\ttil_{L}, \widetilde{b}_{L})^{T}$, transforming
under the Twin-sector symmetries, and the Yukawa couplings are defined
at tree-level as
\be
y_{t}(m_{*})=\f{y_{L}y_{R}f}{\sqrt{M_{S}^{2}+y_{R}^{2}f^{2}}}\quad
\textrm{and}\quad  
\widetilde{y}_{t}(m_{*})=\f{y_{\widetilde{L}}y_{\widetilde{R}}f}{\sqrt{\widetilde{M}_{S}^{2}+y_{\widetilde{R}}^{2}f^{2}}}\,.
\ee

The Wilson coefficients at the scale $m_{*}$ have the boundary values:
\be\l{Wcoefficients1bis}
c_{H}(m_{*})=c_{H\pit}(m_{*})=c_{\pit}(m_{*})=d_H(m_*)=1\, ,
\ee
while the $Z_2$ symmetry enforces
\be
y_{t}(m_{*})=\widetilde{y}_{t}(m_{*})\quad\textrm{and}\quad \alpha_s(m_{*})=\widetilde{\alpha_s}(m_{*})
\ee

We now expand the effective lagrangian in the background of the Higgs
field $h=h_{c}+\hat{h}$ to obtain:
\be
\La_{S}=Z_{\pi}(h_{c})\(\f{1}{2} \demub \pi^{0} \demua \pi^{0}+\demub \pi^{+} \demua \pi^{-}\)+\f{1}{2}Z_{\hat{h}}(h_{c}) \demub \hat{h} \demua \hat{h}+\f{1}{2}Z_{\pit}(h_{c}) \demub \pit \demua \pit
\,,
\ee
where we defined
\be\l{scalarsbackground}
Z_{\pi}(h_{c})=1\,,\qquad Z_{\hat{h}}(h_{c})=1+c_{H}\f{h_{c}^{2}}{f^{2}}+d_H \frac{h_c^4}{f^4}\,,\qquad Z_{\pit}(h_{c})=1 \,,
\ee
and
\begin{equation}
\begin{split}
\La_{F}
=&\dst \sum_{\psi}iZ_{\psi}(h_{c})\ovl{\psi}\dsl \psi -m_{t}(h_{c})\ovl{t}t -m_{b}(h_{c})\ovl{b}b-m_{\ttil }(h_{c})\ovl{\ttil }\ttil-m_{\widetilde{b}}(h_{c})\ovl{\widetilde{b}}\widetilde{b}\\
& -\f{y_{t}(h_{c})}{\sqrt{2}}\hat{h}\ovl{t}t-\f{iy_{t}(h_{c})}{\sqrt{2}}\pi^{0}\ovl{t}\gamma^{5}t-\f{iy_{t}(h_{c})}{2}\pi^{-}\ovl{b}\(1+\gamma^{5}\)t+\f{iy_{t}(h_{c})}{2}\pi^{+}\ovl{t}\(1-\gamma^{5}\)b\\[0.15cm]
& +\f{ \widetilde{y}_{t}(h_{c})}{\sqrt{2}}\hat{h}\ovl{\ttil
}\ttil-\f{i y_{\pit\ttil \ttil }(h_{c})}{\sqrt{2}}\pit^{0}\ovl{\ttil
}\gamma^{5}\ttil-\f{i y_{\pit\ttil \widetilde{b}
  }(h_{c})}{2}\pit^{-}\ovl{\widetilde{b}}\(1+\gamma^{5}\)\ttil +\f{i
  y_{\pit\ttil \widetilde{b} }(h_{c})}{2}\pit^{+}\ovl{\ttil
}\(1-\gamma^{5}\)\widetilde{b}\, .
\end{split}
\end{equation}
Here $\psi$ runs over all the Weyl fermions $\{t_{L},t_{R},b_{L},b_{R},\ttil_{L},\ttil_{R},\widetilde{b}_{L},\widetilde{b}_{R}\}$ and we defined
\be\l{fermionbackground}
\bry{l}
\dst Z_{\psi}(h_{c})=1\,,\qquad y_{t}(h_{c})=y_{t}\,,\qquad m_{t}(h_{c})=\f{y_{t}h_{c}}{\sqrt{2}}\,,\qquad m_{b}(h_{c})=0\,,\\[0.5cm]
\dst  y_{\pit\ttil \widetilde{b} }(h_{c})=\widetilde{y}_{t}\,,\qquad \widetilde{y}_{t}(h_{c})=\f{\widetilde{y}_{t}h_{c}}{f}\f{1}{\sqrt{1-\f{h_{c}^{2}}{f^{2}}}}\,,\qquad m_{\ttil }(h_{c})=\f{ \widetilde{y}_{t}f}{\sqrt{2}}\sqrt{1-\f{h_{c}^{2}}{f^{2}}}\qquad\, m_{\widetilde{b}}(h_{c})=0\,.
\ery
\ee

We can now compute the running masses of the top and Twin top in the
background:
\bes\l{renmassesfinal}
\begin{align}
& \dst M_{t}(h_{c},t) \approx m_{t}(h_{c})\left[1+\(\f{g_{S}^{2}}{4\pi^{2}}-\f{3y_{t}(h_{c})^{2}}{4(4\pi)^{2}Z_{\hat{h}}(h_{c})}\)t+\frac{22 \,g_{S}^{4}}{(4\pi)^4}t^2\right]\,,\l{renmasstopfinal}\vspace{2mm}\\
& \dst M_{\ttil}(h_{c},t) \approx m_{\ttil}(h_{c})\left[1+\(\f{\widetilde{g}_{S}^{2}}{4\pi^{2}}-\f{3\widetilde{y}_{t}(h_{c})^{2}}{4(4\pi)^{2}Z_{\hat{h}}(h_{c})}\)t+\frac{22\, \widetilde{g}_{S}^{4}}{(4\pi)^4}t^2\right]\,,\l{renmassTwintopfinal}
\end{align}
\ees
with $t=\log(m_{*}^{2}/\mu^{2})$.  The values of the parameters in
the Higgs background are simply read
off from the effective lagrangian at the scale $m_{*}$.  Notice that
in the background, only $y_{t}$ and $\widetilde{y}_{t}$ are
scale-dependent, while $h_{c}$ is ``frozen''; hence the running of
the quark mass is identical to that of the corresponding Yukawa coupling.

We can now substitute Eqs.~\eqref{renmassesfinal} into the RG equation \eqref{CWimproveda}, integrate the result and expand at the required order in $h_{c}/f$ to get
\begin{equation}
\l{CWimproved}
\begin{split}
V_{f}(h_{c},t)^{\text{impr}} =&\, \f{N_{c}}{64\pi^{2}}\Bigg\{\bigg[y_{t}^{4}h_{c}^{4}+\widetilde{y}_{t}^{4}f^{4}\(1-\f{h_{c}^{2}}{f^{2}}\)^{2}\bigg]t+\f{y_{t}^{4}h_{c}^{4}}{2}\(\f{g_{S}^{2}}{\pi^{2}}-\f{3y_{t}^{2}}{(4\pi)^{2}}\)t^{2}\\
& +\f{23 }{96\pi^{4}}\bigg[g_{S}^{4}y_{t}^{4}h_{c}^{4}+\widetilde{g}_{S}^{4}\widetilde{y}_{t}^{4}f^{4}\(1-\f{h_{c}^{2}}{f^{2}}\)^{2}\bigg]t^{3}\\[0.1cm]
& + \f{\widetilde{y}_{t}^{4}f^{4}}{32\pi^{2}}\(1-\f{h_{c}^{2}}{f^{2}}\)^{2}\bigg[16\widetilde{g}_{S}^{2}-3\widetilde{y}_{t}^{2}\f{h_{c}^{2}}{f^{2}}\(1-\(1+c_{H}\)\f{h_{c}^{2}}{f^{2}}\)\bigg]t^{2}\Bigg\}\,.
\end{split}
\end{equation}
This is the improved effective potential written in terms of parameters computed at the scale $m_{*}$.
We can now treat $h_{c}$ as a quantum field, fluctuating arount its minimum $h_{c}=h+\<h\>$. We then have to take into account that the potential \eqref{CWimproved} has been computed in a non-canonical basis for $h_{c}$, where its kinetic term coefficient, including only the leading logatithmic running, is
\be\l{higgswfr}
Z_{h_{c}}(t)=1+\f{3y_{t}^{2}}{(4\pi)^{2}}t\,.
\ee
Notice that only the top contributes to the one-loop running of the $h_{c}$ wavefunction, since the Twin top only couples quadratically to $h_{c}$, giving rise to a one-loop contribution that is not logarithmically divergent. After normalizing $h_{c}$ using Eq.~\eqref{higgswfr}, including a $Z_{2}$ breaking mass term necessary to achieve a viable minimum and minimizing the potential, $h_{c}$ acquires a vacuum expectation value $h_{c}=h+\<h\>$. This makes the wavefunction of $h$ (the Higgs field in the minimum) non-canonical again, due to the presence of $c_{H}$:
\be
Z_{h}(\<h\>)=1+c_{H}\f{\<h\>^{2}}{f^{2}}\,.
\ee
After normalizing $h$ taking into account also this last effect, and
including the LO gauge contribution, the desired contribution to the
Higgs mass reads:
\begin{equation}
\l{hmass2}
\begin{split}
\delta m_{h}^{2}|_{\text{IR}} =
 \f{3v^{2}}{8\pi^{2}} \Bigg\{ 
&\(y_{t}^{4} \, t+y_{\tilde t}^{4}\, \tilde t\)\({\red 1}-{\color{verdes}
  c_{H}\xi}+{\blue \(c_{H}^2-d_{H}\)\xi ^2}\) - 
{\red \f{1}{16} \left[\left(3g_2^4 + 2g_1^2 g_2^2 + g_1^4 \right)t + 3\t{g}_2^4 \, \tilde t \right] } \\
&+\f{1}{32\pi^{2}}\(16y_{t}^{4}g_{S}^{2}\, t^2 + 16y_{\tilde t}^{4}\widetilde{g}_{S}^{2} \, \tilde t^2 \)\({\color{verdes}1}-{\blue c_{H}\xi}\) \\[0.2cm]
& +\f{1}{32\pi^{2}} \big(
-{\color{verdes} 15y_{t}^{6} \, t^2 -12y_{t}^{2}\widetilde{y}_{t}^{4}\, \tilde t^2 + 3y_{\tilde t }^{6} \, \tilde t^2\(1+c_{H}\)}\big)\\[0.1cm]
&  + {\blue \f{23}{96 \pi^4} \(y_{t}^{4}g_{S}^{4}\, t^3 +y_{\tilde t}^{4}\widetilde{g}_{S}^{4} \, \tilde t^3\)}\Bigg\}\,,
\end{split}
\end{equation}
where we have highlighted the various contributions present in
Eq.~\eqref{potexpansion} with the corresponding colors and set
$t=\log m_{*}^{2}/m_t^{2}$, $\tilde t = \log m_{*}^{2}/m_{\tilde t}^{2}$.
The Higgs vev has been defined by fixing the $W$ mass according to
\be
m_{W}^{2}=\f{g^{2}\<h\>^{2}}{4}\,,\qquad \Longrightarrow
\<h\>^{2}=v^{2}=\f{1}{\sqrt{2}G_{F}}=246\text{ GeV}\, .
\ee

A numerical determination of the IR contribution $\delta m_h^2|_{IR}$ can be obtained by making use of the experimental value of the top quark mass to fix $y_{t}(m_{*})$ and $\widetilde{y}_{t}(m_{*})$. In fact, the 1-loop RG equation for $y_{t}$ is decoupled from the Twin sector at the order we are interested in and can be easily solved. Including only the orders required to match our calculation of the Higgs mass squared in Eq.~\eqref{hmass2} we get:
\begin{equation}
\begin{aligned}
& \dst y_{t}(m_*) = y_{\tilde t}(m_{*})= y_{t}(m_{t})\left[1 - \left( \frac{ g_{S}(m_{t})^{2}}{4\pi^{2}}- \frac{9  y_{t}(m_{t})^{2}}{64\pi^{2}}  \right) \log\frac{m_{*}^{2}}{m_{t}^{2}}+ \frac{22g_{S}(m_{t})^{4}}{(4\pi)^{4}}\log^{2}\frac{m_{*}^{2}}{m_{t}^{2}}\right]\, ,\vspace{2mm}\\
& \dst g_{S}(m_*) = \widetilde{g}_{S}(m_{*})= g_{S}(m_{t})\left[1 - \f{7g_{S}(m_{t})^{2}}{32\pi^{2}} \log\frac{m_{*}^{2}}{m_{t}^{2}}\right]\, ,
\end{aligned}
\end{equation}
which fixes $y_{t}(m_{*})$ in terms of $y_{t}(m_{t})$. As an input to our numerical analysis we use the PDG combination for the top quark pole mass $m_t = 173.21\pm 0.51 \pm 0.71\,$GeV~\cite{Olive:2016xmw}. 
This is converted into the top Yukawa coupling in the  $\overline{\text{MS}}$ scheme $y_t^{\overline{\text{MS}}}(m_t) = 0.936 \pm 0.005$ by making use of Eq.~(62) of Ref.~\cite{Degrassi:2012ry}. We then use $y_{t}(m_{t})=y_t^{\overline{\text{MS}}}(m_t)$. For the strong interaction, we run the parameter measured at the scale of the $Z$ boson
mass, $g_{S}(m_{Z})\sim 1.22$, to the scale $m_{t}$ to obtain $g_{S}(m_{t})$.

\section{Operator analysis of the heavy-vector contribution to $\delta g_{Lb}$}
\label{AppVectorDeltag}

In this appendix we discuss the UV threshold contribution to $\delta g_{Lb}$ generated by the tree-level exchange of the composite vectors $\rho$ (adjoint of
$SO(7)$) and $\rho^X$ (singlet of $SO(7)$) at zero transferred momentum.  
This effect arises at leading order from diagrams with a loop of heavy fermions, as  in Figure \ref{fig:DeltaGDivergent}.
Our simple effective operator analysis will show that the contribution of the $\rho$ identically vanishes, in agreement with the explicit calculation in the simplified model.

An adjoint of $SO(7)$ decomposes under the custodial $SU(2)_L \times SU(2)_R$ as:
\begin{equation}
\bold{21} = ({\bf{3}},{\bf{1}}) + ({\bf{1}},{\bf{3}}) + 3 \times ({\bf{2}},{\bf{2}}) + 3 \times ({\bf{1}},{\bf{1}}).
\end{equation}
The first two representations contain the vector resonances that are typically predicted by ordinary CH models, namely $\rho_L$ and $\rho_R$. 
They mix at tree-level with the $Z$ boson and in general contribute to $\delta g_{Lb}$.  The remaining resonances do not have the right quantum numbers to both mix 
with the $Z$ boson and couple to the left-handed bottom quark due to isospin conservation. 
As a result, only the components $\rho_L$ and $\rho_R$ inside the $\bold{21}$ can give a contribution to $\delta g_{Lb}$ at the 1-loop level.

In order to analyze such effect, we make use of an operator approach. We classify the operators that can be generated at the scale $m_*$ by integrating out the composite 
states,  focusing on those which can modify the $Zb\bar b$ vertex at zero transferred momentum. 
In general,  since an exact $P_{LR}$ invariance  implies vanishing correction to $g_{Lb}$ at zero transferred momentum, any $\delta g_{Lb}$ 
must be generated proportional to some spurionic coupling breaking this symmetry. In our model, the only coupling breaking $P_{LR}$ in the fermion sector is $y_L$, 
and a non-vanishing $\delta g_{Lb}$ arises at order $y_L^4$.
The effective operators can be constructed using the CCWZ formalism in terms of the covariant spurion
\begin{equation}\label{Invariant}
\chi_L = \Sigma^\dagger \Delta^\dagger \Delta \Sigma,
\end{equation}
where $\Delta$ is defined in Eq.~\eqref{SpurionSM}. By construction $\chi_L$ is an hermitian complex matrix.
Under the action of an element $g \in SO(8)$, it transforms as a ${\bf 21}_a+ {\bf 27}_s + {\bf 7}+ {\bf 1} +{\bf 1}$ of $SO(7)$ 
(where the ${\bf 7}$ is complex), and its formal transformation rule is 
\begin{equation}\label{InvariantTransf}
\chi_{L} \rightarrow h({\bf \Pi},g) \, \chi_{L} \, h^{\dagger}({\bf \Pi},g), \qquad\qquad h \in SO(7)\, .
\end{equation}
As a second ingredient to build the effective operators, we uplift the elementary doublet $q_L$ into a ${\bf 7} + {\bf 1}$ representation of $SO(7)$ 
by dressing it with NGBs:
\begin{equation}\label{EffField}
Q_L = (\Sigma^\dagger \Delta^\dagger q_L).
\end{equation}
We will denote with $Q_L^{(7)}$ and $Q_L^{(1)}$ respectively the septuplet and singlet components of $Q_L$. 
Since  $Q_L^{(1)}$ does not contain $b_L$ (it depends only on $t_L$), only $Q_L^{(7)}$ is of interest for the present analysis.
Under an $SO(8)$ transformation
\begin{equation}\label{EffFieldTransf}
Q_L^{(7)} \rightarrow h({\bf \Pi}, g) Q_L^{(7)} \, .
\end{equation}

The effective operators contributing to $\delta g_{Lb}$ can be thus constructed in terms of $\chi_L$, $d_\mu$ and $Q_L^{(7)}$.
We find that the exchange of $\rho_\mu$ in the diagram of Figure \ref{fig:DeltaGDivergent} can generate two independent operators,
\begin{equation}
\label{eq:op21}
O_{21} = \bar Q_L^{(7)} \gamma^\mu T^a Q_L^{(7)} \, \Tr\!\!\left( d_\mu \chi_L T^a\right)\, , \qquad\qquad
O'_{21} = \bar Q_L^{(7)} \gamma^\mu T^a Q_L^{(7)}  \left( d_\mu T^a\chi_L \right)_{88}\, ,
\end{equation}
where $T^a$ is an $SO(8)$ generator in the adjoint of $SO(7)$;
the exchange of $\rho^X_\mu$ gives rise to other two:~\footnote{Additional structures constructed in terms of $d_\mu$ and $\chi_L$
can be rewritten in terms of those appearing in eqs.~\eqref{eq:op21} and \eqref{eq:op1}, hence they do not generate new linearly independent operators.
Notice that $\Tr\!( d_\mu \chi_L T^a) \propto f^{a \hat a \hat b} d^{\hat a}_\mu (\chi_L^{(7)})^{\hat b}$,  
$(d_\mu T^a \chi_L)_{88} \propto f^{a \hat a \hat b} d^{\hat a}_\mu (\chi_L^{(7)*})^{\hat b}$, 
$( d_\mu \chi_L )_{88} =  - d^{\hat a}_\mu (\chi_L^{(7)})^{\hat a}$,
$\Tr\! ( d_\mu \chi_L ) = - d^{\hat a}_\mu (\chi_L^{(7)})^{\hat a} + d^{\hat a}_\mu (\chi_L^{(7)*})^{\hat a}$,
where $\chi_L^{(7)}$ denotes the component of $\chi_L$ transforming as a (complex) fundamental of $SO(7)$.
A similar classification in the context of $SO(5)/SO(4)$ models in Ref.~\cite{Grojean:2013qca} found only one operator, corresponding to the linear combination 
$O'_{1} - O_{1}$.
}
\begin{equation}
\label{eq:op1}
O_{1} = \bar Q_L^{(7)} \gamma^\mu Q_L^{(7)} \, \Tr\!\!\left( d_\mu \chi_L \right)\, , \qquad\qquad
O'_{1} = \bar Q_L^{(7)} \gamma^\mu Q_L^{(7)} \left( d_\mu \chi_L \right)_{88} \, .
\end{equation}
Simple inspection reveals that only the septuplet component of $\chi_L$ contributes in the above equations.
One can easily check that the operators of Eq.~\eqref{eq:op21} give a vanishing contribution to $\delta g_{Lb}$.
In particular, the terms generated by the exchange of the $(2,2)$ and $(1,1)$ components of the $\rho$ give (as expected) an identically vanishing contribution.
Those arising from  $\rho_L$ and $\rho_R$ (obtained by setting $T^a$ in Eq.~\eqref{eq:op21} equal to respectively one of the $(3,1)$ and $(1,3)$ generators) give instead an 
equal and opposite  correction to $g_{Lb}$. This is in agreement with the results of a direct calculation in the simplified model, from 
which one finds that  the contributions from  $\rho_L$ and $\rho_R$ cancel each other.
Finally, a non-vanishing $\delta g_{Lb}$ arises from the operators of Eq.~\eqref{eq:op1} generated by the exchange of $\rho^X$. 
Upon expanding in powers of the Higgs doublet, 
$O_{1}$ and $O'_{1}$ both match the dimension-6 operator $O_{Hq}$ of Eq.~\eqref{eq:OHpsi} and differ only by higher-order terms.

\section{Explicit formulae for the EWPO}\label{AppFormulae}

In this appendix we report the results of our calculation of the electroweak precision observables, in particular we collect here the explicit expression of
the coefficients $a_{UV} $, $a_{IR}$ of Eq.~\eqref{eq:dTpsi} and $b_{UV}$, $c_{UV}$, $b_{IR}$ of Eq.~\eqref{eq:dgLb1loop}. 

Let us start considering the $\widehat{T}$ parameter. For convenience, we split the UV contribution into two parts, 
re-defining $a_{UV}$ as:
\begin{equation}\label{Splita1}
a_{UV} = a_{UV}^{Fin} + a_{UV} ^{Log} \,\log \!\left( M_\Psi^2 \over M_S^2 + f^2 y_R^2  \right).
\end{equation}
The coefficients $a_{UV}^{Fin}$ and $a_{UV} ^{Log}$ are obtained through a straightforward calculation, but their expressions are complicated functions of the Lagrangian 
parameters.
We thus show them only in the  limit $c_L = c_R \equiv c$ and $M_\Psi = M_S \equiv M$, for simplicity.
For $a_{IR}$ we give instead the complete expression. 
We find:
\begin{equation}
\begin{split}
a_{UV}^{Fin} = & \, \frac{1}{12} \left(-\frac{12 M^4}{f^4 y_R^4}+\frac{6 f^2 M^2 y_R^2}{\left(f^2 y_R^2+M^2\right){}^2}+\frac{9 M^6}{\left(f^2 y_R^2+M^2\right){}^3}-8\right) \\[0.1cm]
                    & + \frac{c \left(-5 f^8 y_R^8-2 f^6 M^2 y_R^6+7 f^4 M^4 y_R^4+12 f^2 M^6 y_R^2+4 M^8\right)}{\sqrt{2} f^4 y_R^4 \left(f^2 y_R^2+M^2\right)^2} \\[0.1cm]
                    & +\frac{c^2 \left(f^2 y_R^2+2 M^2\right)^2 \left(3 f^2 y_R^2-5 M^2\right)}{2 f^4 y_R^4 \left(f^2 y_R^2+M^2\right)},\\[0.5cm]
a_{UV}^{Log} = & \,\frac{f^6 y_R^6-f^4 M^2 y_R^4-f^2 M^4 y_R^2-2 M^6}{2 f^6 y_R^6} \\
                     &  + \frac{\sqrt{2} c \left(2 f^6 y_R^6-3 f^4 M^2 y_R^4+3 f^2 M^4 y_R^2+2 M^6\right)}{f^6 y_R^6}
                        + \frac{c^2 \left(f^2 M^4 y_R^2-10 M^6\right)}{f^6 y_R^6}, \\[0.5cm]
a_{IR}=&  \,{1\over 2} + {M_S^2 M_\Psi^2 \over 2(M_S^2 + f^2 y_R^2)^2} + \sqrt{2} ~{c_L M_S M_\Psi + 2 c_R f^2 y_R^2\over M_S^2 + f^2 y_R^2}.
\end{split}
\end{equation}

\begin{figure}[t!]
\begin{center}
\includegraphics[width=1\textwidth]{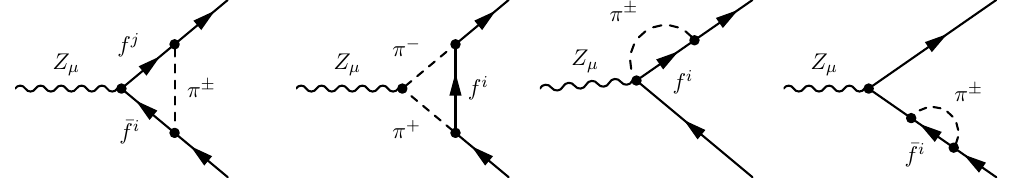}
\end{center}
\caption{\it Diagrams with a loop of fermions and NGBs contributing to the $Z b_L \bar{b}_L$ vertex. Here $f^i$ indicates any fermion (both heavy and light) 
in our simplified model; $\pi^{\pm}$ are the charged NGBs.}
\label{fig:DeltaGLoop}
\end{figure}

\begin{figure}[t!]
\begin{center}
\includegraphics[width=0.35\textwidth]{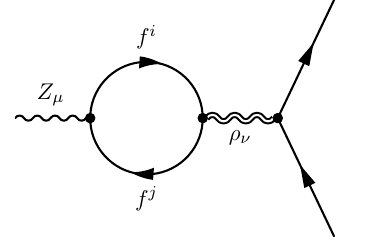}
\end{center}
\caption{\it The diagram contributing to $Zb_L \bar{b}_L$ with a loop of fermions and tree-level exchange of a heavy vector.
Here $f^i$ and $f^j$ denote generic fermions in the theory, both heavy and light, whereas $\rho_\nu$ indicates the heavy vector.}
\label{fig:DeltaGDivergent}
\end{figure}

The derivation of $\delta g_{b_L}$ at 1-loop level is more involved and requires the computation of a series of diagrams. As explained in the text, we focus on those
featuring a loop of fermions and NGBs (see Figure \ref{fig:DeltaGLoop}), and that one with a loop of fermion and the tree-level exchange of a heavy vector (see 
Figure \ref{fig:DeltaGDivergent}).
%
%
The coefficients $c_{UV}$ is generated only by the latter diagram; we find:
\begin{equation}\label{cUV}
c_{UV} = \alpha_{7L} (\alpha_{1R} + \alpha_{7R}) (1+\sqrt{2} c_R) {g^2_{\rho^X} f^2 \over M_{\rho^{X}}^2} { M_{\Psi}^2 \over 2(M_\Psi^2 + y_L^2 f^2)}.
\end{equation}
We remind the reader that in our numerical analysis we use $M_{\rho^{X}}/(g_{\rho^{X}}f)=1$, see footnote \ref{footnotegrho}.
We re-define the other two coefficients as 
\begin{equation}
\begin{split}
b_{IR} & = \delta_{IR} + \bar\delta_{IR} \\
b_{UV} & = \left( \delta_{UV}^{Fin}+ \bar\delta_{UV}^{Fin} \right) + \left( \delta_{UV} ^{Log}  + \bar\delta_{UV} ^{Log}  \right)\,\log\!\left( \frac{M_\Psi^2}{M_S^2 + f^2 y_R^2}\right)\, ,
\end{split}
\end{equation}
where $\delta_{IR}$, $\delta_{UV}^{Fin}$ and $\delta_{UV} ^{Log}$ are generated by the diagrams in Figure \ref{fig:DeltaGLoop} only, 
whereas $\bar\delta_{IR}$, $\bar\delta_{UV}^{Fin}$ and $\bar\delta_{UV}^{Log}$ parametrize 
the correction due to the tree-level exchange of a heavy spin-1 singlet in Figure \ref{fig:DeltaGDivergent}.
As before, we report the expression of the UV parameters in the limit $c_L = c_R \equiv c$, $M_\Psi = M_S \equiv M$, for simplicity; in the case of the coefficients with a bar,
generated by the diagram of Figure \ref{fig:DeltaGDivergent}, we further set $\alpha_{7L}=\alpha_{1L}$ and $\alpha_{7R}=\alpha_{1R}$. We find:
\begin{equation}
\begin{split}
\delta_{UV}^{Fin} = & \, \frac{-2 f^6 y_R^6-4 f^4 M^2 y_R^4-4 f^2 M^4 y_R^2+M^6}{12 \left(f^2 y_R^2+M^2\right){}^3}
                               - \frac{c \left(f^6 y_R^6+4 f^4 M^2 y_R^4-2 f^2 M^4 y_R^2+M^6\right)}{6 \sqrt{2} f^2 y_R^2 \left(f^2 y_R^2+M^2\right){}^2} \\[0.1cm]
                           & + \frac{1}{6} c^2 M^2 \left(\frac{3}{f^2 y_R^2+M^2}-\frac{2}{f^2 y_R^2}\right)-\frac{c^3 M^2}{3 \sqrt{2} f^2 y_R^2},\\[0.4cm]
\delta_{UV}^{Log} = & \, \frac{1}{6} -\frac{M^2}{6 f^2 y_R^2}-\frac{c^3 M^4}{3 \sqrt{2} f^4 y_R^4}-\frac{c^2 M^4}{3 f^4 y_R^4}
                                 -\frac{c \left(-f^4 y_R^4+2 f^2 M^2 y_R^2+M^4\right)}{6 \sqrt{2} f^4 y_R^4},\\[0.4cm]
\bar\delta_{UV}^{Fin} = & \,\frac{f^2 M^2 \alpha _{1L} g_{\rho ^X}^2 \left(f^4 y_R^4 \left(\alpha _{1L}+2 \alpha _{1R}\right) +f^2 M^2 y_R^2 \left(3 \alpha _{1L}+8 \alpha _{1R}\right)
                                      +2 M^4 \left(\alpha _{1L}+\alpha _{1R}\right)\right)}{4 M_{\rho^{X}}^2 \left(f^2 y_L^2+M^2\right) \left(f^2 y_R^2+M^2\right){}^2}\\[0.1cm]
                                  &+ \frac{c f^2 M^2 \alpha _{1L} g_{\rho ^X}^2 \left(f^2 y_R^2 \left(2 \alpha _{1L}+\alpha _{1R}\right)
                                    +2 M^2 \alpha _{1L}\right)}{\sqrt{2} M_{\rho^{X}}^2 \left(f^2 y_L^2+M^2\right) \left(f^2 y_R^2+M^2\right)},\\[0.4cm]
\bar\delta_{UV}^{Log} = & \, \frac{c M^4 \alpha _{1L} g_{\rho ^X}^2 \left(\alpha _{1L}-\alpha _{1R}\right)}{\sqrt{2} y_R^2 M_{\rho^{X}}^2 \left(f^2 y_L^2+M^2\right)}
                                   +\frac{f^2 M^2 \alpha _{1L} \alpha _{1R} g_{\rho ^X}^2}{2 M_{\rho^{X}}^2 \left(f^2 y_L^2+M^2\right)}\, .
\end{split}
\end{equation}
For the IR coefficients we give instead the full expressions. We find:
\begin{equation}
\begin{split}
\delta_{IR} & =  {1\over 6} + {M_S^2 M_\Psi^2 \over 6(M_S^2 + f^2 y_R^2)^2} + \sqrt{2} ~{c_L M_S M_\Psi \over 3(M_S^2 + f^2 y_R^2)}
                      + \sqrt{2}~{ c_R f^2 y_R^2\over 12(M_S^2 + f^2 y_R^2)}\, ,\\[0.2cm]
\bar\delta_{IR} & = \alpha_{7L}\alpha_{1R} {g_{\rho^X}^2\over M_{\rho^{X}}^2}\frac{ f^4 M_{\Psi }^2y_R^2 }{2\left(f^2 y_L^2+M_{\Psi }^2\right) \left(f^2 y_R^2+M_S^2\right)}\, .
\end{split}
\end{equation}
Notice that the IR corrections $a_{IR}$ and $b_{IR}$ are related to each other and parametrize the running of the effective coefficients $\bar c_{Hq}$, $\bar c_{Hq}'$ and 
$\bar c_{Ht}$, as explained  in the main text. 

Finally, we report the contribution to $\widehat{S}$ generated in our simplified model by loops of heavy fermions. 
We do not include this correction in our electroweak fit, 
because in the perturbative region of the parameter space it is sub-dominant with respect to the tree-level shift of Eq.~\eqref{eq:dSrho}. 
Rather, we use this computation as an additional way to estimate the perturbativity bound, as discussed in Sec.~\ref{sec:perturbativity}. 
Analogously to what we did for $\widehat{T}$ and $\delta g_{Lb}$, we parametrize the fermionic contribution to $\widehat{S}$ as:
\begin{equation}\label{DSPsi}
\begin{split}
\Delta \widehat{S}_{\Psi} = & \,  {g_2 ^2 \over 8 \pi^2} \xi \left[ (1- c_L^2 -c_R^2)\log {m_*^2 \over M_\Psi^2 }
                                            +(1- \widetilde{c}_L^2 -\widetilde{c}_R^2) \log {m_*^2 \over \widetilde{M}_\Psi^2 } \right] \\[0.1cm]
                                         & + {g_2^2\over 16 \pi^2}  \xi \left[ s_{UV}^{Fin} + {s}_{UV}^{Log} \log\!\left( \frac{M_\Psi^2}{M_S^2 + f^2 y_R^2} \right)
                                             + \widetilde{s}_{UV}^{Fin} + \widetilde{s}_{UV}^{Log} \log\!\left( \frac{\tilde M_\Psi^2}{\tilde M_S^2 + f^2 \tilde y_R^2} \right)\right] \\[0.1cm]
                                          &+ {g_2^2\over 16 \pi^2}  \xi s_{IR} {y_L^2 f^2 \over M_{\Psi}^2}\log {M_1^2 \over m_t^2  }  \, .
\end{split}
\end{equation}
Terms in the first line are logarithmically sensitive to the UV cut-off,  the second line contains the UV threshold corrections, while the IR running appears in the
third line.
The UV thresholds include a contribution from the Twin composites $\widetilde{\Psi}_7$ and $\widetilde{\Psi}_1$, parametrized by $\widetilde{s}_{UV}^{Fin}$ and 
$\widetilde{s}_{UV}^{Log}$.  At leading order in $y_L$, by virtue of the Twin parity invariance of the strong sector,
such a contribution can be obtained from that of $\Psi_7$ and $\Psi_1$ (i.e.~from $s_{UV}^{Fin}$ and $s_{UV}^{Log}$) by simply interchanging the tilded quantities with the 
un-tilded ones. Higher orders in $y_L$ break this symmetry and generate different corrections in the two sectors. We performed the computation of the UV coefficients 
for $y_L=0$, whereas $s_{IR}$ is derived up to order $y_L^2$. We find:
\begin{equation}
\begin{split}
s_{UV}^{Fin} = &\,  \frac{1}{2}-\frac{6 c_L c_R M_S M_{\Psi } \left(f^2 y_R^2+M_S^2+M_{\Psi }^2\right)}{ \left(f^2 y_R^2+M_S^2-M_{\Psi }^2\right)^2}
                           +{(c_R^2 +c_L^2)\over 6}\left(\frac{24 M_S^2 M_{\Psi }^2}{\left(f^2 y_R^2+M_S^2-M_{\Psi }^2\right)^2}-7\right), \\[0.5cm]
s_{UV}^{Log} = & \, -{2\left(M_S^2+f^2 y_R^2\right)\over \left(M_S^2 -M_\Psi^2+f^2 y_R^2\right)^3}\Big[6 c_L c_R M_S M_{\Psi }^3
                        +c_R^2 M_S^2 \left(f^2 y_R^2+M_S^2-3 M_{\Psi }^2\right)\\[0.1cm]
                    & + c_L^2 \left(f^2 y_R^2+M_S^2\right) \left(f^2 y_R^2+M_S^2-3 M_{\Psi }^2\right)\Big], \\[0.5cm]
s_{IR} = & \, \frac{M_S^2 M_{\Psi }^4-f^2 y_L^2 \left(\left(f^2 y_R^2+M_S^2\right) \left(M_S^2-f^2 y_R^2\right)+M_S^2 M_{\Psi }^2\right)
               +M_{\Psi }^2 \left(f^2 y_R^2+M_S^2\right){}^2}{6 M_{\Psi }^2 \left(f^2 y_R^2+M_S^2\right)^2}\\[0.1cm]
           & - c_R\frac{2 \sqrt{2}f^2   y_R^2 \left(M_{\Psi }^2-f^2y_L^2\right) }{3  M_{\Psi }^2 \left(f^2 y_R^2+M_S^2\right)}
              -c_L \frac{\sqrt{2}  M_S \left(M_{\Psi }^2-f^2 y_L^2\right) }{3 M_{\Psi } \left(f^2 y_R^2+M_S^2\right)}\, .
\end{split}
\end{equation} 

\section{The EW fit}\label{ewfit}

For our analysis of the electroweak observables we make use of the fit to the parameters $\epsilon_{1,2,3,b}$~\cite{Altarelli:1990zd,Altarelli:1991fk,Altarelli:1993sz} 
performed in Ref.~\cite{Ciuchini:2014dea}  (see also Ref.~\cite{Ciuchini:2013vb}). 
The central values there obtained for the shifts $\Delta \epsilon_i \equiv \epsilon_i - \epsilon_i^{SM}$ and the corresponding correlation matrix are:
\begin{equation}
\label{eq:fitofepsilons}
\begin{split}
\Delta \epsilon_1 = & \displaystyle 0.0007 \pm 0.0010\\
\Delta \epsilon_2 = & \displaystyle -0.0001 \pm 0.0009\\
\Delta \epsilon_3 = & \displaystyle 0.0006 \pm 0.0009\\
\Delta \epsilon_b = & \displaystyle 0.0003 \pm 0.0013\\
\end{split}
\hspace{1cm}
\rho = 
\begin{pmatrix}
 1 & 0.8 & 0.86 & -0.33 \\
 0.8 & 1 & 0.51 & -0.32 \\
 0.86 & 0.51 & 1 & -0.22 \\
 -0.33 & -0.32 & -0.22 & 1
\end{pmatrix}\, .
\end{equation}
We can directly relate $\Delta \epsilon_1$ to $\Delta \widehat{T}$ and $\Delta \epsilon_3$ to $\Delta \widehat{S}$ by using the results of 
Refs.~\cite{Contino:2015mha,Barbieri:2004p1607}, and furthermore $\Delta \epsilon_b = -2 \delta g_{b_L}$. 
We set $\Delta \epsilon_2 = 0$ in our study, since its effect is sub-dominant in our model as well as
in CH models~\cite{Barbieri:2004p1607}.
We thus make use of Eq.~\eqref{eq:fitofepsilons} to perform a $\chi^2$ test of the compatibility of our predictions with the experimental constraints. 
The $\chi^2$ function is defined as usual:
\begin{equation}
\chi^2 = \sum_{ij} (\Delta \epsilon_i - \mu_i) (\sigma^2)_{ij}^{-1} (\Delta \epsilon_j - \mu_j), \qquad\quad (\sigma)^2_{ij} = \sigma_i \rho_{ij} \sigma_j,
\end{equation}
where $\mu_i$ and $\sigma_i$ denote respectively the mean values and the standard deviations of Eq.~\eqref{eq:fitofepsilons}, while
$\Delta \epsilon_i$ indicates the theoretical prediction for each EW observable computed in terms of the Lagrangian parameters. 
After deriving the $\chi^2$, we perform a fit by scanning over the points in our parameter space keeping only those for which 
$\Delta \chi^2 \equiv \chi^2 - \chi^2_{min} < 7.82$, the latter condition corresponding to the $95\%$ Confidence Level with 3 degrees of freedom. 
Using this procedure, we  convert the experimental constraints into bounds over the plane $(M_\Psi, \xi)$. 

\section{Estimates of the perturbativity bound}
\label{Apppertbound}

This appendix contains details on the derivation of the perturbative
limits discussed in Section \ref{sec:perturbativity}.
As explained there,  we consider the processes $\pi^{a}\pi^{b}\to \pi^{c}\pi^{d}$ and $\pi^{a}\pi^{b}\to \overline{\psi}^{c}\psi^{d}$, 
where $\psi = \{ \Psi_{\bf{7}}, \tilde\Psi_{\bf{7}} \}$ and all indices transform under the fundamental representation of the unbroken $SO(7)$.
In order to better monitor how the results depend on the multiplicity
of NGBs and fermions, we perform the calculation for a generic $SO(N)/SO(N-1)$
coset with $N_f$ composite fermions $\psi$ in the fundamental of
$SO(N-1)$. Taking $N=8$ and $N_f = 2\times 3 = 6$ thus reproduces the
simplified model of Section \ref{sec:EffLag}.

The perturbative limits are obtained by first expressing the scattering amplitudes in terms of components with definite $SO(N-1)$ quantum numbers.
In the case of $SO(7)$ the product of two fundamentals decomposes as ${\mathbf 7}\otimes {\mathbf 7}={\mathbf 1}\oplus{\mathbf {21}}_{a}\oplus{\mathbf {27}}_{s}$, 
where the indices $a$ and $s$ label respectively the  anti-symmetric and symmetric two-index representations. 
A completely analogous decomposition holds in the general case of $SO(N)/SO(N-1)$,\footnote{One has 
${\mathbf N}\otimes {\mathbf N}={\mathbf 1}\oplus [{\mathbf {N(N-1)/2}}]_{a}\oplus [{\mathbf {N(N+1)/2 -1}}]_{s}$.}
but for simplicity  we will use the $SO(7)$ notation in the following to label the various components.
The leading tree-level contributions  to the scattering amplitudes arise 
from the contact interaction generated by the expansion of the NGB kinetic term of Eq.~\eqref{Lsigmamodel} and from the NGB-fermion interactions of Eq.~\eqref{LCompF}. 
The structure of the corresponding vertices implies that the four-NGB amplitude has components in all three irreducible representations of $SO(N-1)$
and contains all partial waves. The amplitude with two NGBs and two fermions, instead, has only the anti-symmetric component of $SO(N-1)$
and starts with the $p$-wave.
At energies much larger than all masses the amplitudes read
\begin{equation}
\begin{split}
\mathcal M(\pi^{a}\pi^{b}\to \pi^{c}\pi^{d}) & = \frac{s}{f^{2}}\delta^{ab}\delta^{cd}+\frac{t}{f^{2}}\delta^{ac}\delta^{bd}+\frac{u}{f^{2}}\delta^{ad}\delta^{bc}\,,\\[0.1cm]
\mathcal M(\pi^{a}\pi^{b}\to \overline{\Psi}_{7_L}^{c}\Psi_{7_L}^{d}) & =\frac{s}{2f^{2}}\sin\theta (\delta^{ac}\delta^{bd}-\delta^{ad}\delta^{bc})\,,\\[0.1cm]
\mathcal M(\pi^{a}\pi^{b}\to \overline{\Psi}_{7_R}^{c}\Psi_{7_R}^{d}) & =\frac{s}{2f^{2}}\sin\theta(\delta^{ac}\delta^{bd}-\delta^{ad}\delta^{bc})\,.
\end{split}
\end{equation}
They decompose into irreducible representations of $SO(N-1)$ as follows:
\begin{equation}
\begin{split}
\mathcal M^{(\mathbf{1})}(\pi^{a}\pi^{b}\to \pi^{c}\pi^{d}) & =(N-2)\frac{s}{f^{2}}\,, \\
\mathcal M^{(\mathbf{21})}(\pi^{a}\pi^{b}\to \pi^{c}\pi^{d}) & =\frac{s}{f^{2}}\cos\theta\,, \\
\mathcal M^{(\mathbf{27})}(\pi^{a}\pi^{b}\to \pi^{c}\pi^{d}) & =-\frac{s}{f^{2}}\,,
\end{split}
\qquad
\begin{split}
\mathcal M^{(\mathbf{21})}(\pi^{a}\pi^{b}\to \overline{\Psi}_{7_L}^{c}\Psi_{7_L}^{d}) & =\frac{s}{2f^{2}}\sin\theta \,,\\
\mathcal M^{(\mathbf{21})}(\pi^{a}\pi^{b}\to  \overline{\Psi}_{7_R}^{c}\Psi_{7_R}^{d}) & =\frac{s}{2f^{2}}\sin\theta\,.
\end{split}
\end{equation}
Performing a partial wave decomposition we get
\begin{equation}
\mathcal{M}^{(\mathbf{r})}=\sum_{\lambda_{i},\lambda_{f}}\mathcal{M}^{(\mathbf{r})}_{\lambda_{i},\lambda_{f}}=16\pi k^{(i)}k^{(f)} \sum_{j=0}^{\infty}a_{j}^{(\mathbf{r})}(2j+1)\sum_{\lambda_{i},\lambda_{f}}D^{j}_{\lambda_{i},\lambda_{f}}(\theta)\,,
\end{equation}
where $\lambda_{i},\lambda_{f}$ are the initial and final state total helicities, and 
$k^{(i)}(k^{(f)})$ is equal to either $1$ or $\sqrt{2}$ depending on whether the two  particles in the initial (final) state are distinguishable or identical respectively.
In the above equation $\mathcal{M}^{(\mathbf{r})}$ should be considered as a matrix acting on the space of different channels.
The coefficients $a_{j}^{(\mathbf{r})}$ are given by
\begin{equation}
a_{j}^{(\mathbf{r})}=\frac{1}{32\pi k^{(i)}k^{(f)}}\int_{0}^{\pi}d\theta \sum_{\lambda_{i},\lambda_{f}}\mathcal{D}^{j}_{\lambda_{i},\lambda_{f}}(\theta)\mathcal{M}^{(\mathbf{r})}_{\lambda_{i},\lambda_{f}}\,.
\end{equation}
and act as matrices  on the space of (elastic and inelastic) channels with 
total angular momentum $j$ and $SO(N-1)$ irreducible representations $\mathbf{r}$.
They can be rewritten as a function of the scattering phase as
\begin{equation}
a_j^{(\mathbf{r})} = {e^{2i \delta_j^{(\mathbf{r})}}-1\over 2i} \sim  \delta_j^{(\mathbf{r})}.
\end{equation}
Our NDA estimate of the perturbativity bound is derived by requiring this phase to be smaller than maximal:
\begin{equation}\label{Constraint}
|\delta_j^{(\mathbf{r})}| < {\pi \over 2}\qquad \Longrightarrow\qquad |a_j^{(\mathbf{r})}| < {\pi \over 2}
\end{equation}

Let us consider first the case $\mathbf{r} = \mathbf{1}$, corresponding to the amplitude singlet of $SO(N-1)$. The only contribution comes from the four-NGB  channel.
Since the helicities of the initial and final states are all zeros, in this particular case the Wigner functions $\mathcal{D}^{j}_{\lambda_{i},\lambda_{f}}(\theta)$ reduce to 
the Legendre polynomials:
\begin{equation}
a^{(\mathbf{1})}_j = {1\over 64 \pi} \int^\pi_0 d \theta P_j(\cos \theta) \mathcal{M}^{(\mathbf{1})}.
\end{equation}
The first and strongest perturbativity constraint comes from the $s$-wave amplitude, which corresponds to $j= 0$. We find:
\begin{equation}
\label{eq:a01}
a_0^{(\mathbf{1})} = {N-2\over 32 \pi} {s\over f^2}\,,
\end{equation}
where $N=8$ in our case.
From Eqs.~\eqref{Constraint} and~\eqref{eq:a01}, one obtains the constraint of Eq.~\eqref{FirstLimit}.

We analyze now the constraint from the scattering in the anti-symmetric representation, $\mathbf{r}=\mathbf{21}$. In this case, both the NGB and the fermion channels 
contribute; the process $\pi\pi\to \pi\pi$ is however independent of the fermion and Goldstone multiplicities and can be neglected in the limit of $N$ and $N_f$. 
The process involving fermions is a function of $N_f$ and generates a perturbative limit which is comparable and complementary to the previous one. We have:
\begin{equation}
a_j^{(\mathbf{21})} =\sum_{\lambda_{f}=\pm 1} {1\over 32 \pi} \int^\pi_0 d \theta ~\mathcal{D}^j_{0,\lambda_{f}} (\theta)\mathcal{M}^{(\mathbf{21})}_{0,\lambda_{f}}.
\end{equation}
As anticipated, this equation vanishes for $j=0$, so that the strongest constraint is now derived for $p$-wave scattering, with $j=1$. We have
\begin{equation}
\label{eq:a121}
a_1^{(\mathbf{21})} = \frac{N_f}{24 \sqrt{2} \pi} \frac{s}{f^2}\, .
\end{equation}
From Eqs.~\eqref{Constraint} and~\eqref{eq:a121} follows the constraint of Eq.~\eqref{SecondLimit}.

\bibliographystyle{mine}
\bibliography{Paper}

\end{document}